\documentclass{vldb}
\usepackage{amsmath}  
\usepackage{url}

\usepackage[noconfig]{ntheorem}
\newdef{definition}{Definition}
\newdef{example}{Example}

\newtheorem{theorem}{Theorem}
\newtheorem{proposition}{Proposition}

\usepackage{bbold}

\usepackage{float}
\usepackage{xspace}
\usepackage[x11names,svgnames]{xcolor}
\usepackage{xcolor-solarized}
\usepackage{colonequals}
\usepackage{multirow}
\usepackage{amssymb}
\usepackage{tikz}
\usetikzlibrary{decorations}
\usepgflibrary{decorations.pathmorphing}
\usetikzlibrary{arrows}
\usetikzlibrary{patterns}
\usetikzlibrary{calc}

\newcommand{\C}{\mathcal{C}}
\newcommand{\N}{\mathcal{N}}
\renewcommand{\S}{\mathcal{S}}

\newcommand{\I}{\mathcal{I}}
\renewcommand{\L}{\mathcal{L}}
\newcommand{\U}{\mathcal{U}}
\newcommand{\B}{{\ensuremath{\_\mathsf{b}}}}
\renewcommand{\P}{\mathcal{P}}
\newcommand{\Inv}{\mathit{Inv}}
\newcommand{\minus}{\mathop{\setminus}}

\newcommand{\BisimRefine}{\mathit{BisimRefine}}

\newcommand{\recolor}{\mathit{recolor}}
\newcommand{\reweight}{\mathit{reweight}}
\newcommand{\outcolor}{\mathit{out\text{-}color}}

\newcommand{\out}{\mathit{out}}
\newcommand{\dist}{\mathit{dist}}

\newcommand{\Literals}{\mathit{Literals}}
\newcommand{\Uris}{\mathit{URIs}}
\newcommand{\Blank}{\mathit{Blank}}
\newcommand{\Blanks}{\mathit{Blanks}}

\newcommand{\Unaligned}{\mathit{Unaligned}}

\newcommand{\Bisim}{\mathsf{Bisim}}
\newcommand{\Propagate}{\mathsf{Propagate}}
\newcommand{\Deblank}{\mathsf{Deblank}}
\newcommand{\Trivial}{\mathsf{Trivial}}
\newcommand{\Hybrid}{\mathsf{Hybrid}}
\newcommand{\Dist}{\mathsf{Dist}}
\newcommand{\Edit}{\mathsf{Edit}}
\newcommand{\Overlap}{\mathsf{Overlap}}
\newcommand{\OverlapMatch}{\mathsf{OverlapMatch}}
\newcommand{\freq}{\mathit{freq}}
\newcommand{\overlap}{\mathit{overlap}}
\newcommand{\diff}{\mathit{diff}}
\newcommand{\Align}{\mathit{Align}}

\newcommand{\Enrich}{\mathsf{Enrich}}

\floatstyle{ruled}
\newfloat{float@algorithm}{H}{loa}
\floatname{float@algorithm}{Algorithm}
\newcounter{LineCounter@algorithm} 

\newenvironment{BasicCommands@algorithm}{%
  \newcommand{\TAB}{\makebox[4ex][r]{}}%
  \newcommand{\FUNCTION}{\textbf{function}\xspace}%
  \newcommand{\IF}{\textbf{if}\xspace}%
  \newcommand{\THEN}{\textbf{then}\xspace}%
  \newcommand{\UNTIL}{\textbf{until}\xspace}%
  \newcommand{\FOR}{\textbf{for}\xspace}%
  \newcommand{\DO}{\textbf{do}\xspace}%
  \newcommand{\RETURN}{\textbf{return}\xspace}%
}{%
}

\newenvironment{algorithm*}%
{%
\begin{BasicCommands@algorithm}%
\list{}{\itemindent 0em%
        \listparindent\itemindent
        \rightmargin  \leftmargin}%
\item\relax
}{%
\endlist
\end{BasicCommands@algorithm}%
}

\newenvironment{algorithm}%
{%
  \newcommand{\ResetLineCounter}{%
    \setcounter{LineCounter@algorithm}{0}%
  }%
  \newcommand{\LN}{%
    \makebox[0pt][l]{%
      \makebox[0pt][l]{%
        \addtocounter{LineCounter@algorithm}{1}%
      }%
      \makebox[10.75pt][r]{
        \fontsize{7}{10}%
        \selectfont%
        \arabic{LineCounter@algorithm}:%
      }%
    }%
    \TAB%
  }%
  \begin{BasicCommands@algorithm}%
  \begin{float@algorithm}%
    \ResetLineCounter%
}%
{%
  \end{float@algorithm}%
  \end{BasicCommands@algorithm}%
}

\def\qed {{                
   \parfillskip=0pt        
   \widowpenalty=10000     
   \displaywidowpenalty=10000  
   \finalhyphendemerits=0  
   \leavevmode             
   \unskip                 
   \nobreak                
   \hfil                   
   \penalty50              
   \hskip.2em              
   \null                   
   \hfill                  
   $\square$
   \par}}                  

\newcommand{\eat}[1]{}

\colorlet{Ca}{solarized-violet!66!white}
\colorlet{Cb}{solarized-yellow!66!white}
\colorlet{Cu}{solarized-blue!33!white}
\colorlet{Cw}{solarized-green!33!white}
\colorlet{Cv}{solarized-red!33!white}
\colorlet{Cp}{solarized-orange!50!white}
\colorlet{Cr}{solarized-magenta!50!white}
\colorlet{Cq}{solarized-cyan!50!white}
\colorlet{CB}{solarized-base03}
\colorlet{Ha}{solarized-blue!50!solarized-base03}
\colorlet{Haa}{Ha}
\colorlet{Hb}{solarized-magenta!50!solarized-base03}
\colorlet{Hc}{solarized-green!50!solarized-base03}
\colorlet{Hct}{solarized-green!50!solarized-base3}
\colorlet{Hd}{solarized-yellow!75!solarized-base03}
\colorlet{Hdt}{solarized-yellow!50!solarized-base3}
\colorlet{He}{solarized-red!50!solarized-base3}
\colorlet{Hf}{solarized-green!50!solarized-base3}
\colorlet{Hg}{solarized-orange!50!solarized-base3}
\colorlet{Hgg}{solarized-orange!50!solarized-base3}
\colorlet{Hh}{solarized-orange!50!solarized-base03}
\colorlet{Hht}{solarized-orange!50!solarized-base3}
\colorlet{Hbb}{solarized-red!50!solarized-base03}
\colorlet{Hi}{solarized-green!50!solarized-base2}
\colorlet{Hit}{solarized-green!50!solarized-base01}
\colorlet{Hcc}{solarized-green!75!solarized-base03}
\colorlet{Hj}{solarized-violet!50!solarized-base2}

\colorlet{Fdbfill}{solarized-base3!25!white}
\colorlet{Fdbdraw}{solarized-base00}
\colorlet{Ftrivial}{solarized-violet}
\colorlet{Fbisimulation}{solarized-green}
\colorlet{Fsimilarity}{solarized-red}
\colorlet{Fdeblank}{solarized-cyan!40!solarized-blue}
\colorlet{Fhybrid}{solarized-cyan!40!solarized-green}
\colorlet{Fedit}{solarized-yellow}
\colorlet{Fpropagation}{solarized-red}

\colorlet{Ww}{solarized-violet}
\colorlet{Wu}{solarized-blue}
\colorlet{Wa}{solarized-cyan!75!solarized-base3}
\colorlet{Wb}{solarized-green}
\colorlet{Wc}{solarized-yellow!75!yellow}
\colorlet{Wabc}{solarized-orange}
\colorlet{Wv}{solarized-red!75!solarized-magenta}

\tikzset{ETrivial/.style={fill=solarized-magenta!50!solarized-violet}}
\tikzset{ETrivialP/.style={pattern=north west lines, pattern color=solarized-base03}}
\tikzset{EHybrid/.style={fill=solarized-green!65!solarized-yellow}}
\tikzset{EHybridP/.style={pattern=north east lines, pattern color=solarized-base03}}
\tikzset{EOverlap/.style={fill=solarized-yellow!75!solarized-orange}}
\tikzset{EOverlapP/.style={pattern=horizontal lines, pattern color=solarized-base03}}

\tikzset{Eblanks/.style={fill=solarized-base02}}
\tikzset{EblanksP/.style={pattern=crosshatch, pattern color=solarized-base1}}
\tikzset{Euris/.style={fill=solarized-violet!50!solarized-blue}}
\tikzset{EurisP/.style={pattern=north west lines, pattern color=solarized-base2}}
\tikzset{Eliterals/.style={fill=solarized-cyan!50!solarized-green}}
\tikzset{EliteralsP/.style={pattern=north east lines, pattern color=solarized-base2}}
\tikzset{Eedges/.style={fill=solarized-red!50!solarized-orange}}
\tikzset{EedgesP/.style={pattern=horizontal lines, pattern color=solarized-base2}}

\colorlet{EmaxR}{green!75!black}
\colorlet{EminR}{red!75!black}
\colorlet{EmaxA}{solarized-blue}
\colorlet{EminA}{solarized-base3}

\tikzset{Ehybrid/.style={fill=solarized-cyan}}
\tikzset{EhybridP/.style={pattern=north east lines, pattern color=solarized-base2}}
\tikzset{Eoverlap/.style={fill=solarized-yellow}}
\tikzset{EoverlapP/.style={pattern=north west lines, pattern color=solarized-base2}}
\tikzset{EGtoPdb/.style={fill=solarized-red}}
\tikzset{EGtoPdbP/.style={pattern=north east lines, pattern color=solarized-base02}}
\tikzset{Etotal/.style={fill=solarized-violet!75!solarized-magenta}}
\tikzset{EtotalP/.style={pattern=north west lines, pattern color=solarized-base02}}

\tikzset{Eexact/.style={fill=solarized-cyan!50!solarized-green}}
\tikzset{EexactP/.style={pattern= north east lines , pattern color= solarized-base2}}
\tikzset{Einclusive/.style={fill=solarized-violet}}
\tikzset{EinclusiveP/.style={pattern= north west lines , pattern color= solarized-base2}}
\tikzset{Efalse/.style={fill=solarized-yellow!75!solarized-orange}}
\tikzset{EfalseP/.style={pattern= crosshatch dots , pattern color= solarized-base2}}
\tikzset{Emissing/.style={fill=solarized-magenta!50!solarized-red}}
\tikzset{EmissingP/.style={pattern= crosshatch dots , pattern color=solarized-base02}}

\colorlet{Fdbfill}{solarized-base3!25!white}
\colorlet{Fdbdraw}{solarized-base00}
\colorlet{Ftrivial}{blue!75!black}
\colorlet{Fbisimulation}{green!75!blue!75!black}
\colorlet{Fsimilarity}{red!85!black}
\colorlet{Fdeblank}{orange!75!black}
\colorlet{Fhybrid}{green!75!blue!75!black}
\colorlet{Fedit}{violet!75!black}
\colorlet{Fpropagation}{red!85!black}

\tikzset{poz/.style={coordinate}}
\tikzset{aux/.style={gray, line width=0.25pt, densely dotted}}
\tikzset{pt/.style={gray, circle, inner sep=0.05cm,fill=gray}}
\newcommand{\poz}[2]{
  \path #2 node[poz] (#1) {};
}
\newcommand{\makeTokenAnchors}[1]{
  \poz{#1_token_tl}{($(#1)!0.2cm!30:(#1_parent)$)}
  \poz{#1_token_bl}{($(#1)!0.2cm!-30:(#1_gravity)$)}
  \poz{#1_token_cl}{($(#1_token_tl)!.5!(#1_token_bl)$)}
  \poz{#1_token_ml}{($(#1)!.2cm!(#1_token_cl)$)}
  \poz{#1_token_br}{($(#1)!0.2cm!30:(#1_gravity)$)}
  \poz{#1_token_tr}{($(#1)!0.2cm!-30:(#1_parent)$)}
  \poz{#1_token_cr}{($(#1_token_tr)!.5!(#1_token_br)$)}
  \poz{#1_token_mr}{($(#1)!.2cm!(#1_token_cr)$)}
  \poz{#1_token_top}{($(#1_token_tl)!0.5!(#1_token_tr)$)}
  \poz{#1_token_btm}{($(#1_token_bl)!0.5!(#1_token_br)$)}
}
\newcommand{\makeBatonAnchors}[2]{
  \poz{#1_baton_top}{($(#1)!.25!(#2)$)}
  \poz{#1_baton_mid}{($(#1)!.50!(#2)$)}
  \poz{#1_baton_btm}{($(#1)!.75!(#2)$)}
  \poz{#1_baton_br}{($(#1_baton_btm)!0.05cm!90:(#1)$)}
  \poz{#1_baton_bl}{($(#1_baton_btm)!0.05cm!270:(#1)$)}
  \poz{#1_baton_mr}{($(#1_baton_mid)!0.15cm!90:(#1)$)}
  \poz{#1_baton_ml}{($(#1_baton_mid)!0.15cm!270:(#1)$)}
  \poz{#1_baton_tr}{($(#1_baton_top)!0.05cm!90:(#1)$)}
  \poz{#1_baton_tl}{($(#1_baton_top)!0.05cm!270:(#1)$)}
}
\newcommand{\korzen}[2]{
  \poz{#1_parent}{(0,1cm)}
  \poz{#1}{(0,0)}
  \poz{#1_gravity}{($(#1)!1cm!-180+#2:(#1_parent)$)}
  \makeTokenAnchors{#1}
}
\newcommand{\child}[4]{
  \poz{#1_parent}{(#2)}
  \poz{#1}{($(#2)+(#3-90:1cm)$)}
  \poz{#1_gravity}{($(#1)!1cm!-180+#4:(#1_parent)$)}
  \makeTokenAnchors{#1}
  \makeBatonAnchors{#1}{#2}
}
\newcommand{\token}[3]{
  \fill[fill=#2] plot[smooth cycle, tension=0.35] coordinates { 
    (#1_token_tl) (#1_token_ml) (#1_token_bl) (#1_token_br) (#1_token_mr) (#1_token_tr)
  };
  \path (#1_token_mr) -- (#1_token_ml) node[pos=0.5,sloped] {\small #3};
}
\newcommand{\batton}[3]{
  \fill[fill=#2] plot[smooth cycle, tension=0.35] coordinates { 
    (#1_baton_bl) (#1_baton_ml) (#1_baton_tl) (#1_baton_tr) (#1_baton_mr) (#1_baton_br)
  };

  \path (#1_baton_ml) -- (#1_baton_mr) node[pos=0.5,sloped] {\small #3};
}
\newcommand{\offset}[4]{
  \poz{#1}{($(#2)!1+#4!(#2_token_#3)$)}
}
\newcommand{\hull}[3]{
  \def\hullCoordinates{}
  \foreach[count=\x] \n/\d in {#1} {
    \offset{hull_\x}{\n}{\d}{#2}
    \xdef\hullCoordinates{\hullCoordinates (hull_\x)}
  }

  \fill[fill=#3] plot[smooth cycle, tension=0.375] coordinates { \hullCoordinates };
}
\def\HullA{\hull{w/tl,w/ml,B0/ml,B0/bl,B0/br,B0/mr,w/mr,w/tr}{.3}{Ha}}
\def\HullB{\hull{u/tl,a1/ml,a1/bl,a1/br,w/bl,w/br,B2/bl,B2/br,b1/bl,b1/br,b1/mr,u/tr}{.3}{Hb}}
\def\HullC{\hull{B1/tl,B1/ml,b2/ml,b2/bl,b2/br,u/bl,u/br,u/mr,B1/mr,B1/tr}{.3}{Hc}}
\def\HullCTilt{\hull{B1/tl,b2/ml,b2/bl,b2/br,u/bl,u/br,u/mr,B1/mr,B1/tr}{.3}{Hc}}

\def\HullD{\hull{B2/tl,B2/ml,a2/ml,a2/bl,a2/br,a2/mr,B2/mr,B2/tr}{.3}{Hd}}
\def\HullDT{\hull{B2/tl,B2/ml,a2/ml,a2/bl,a2/br,a2/mr,B2/mr,B2/tr}{.9}{Hdt}}
\def\HullAA{\hull{w/tl,w/ml,B1/ml,B1/bl,B2/br,B2/mr,w/mr,w/tr}{.3}{Haa}}
\def\HullE{\hull{w/tl,B1/ml,b2/ml,b2/bl,b2/br,a2/bl,a2/br,a2/mr,B2/mr,w/tr}{.9}{He}}
\def\HullF{\hull{u/tl,a1/ml,a1/bl,B0/bl,a2/br,b1/br,b1/mr,u/tr}{.9}{Hf}}
\def\HullG{\hull{B1/tl,b2/ml,a1/bl,a1/br,w/bl,w/br,B2/bl,B2/br,b1/bl,b1/br,b1/mr,B1/tr}{.9}{Hg}}
\def\HullGG{\hull{B1/tl,b2/ml,a1/bl,a1/br,w/bl,w/br,B2/bl,B2/br,b1/bl,b1/br,b1/mr,B1/tr}{.9}{Hgg}}
\def\HullH{\hull{B1/tl,B1/ml,b2/ml,b2/bl,b2/br,u/bl,u/br,u/mr,B1/mr,B1/tr}{.3}{Hh}}

\def\HullBB{\hull{u/tl,a1/ml,a1/bl,a1/br,w/bl,w/br,B2/bl,B2/br,b1/bl,b1/br,b1/mr,u/tr}{.3}{Hbb}}
\def\HullI{\hull{u/tl,a1/ml,a1/btm,a2/bl,a2/br,a2/mr,b1/br,b1/mr,u/tr}{.9}{Hi}}
\def\HullIT{\hull{u/tl,a1/ml,a1/btm,a2/bl,a2/br,a2/mr,b1/br,b1/mr,u/tr}{1.5}{Hit}}
\def\HullJ{\hull{B1/tl,b2/ml,a1/bl,a1/br,a2/bl,a2/br,b1/br,B1/tr}{1.5}{Hj}}
\def\HullCCTilt{\hull{B1/tl,b2/ml,b2/bl,b2/br,u/bl,u/br,u/mr,B1/mr,B1/tr}{.3}{Hcc}}

\begin{document}
\thispagestyle{empty}
\title{RDF Graph Alignment with Bisimulation}
\numberofauthors{1}
\author{
\alignauthor
Peter Buneman${}^\mathsf{1}$\qquad
S\l{}awek Staworko${}^\mathsf{1,2,3}$\\
\affaddr{${}^\mathsf{1}$University of Edinburgh}\qquad
\affaddr{${}^\mathsf{2}$University of Lille}\qquad
\affaddr{${}^\mathsf{3}$LINKS, INRIA Nord-Europe}
\email{{\large\{}opb,sstawork{\large\}}@inf.ed.ac.uk}
}

\maketitle
\thispagestyle{empty}

\begin{abstract}
  We investigate the problem of aligning two RDF databases, an
  essential problem in understanding the evolution of ontologies. Our
  approaches address three fundamental challenges: 1) the use of
  ``blank'' (null) names, 2) ontology changes in which different names
  are used to identify the same entity, and 3) small changes in the
  data values as well as small changes in the graph structure of the
  RDF database. We propose approaches inspired by the classical notion
  of graph bisimulation and extend them to capture the natural metrics
  of edit distance on the data values and the graph structure. We
  evaluate our methods on three evolving curated data sets.
  Overall, our results
  show that the proposed methods perform  well and are scalable.
 
\end{abstract}

\section{Introduction}
Identifying references to the same real-life entity is one of the most
fundamental concerns in databases. It plays an important if not
crucial role in virtually all non-trivial data processing tasks from
computing join of two tables to removing duplicate entries in data
cleaning~\cite{RaDo00} to combining data objects in multiple databases in
data integration~\cite{HaRaOr06}. This problem comes in a number of flavors
depending on the type of data used to identify the entity represented
by a given data object. Ideally, as in the case of a well-designed
stand-alone database, a consistent system of unique identifiers
supports the linkage of objects in a manner that is reliable and
efficient. However, independent databases may use different and often
incompatible schemes of unique identifiers. Consequently, linking
their contents may require using other methods, based on data values
and the structure of the databases to match corresponding identifiers~\cite{KoRa10}.

In this paper we study an instance of this problem that arises in the
context of evolving RDF graphs: for two consecutive versions of an RDF
graph we wish to construct an \emph{alignment} that connects pairs of
nodes in the two versions that represent the same entity.  RDF is
essentially an edge-labeled graph that uses \emph{URIs} (Unique
Resource Identifiers) as nodes and edge labels but also has
\emph{blank} nodes, which are not persistent identifiers, as well as
\emph{literal} nodes, which store (unique) data strings. Because of
the varied types of nodes aligning two RDF graphs presents a number of
interesting challenges. While it is reasonable to assume that two
nodes labeled with the same URI represent the same entity, the
converse is not necessarily true. Indeed, the same entity may be
represented in different versions with different identifiers, for
instance, as a result of changing the scheme of attributing URIs. Even
more problematic are blank nodes, which although discouraged are often
misused when using reification for purpose of representing data
structures such as lists and records~\cite{HAAP14}. Because blank
nodes are not persistent identifiers, we require methods to establish
an identity for a blank node based on a description
by its neighborhood in the graph. This however is a nontrivial task
because both the data values and the connections may undergo 
modifications in the subsequent version of the RDF graph. Finally, 
constructing an alignment between two RDF graphs
presents a significant computational challenge: RDF graphs tend to be
large, which quickly renders infeasible any method that attempts
to perform pairwise comparison between {\em all} pairs of nodes of the
two graphs.

We investigate a number of methods of aligning RDF graphs inspired by
the classical notion of bisimulation for graphs. In essence, two nodes
are bisimilar if they cannot be distinguished from each other by
structural comparison of their outbound neighborhoods in which the
nodes reachable from the bisimilar nodes are also bisimilar. What
makes bisimulation particularly interesting is its computational
properties: it is well-known that bisimulation can be computed in
sub-quadratic time~\cite{PaTa87} but the basic partition refinement
approach, while having quadratic worst time complexity scales well in
practice~\cite{SNLPZ13}, and we have chosen it as a basis of RDF
alignment algorithms.
\begin{example}
  \label{ex:1}
  \begin{figure*}[htb]
    \centering
    \begin{tikzpicture}[>=latex]
      \path[use as bounding box] (-7,-4.8) rectangle (7,3.25);

      \node at (-6,3) {\textsl{version 1}};
      \draw[thin, draw=Fdbdraw, fill=Fdbfill] 
      plot[smooth cycle, tension=0.45] coordinates {
        (-7.55,0)
        (-5.25,2.1)
        (-3.5,2.45)
        (-1.25,2.25)
        (-0.5,1.5)
        (-0.1,-0.15)
        (-0.5,-2)
        (-1.5,-3)
        (-3.25,-3.25)
        (-5.25,-2.5)
      };

      \node at (6,3) {\textsl{version 2}};
      \draw[thin, draw=Fdbdraw, fill=Fdbfill]
      plot[smooth cycle, tension=0.45] coordinates {
        (7.55,0)
        (5.25,2.1)
        (3.5,2.45)
        (1.25,2.25)
        (0.5,1.5)
        (0.1,-0.15)
        (0.5,-2.25)
        (1.5,-2.8)
        (3.25,-2.85)
        (5,-2.25)
      };

      \node[right] (ss_1)     at (-7.5,0)    {\sf ss}; 
      \path (ss_1) node[above] (above_ss_1) {};
      \node[right] (_b_1)     at (-5.25,1.65)  {$\B_1$};
      \path (_b_1) node[above=6pt] (above_b_1) {};
      \node[right] (uoe_1)    at (-5.25,0)    {\sf ed-uni};
      \path (uoe_1) node[below=6pt] (below_uoe_1) {};
      \node[right] (_b_2)     at (-5.25,-2.1) {$\B_2$};
      \path (_b_2) node[below=7pt] (below_b_2) {};
      \node[right] (EH8_1)    at (-2.85,2.1)    {\it ``EH8''};
      \node[right] (Edin_1)   at (-2.85,1.05)    {\it ``Edinburgh''};
      \node[right, text width=6.25em, text centered] 
                   (UoE_1)    at (-2.85,-0.15)    {\it ``University~of Edinburgh''};
      \node[right] (Staw_1)   at (-2.85,-1.5)   {\it ``Staworko''};
      \node[right] (Slawek_1) at (-2.85,-2.25)   {\it ``S\l{}awek''};
      \node[right] (_Pawel_1)  at (-2.85,-2.85)   {\it ``Pawe\l{}''};      
      \node at (-2.75,-2.875) (Pawel_1){};    

      \draw (ss_1) edge[->, semithick] node[above,sloped] {\sf addr} (_b_1)
      edge[->, semithick] node[above=-2pt,sloped] {\sf employer} (uoe_1)
                   edge[->, semithick] node[below,sloped] {\sf name} (_b_2);

      \draw (_b_1) edge[->, semithick] node[above,sloped] {\sf zip} (EH8_1)
                   edge[->, semithick] node[below,sloped] {\sf city} (Edin_1);

      \draw (uoe_1) edge[->, semithick] node[below,sloped] {\sf name} (UoE_1)
                    edge[->, semithick] node[above,sloped,pos=0.3] {\sf city} (Edin_1);

      \draw (_b_2) edge[->, semithick] node[below=-2pt,sloped,pos=0.65] {\sf first} (Slawek_1)   
                   edge[->, semithick] node[below,sloped] {\sf middle} (Pawel_1)
                   edge[->, semithick] node[above,sloped] {\sf last} (Staw_1);

      \begin{scope}[xshift=0]
      \node[left] (ss_2)     at (7.5,0)    {\sf ss};
      \path (ss_2) node[above] (above_ss_2) {};
      \node[left] (_b_3)     at (5.25,1.65)  {$\B_3$};
      \path (_b_3) node[above=3pt] (above_b_3) {};
      \node[left] (uoe_2)    at (5.25,0)    {\sf uoe};
      \path (uoe_2) node[below=2pt] (below_uoe_2) {};
      \node[left] (_b_4)     at (5.25,-1.71) {$\B_4$};
      \path (_b_4) node[below=3pt] (below_b_4) {};
      \node[left] (EH8_2)    at (2.85,2.1)    {\it ``EH8''};
      \node[left] (Edin_2)   at (2.85,1.05)    {\it ``Edinburgh''};
      \node[left, text width=6.25em, text centered] 
                  (UoE_2)    at (2.85,-0.15)    {\it ``University~of Edinburgh''};
      \node[left] (Slawek_2) at (2.85,-2.25)   {\it ``S\l{}awomir''};
      \node[left] (Staw_2)   at (2.85,-1.5)   {\it ``Staworko''};

      \draw (ss_2) edge[->, semithick] node[above,sloped] {\sf addr} (_b_3)
                   edge[->, semithick] node[above=-2pt,sloped] {\sf employer} (uoe_2)
                   edge[->, semithick] node[below,sloped] {\sf name} (_b_4);

      \draw (_b_3) edge[->, semithick] node[above,sloped] {\sf zip} (EH8_2)
                   edge[->, semithick] node[below,sloped] {\sf city} (Edin_2);

      \draw (uoe_2) edge[->, semithick] node[below,sloped] {\sf name} (UoE_2)
                    edge[->, semithick] node[above,sloped,pos=0.3] {\sf city} (Edin_2);

      \draw (_b_4) edge[->, semithick] node[below,sloped] {\sf first} (Slawek_2)
                   edge[->, semithick] node[above,sloped] {\sf last} (Staw_2);
      \end{scope}
      
      \begin{scope}[thick,o-o,Ftrivial]
        \draw plot[smooth] coordinates {
          (above_ss_1)
          (-6,2.25)
          (-3,3.15) (3,3.15)
          (6,2.25)
          (above_ss_2)
        };
        \draw (EH8_1) -- (EH8_2);
        \draw (Edin_1) -- (Edin_2);
        \draw (UoE_1) -- (UoE_2);
        \draw (Staw_1) -- (Staw_2);
        \draw (-6.3,-3.9) -- (-2.7,-3.9);
      \end{scope}
      \node[below, text width=8em, text centered] at (-4.5,-3.9) {
        trivial alignment (label equality)
      };

      \begin{scope}[thick,dashed,o-o,Fbisimulation]
        \draw plot[smooth] coordinates {
          (above_b_1) 
          (-3.75,2.55)
          (-1.5,2.75)
          (1.5,2.75)
          (3.75,2.55)
          (above_b_3)
        };

        \draw plot[smooth] coordinates {
          (below_uoe_1) 
          (-3.75,-0.85)
          (-1.5,-1)
          (1.5,-1)
          (3.75,-0.85)
          (below_uoe_2)
        };
        \draw (-1.8,-3.9) -- (1.8,-3.9);
      \end{scope}
      \node[below, text width=8em, text centered] at (0,-3.9) {
        bisimulation alignment
      };

      \begin{scope}[thick,densely dotted ,o-o,Fsimilarity]
        \draw plot[smooth] coordinates {
          (below_b_2) 
          (-4.2,-3.25)
          (-2,-3.45)
          (2,-3.45)
          (4.5,-3.25)
          (below_b_4)
        };
        \draw (Slawek_1) -- (Slawek_2);
        \draw (2.7,-3.9) -- (6.3,-3.9);
      \end{scope}
      \node[below, text width=10em, text centered] at (4.5,-3.9) {
        similarity measure\\ alignment };
    \end{tikzpicture}
    \caption{\label{fig:evolving-rdf}Alignment methods on two consecutive versions of an
      evolving RDF graph: {\sf uri}s are typeset in sanserif, {\it
        ``literals''} are in italics surrounded by quotes, and blank nodes
      are $\B$.}
  \end{figure*}
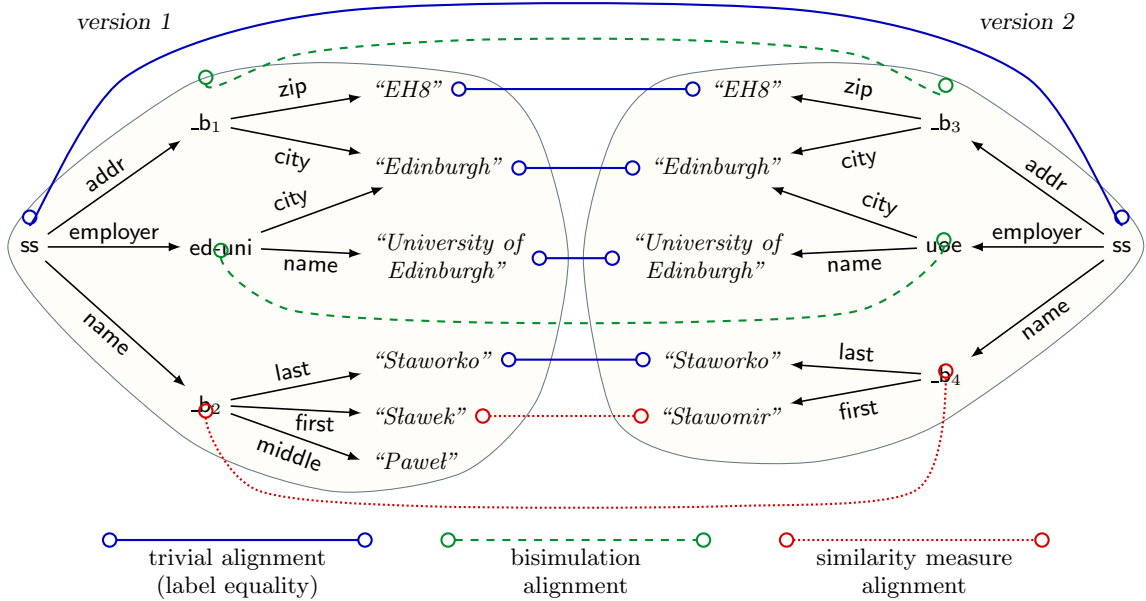
  Consider corrections in an evolving RDF graph presented in
  Figure~\ref{fig:evolving-rdf} containing personal information of one
  of the authors of this paper. The first
  name is changed from the diminutive S\l{}awek to its legal variant
  S\l{}awomir, and an erroneous middle name Pawe\l{} is removed. Also,
  the URI representing the University of Edinburgh is changed from
  \textsf{ed-uni} to \textsf{uoe}.  Note that a majority of literals
  and one URI, \textsf{ss}, can be trivially aligned by testing label
  equality. However, this simple method does not work for the address
  information even though it does not change. Here, address is
  structured as a record represented with a blank node and blank nodes
  are labeled with \emph{local} identifiers that distinguish
  them only in a single version. Bisimulation aligns the blank nodes
  $\B_1$ and $\B_3$ because they represent a record with the same
  information structured in the same manner. Similarly, bisimulation
  aligns the nodes \textsf{ed-uni} and \textsf{uoe}. However,
  bisimulation requires strict similarity in the data values and
  the structure of the graph, and therefore, cannot handle edit
  changes in the data values and the structure. Consequently,
  bisimulation does not align the nodes $\B_2$ and $\B_4$ even though
  there is a significant evidence that they represent the same entity
  (the name of the same person.) \qed
\end{example}

Aligning nodes $\B_2$ and $\B_4$ (Figure~\ref{fig:evolving-rdf}) calls
for similarity methods, and we propose a natural similarity measure
based on the string edit distance on literal nodes and the graph edit
distance for non-literal nodes. While this method can align the nodes
$\B_2$ and $\B_4$, it suffers from high complexity, which springs from
the sizes of the input RDF graphs and the combinatorial nature of the
edit distance problem: indeed, the lower complexity bound for the edit
distance is quadratic for strings~\cite{WoCh76,Hu88} and cubic for
graphs~\cite{JuHe06}.

To overcome this obstacle, we investigate extending the bisimulation
approach to account for such edits. While bisimulation defines a
partition of nodes into clusters of indistinguishable nodes, we
propose an approach that defines a \emph{weighted} partition, where
every node still belongs to exactly one cluster but is additionally
attributed with a \emph{confidence} value. Intuitively, the confidence
value captures the distance of the node from the center of the
cluster, which can be used to approximate the relative distance
between two nodes in the same cluster. The limitation of the
membership of a node to exactly one cluster has both positive and
negative consequences.  While it enables a scalable method for
constructing weighted partitions, the weighted partition only
approximates the goal similarity measure and the resulting alignment
may be incomplete. Our experiments show, however, that the trade-off
is generally positive: with a diligent application of a number of
optimizations we obtain a relatively scalable method for RDF alignment
that fails to align correctly relatively few pairs of nodes. We also
show that any pair aligned with this method is also aligned with the
similarity method we wish to approximate.

The main contributions of the present paper are summarized as follows:
\begin{enumerate}
\itemsep0pt
\item We formalize the problem of RDF graph alignment and present a
  methodology of aligning RDF graphs with partitions of the nodes. 
\item We propose RDF alignment methods based on the standard notion of
  bisimulation for graphs that  handles blank nodes and changes
  in ontology (URI naming schemes). 
\item We propose a natural measure of node similarity that yields an
  RDF alignment method robust under editing operations and extend the
  bisimulation approach to approximate the proposed similarity method.
\item We evaluate the accuracy and effectiveness of these methods
  on widely used databases presented in RDF.
\end{enumerate}

\paragraph*{Related Work}
The similarity measure we define bears some resemblance to the
similarity flooding approach~\cite{MeGaRa02} with an important difference on
on how similarities are propagated: when defining the similarity of
two nodes, the similarity flooding takes a weighted average over the
Cartesian product of sets of outgoing edges of the two nodes while our
approach identifies the optimal matching among the outgoing edges. We
believe this approach to be more appropriate in the context of
evolving graphs and incorporates edit operations on the edges. Still,
the inherent high complexity of both methods limits their scalability,
and the aim of our research is development of scalable methods for
identifying similar objects. 

There is extensive work on entity resolution in the context of
relational databases, however a comparison with that work is
problematic.  Because we are using RDF and because we are placing
predicates on the same footing as other URIs the problem we are
setting ourselves is equivalent to finding an alignment between two
versions in which one (a) changes all the table names and column names
and (b) changes all the key values.  All that is kept are the non-key
data values and the foreign key constraints.  In other words
are trying to  find an alignment between two
versions of a relational database in which one applies a bijective map
to all the column names and to all the key values as well as making
some updates to one of the versions.

There is also extensive literature on graph alignment~\cite{emmert2016fifty}.
Constructing an alignment between two graphs is virtually equivalent
to constructing their \emph{delta}~\cite{ZeTzCh11}, a description of
changes occurring between the two graphs. The existing
research~\cite{PFFK13} focuses mainly on on reporting high-level
changes identifying ontology changes (\texttt{rdfs:type}) and
compactly representing the delta whereas we treat RDF as a stand-alone
data representation system and identify low-level changes occurring on
the atomic level of nodes and their labels. The ability of identifying
ontology changes may potentially allow both directions to reinforce
each other. Handling blank nodes in the context of change detection is
known to be very challenging (graph isomorphism) and a method of
label-invention have been proposed~\cite{TzLaZe12}. This method works
under the assumption that the blank nodes do not form cycles and can
be seen as an adaptation of existing XML archiving
techniques~\cite{BKTT04}. Our work  generalises this
method: we handle cycles, editing operations, and can even identify
ontology changes.

\paragraph*{Organization} The paper is organized as follows. In
Section~\ref{sec:basics} we define basic notions. In
Section~\ref{sec:alignment} we state the problem of RDF graph
alignment and present a number of alignment methods inspired by
bisimulation. In Section~\ref{sec:dissimilarity} we present a natural
measure for node similarity for RDF alignment and propose its
approximation based on an extension of bisimulation robust under
editing operations. Section~\ref{sec:experiments} contains
experimental evaluation of the proposed methods.
We present the conclusions of our study and discuss directions of
future research in Section~\ref{sec:concl}. Because of space
restrictions we omit proofs, some formal definitions, and numerical
values of our experiments; they can be found in the appendix of the
complete version available at \mbox{\url{http://homepages.inf.ed.ac.uk/sstawork/vldb16.pdf}}.

\section{Preliminaries}
\label{sec:basics}
In this section we define the data model for RDF graphs, formalize the
notion of partition, and define the notion of bisimulation for RDF graphs.

\subsection{Data model}
\label{sec:basicsdatamodel}
RDF graphs are typically represented as sets of triples of URIs,
literals, and blank nodes. Because we are dealing with two graphs that
may contain the same URI we need a more general model that uses
node identifiers and treats the URIs and literals as labels: we assume an
enumerable set of node identifiers $\N$ and a set
of labels $\I=\U\cup\L\cup\{\B\}$, which consists of URI labels $\U$,
literal values $\L$, and a special \emph{blank} value $\B$ used to
label blank nodes. We assume that $\U$ and $\L$ are disjoint and
neither contains $\B$.
\begin{definition}
  \label{def:1}
  A (\emph{triple}) \emph{graph} is a tuple $G=(N_G,E_G,\ell_G)$,
  where $N_G\subseteq\N$ is a finite set of nodes, $E_G \subseteq
  N_G\times N_G\times N_G$ is a set node triples (edges), and
  $\ell_G:N_G\rightarrow\I$ is a node labeling function.\qed
\end{definition}
The URIs of $G$, $\Uris(G)$  are those nodes $n\in G$ for which
$\ell(n)\in\U$. $\Literals(G)$ and $\Blanks(G)$ are defined similarly.

\begin{figure}[htb]
\begin{center}
  \begin{tikzpicture}[>=latex]
    \path[use as bounding box] (-2,-0.65) rectangle (2,2);
    \begin{scope}
      \node at (-1.75,1.5)   (w0) {$\mathsf{w}$};
      \node at (-1.75,0)   (b1) {$\B_1$};
      \node at (0,0)    (u2) {$\mathsf{u}$};
      \node at (1.95,0)    (b3) {\textit{``b''}};
      \node at (1.95,1.5)    (a4) {\textit{``a''}};
      \node at (0,1.25)    (b5) {$\B_2$};
      \node at (0,2) (b6) {$\B_3$};
      \draw (w0) edge[semithick, ->] node[above, sloped] {$\mathsf{p}$} (b6);
      \draw (w0) edge[semithick, ->] node[left] {$\mathsf{p}$} (b1);
      \draw (b1) edge[semithick, ->] node[above, sloped] {$\mathsf{q}$} (u2);
      \draw (b1) edge[semithick, ->, bend right] node[below, sloped] {$\mathsf{r}$} (b3);
      \draw (u2) edge[semithick, ->] node[above, sloped] {$\mathsf{p}$} (b5);
      \draw (u2) edge[semithick, ->] node[above, sloped] {$\mathsf{q}$} (a4);
      \draw (u2) edge[semithick, ->] node[above, sloped] {$\mathsf{q}$} (b3);
      \draw (u2) edge[semithick, ->] node[above, sloped] {$\mathsf{r}$} (w0);
      \draw (b5) edge[semithick, ->] node[below, sloped] {$\mathsf{q}$} (a4);
      \draw (b6) edge[semithick, ->] node[above, sloped] {$\mathsf{q}$} (a4);
    \end{scope}

\end{tikzpicture}
\end{center}
\caption{\label{fig:rdf-graph}An RDF graph: {\sf URI}s are typeset in sanserif, {\it
    ``literals''} are in italics surrounded by quotes, and blank nodes
  are $\B$.}
\end{figure}
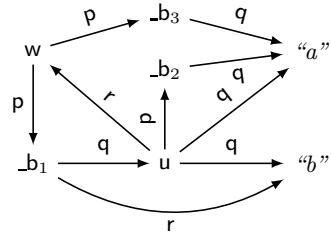

We now define an \emph{RDF graph} (e.g., one of the two versions we
are trying to align) as a  triple graph in which no two nodes have the
same URI or literal label and the labels agree with the usual RDF conventions 
(literal labels only occur as objects and  predicates cannot have
blank labels.) Figure~\ref{fig:rdf-graph}
shows an example of an RDF graph in which nodes are identified by
their label and blank-labeled nodes are decorated with a subscript.

Using node identifiers that are independent of labels allows us to
take two versions of the same RDF graph with possibly overlapping
labels and combine them  without confusing nodes with the
same label. Graphs $G_1=(N_1,E_1,\ell_1)$ and
$G_2=(N_2,E_2,\ell_2)$  are \emph{disjoint} if $N_1\cap
N_2=\emptyset$. Their \emph{disjoint} \emph{union} is $G_1\uplus
G_2=(N_1\cup N_2, E_1\cup E_2, \ell_1\cup\ell_2)$. \eat{Note that with a
simple change of node identifiers we can make any pair of graphs
disjoint while preserving their structure and data stored in the
nodes.}

\subsection{Partitions}
We align two versions of an RDF graph using  equivalence
relations represented by a partitions of the node set of the combined
graph. For our purposes we assign every node a unique color, and the
equivalence classes of a partition are the sets of nodes with the same color.

Formally, we assume an enumerable set of \emph{colors} $\C$, which
allows both node labels and node identifiers to be used as colors as
well as other structures that we can build from these.  A
\emph{partition} of a graph $G$ is a function $\lambda:N_G\rightarrow
\C$ that assigns a color to every node of $G$. Throughout this paper,
we only work with partitions of the same graph and we normally assume
the graph to be known from the context. Note that the node labeling
function $\ell_G$ is also a partition of $G$, which groups nodes by
their labels, and in particular, places all blank nodes in the same
equivalence class.

A partition $\lambda$ defines an equivalence
relation on the nodes of the graph, $R_\lambda=\{(n,m)\in N_G\times
N_G \mid \lambda(n) = \lambda(m)\}$. A partition $\lambda_1$ is
\emph{finer} than a partition $\lambda_2$ if $R_{\lambda_1} \subseteq
R_{\lambda_2}$. Two partitions $\lambda_1$ and $\lambda_2$ are
\emph{equivalent}, in symbols $\lambda_1\equiv \lambda_2$, if
$R_{\lambda_1}=R_{\lambda_2}$.

\subsection{Bisimulation}

Bisimulation is often defined on edge-labeled graphs.  While RDF
graphs are often drawn as such graphs with a triple $(s,p,o)$
represented as an edge $(s,o)$ labeled with $p$, the label $p$ is
itself a node, and should participate in the bisimulation relation.
We therefore  adapt the definition of bisimulation to triple graphs
by treating them as graphs in which the triple $(s,p,o)$ is
represented as an unlabeled edge connecting the node $s$ to 
the {\em pair} $(p,o)$ and define
the 
\emph{outbound neighborhood} of a node $n$ in $G$ is:
\[
\out_G(n)=\{(p,o) \mid (n,p,o) \in E_G\}.
\]

\begin{definition}
  \label{def:bisimulation}
  A binary relation $R\subseteq N_G\times N_G$ is a \emph{simulation}
  on a graph $G=(N_G,E_G,\ell_G)$ if for every $(n,m)\in R$ we have
  $\ell_G(n)=\ell_G(m)$ and for any $(p,o)\in \out_G(n)$ there is
  $(p',o')\in\out_G(m)$ such that $(p,p')\in R$ and $(o,o')\in R$. $R$
  is a \emph{bisimulation} on $G$ if both $R$ and $R^{-1}$ are
  simulations on $G$. Two nodes $n$ and $m$ of $G$ are
  \emph{bisimilar} if there is a bisimulation $R$ on $G$ such that
  $(n,m)\in R$. \qed
\end{definition}
In the graph in Figure~\ref{fig:rdf-graph} the nodes
$\B_2$ and $\B_3$ are bisimilar. Bisimulation identifies pairs of
nodes that are indistinguishable by means of exploration of their
outbound neighborhood, or intuitively, nodes having the same contents,
as it is the case with the nodes $\B_1$ and $\B_3$ in the graph in
Figure~\ref{fig:evolving-rdf}.

The identity relation on the nodes of a graph is always a
bisimulation. If $R_1$ and $R_2$ are bisimulations on $G$, so is their
union $R_1\cup R_2$ is. Since all bisimulations on $G$ are
subsets of the finite Cartesian product $N_G\times N_G$, there exists
a unique maximal bisimulation,  $\Bisim(G)$, on $G$. The maximal
bisimulation on a graph is an 
equivalence relation on nodes of the graph and
thus defines a partition.

\section{RDF graph alignment}
\label{sec:alignment}
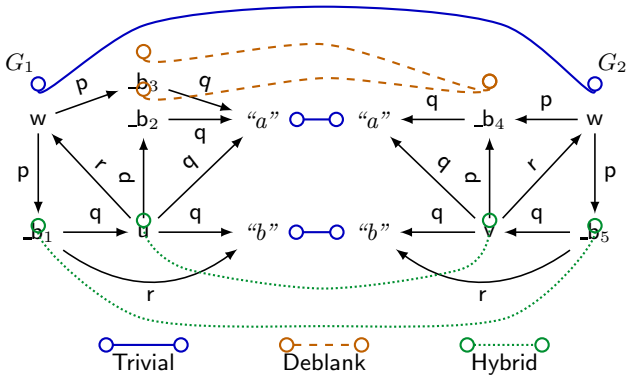
\begin{figure}[htb]
\begin{center}
  \begin{tikzpicture}[>=latex,xscale=.8]
    \path[use as bounding box] (-1,-1.5) rectangle (7,2.85);
    \begin{scope}
    \begin{scope}
      \node at (-2.05, 2.25) {$G_1$};
      \node at (-1.75,1.5)   (w0) {$\mathsf{w}$};
      \node at (-1.75,0)   (b1) {$\B_1$};
      \node at (0,0)    (u2) {$\mathsf{u}$};
      \node at (1.95,0)    (b3) {\textit{``b''}};
      \node at (1.95,1.5)    (a4) {\textit{``a''}};
      \node at (0,1.5)    (b5) {$\B_2$};
      \node at (0,2) (b6) {$\B_3$};
      \draw (w0) edge[semithick, ->] node[above, sloped] {$\mathsf{p}$} (b6);
      \draw (w0) edge[semithick, ->] node[left] {$\mathsf{p}$} (b1);
      \draw (b1) edge[semithick, ->] node[above, sloped] {$\mathsf{q}$} (u2);
      \draw (b1) edge[semithick, ->, bend right] node[below, sloped] {$\mathsf{r}$} (b3);
      \draw (u2) edge[semithick, ->] node[above, sloped] {$\mathsf{p}$} (b5);
      \draw (u2) edge[semithick, ->] node[above, sloped] {$\mathsf{q}$} (a4);
      \draw (u2) edge[semithick, ->] node[above, sloped] {$\mathsf{q}$} (b3);
      \draw (u2) edge[semithick, ->] node[above, sloped] {$\mathsf{r}$} (w0);
      \draw (b5) edge[semithick, ->] node[below, sloped] {$\mathsf{q}$} (a4);
      \draw (b6) edge[semithick, ->] node[above, sloped] {$\mathsf{q}$} (a4);
    \end{scope}

    \begin{scope}[xshift=5.75cm]
      \node at (2.05, 2.25) {$G_2$};
      \node at (1.75,1.5)   (w10) {$\mathsf{w}$};
      \node at (1.75,0)   (b11) {$\B_5$};
      \node at (0,0)    (v12) {$\mathsf{v}$};
      \node at (-1.95,0)    (b13) {\textit{``b''}};
      \node at (-1.95,1.5)    (a14) {\textit{``a''}};
      \node at (0,1.5)    (b15) {$\B_4$};
      \draw (w10) edge[semithick, ->] node[above, sloped] {$\mathsf{p}$} (b15);
      \draw (w10) edge[semithick, ->] node[right] {$\mathsf{p}$} (b11);
      \draw (b11) edge[semithick, ->] node[above, sloped] {$\mathsf{q}$} (v12);
      \draw (b11) edge[semithick, ->, bend left] node[below, sloped] {$\mathsf{r}$} (b13);
      \draw (v12) edge[semithick, ->] node[above, sloped] {$\mathsf{p}$} (b15);
      \draw (v12) edge[semithick, ->] node[above, sloped] {$\mathsf{q}$} (a14);
      \draw (v12) edge[semithick, ->] node[above, sloped] {$\mathsf{q}$} (b13);
      \draw (v12) edge[semithick, ->] node[above, sloped] {$\mathsf{r}$} (w10);
      \draw (b15) edge[semithick, ->] node[above, sloped] {$\mathsf{q}$} (a14);
    \end{scope}

    \begin{scope}[thick,o-o,Ftrivial]
      
      \path (w0) node[above=2pt] (above_w0) {};
      \path (w10) node[above=2pt] (above_w10) {};

      \draw plot[smooth] coordinates {
        (above_w0)
        (0,2.75)
        (3,3)
        (6,2.75)
        (above_w10)
      };
      \draw (a4) -- (a14);
      \draw (b3) -- (b13);

      \draw (-0.75,-1.5) -- (0.75,-1.5);
    \end{scope}
    \node[below] at (0,-1.5) {$\Trivial$};

    \begin{scope}[thick,dashed,o-o,Fdeblank]

      \path (b6) node[above] (above_b6) {};
      \path (b5) node[above] (above_b5) {};
      \path (b15) node[above=3pt] (above_b15) {};
      
      \draw plot[smooth] coordinates {
        (above_b5)
        (3,2.05)
        (above_b15)
      };
      \draw plot[smooth] coordinates {
        (above_b6)
        (3,2.5)
        (above_b15)
      };

      \draw (2.25,-1.5) -- (3.75,-1.5);
    \end{scope}
    \node[below] at (3,-1.5) {$\Deblank$};

    \begin{scope}[thick,densely dotted,o-o,Fhybrid]
      \path (b1) node[below=2pt] (below_b1) {};
      \path (b11) node[below=2pt] (below_b11) {};
      \path (u2) node[below] (below_u2) {};
      \path (v12) node[below] (below_v12) {};

      \draw plot[smooth] coordinates {
        (below_u2)
        (0.75,-0.5)
        (3,-0.75)
        (5.25,-0.5)
        (below_v12)
      };

      \draw plot[smooth] coordinates {
        (below_b1)
        (-0,-1.15)
        (3,-1.25)
        (6,-1.15)
        (below_b11)
      };

      \draw (5.25,-1.5) -- (6.75,-1.5);
    \end{scope}
    \node[below] at (6,-1.5) {$\Hybrid$};
  \end{scope}
\end{tikzpicture}
\end{center}
\caption{\label{fig:rdf-alignment}Progressive alignment of two RDF graphs.}
  
\end{figure}

Throughout most of the development we fix a single \emph{combined}
graph $G=(N_G,E_G,\ell_G)$, which (see
Section~\ref{sec:basicsdatamodel}) is the disjoint union of the source
graph $G_1=(N_1,E_1,\ell_1)$ and target graph $G_2=(N_2,E_2,\ell_2)$
we want to align.Figure~\ref{fig:rdf-alignment} shows such a union
whose evolution can be described as replacing the equivalent blank
nodes $\B_2$ and $\B_3$ with a single blank node $\B_4$ and renaming
the URI $\mathsf{u}$ to $\mathsf{v}$. While the blank node $\B_1$ has
not been modified, in the second graph it has a different identifier
$\B_5$.

\subsection{Alignment by partition}
\label{sec:alignment-by-partition}
Aligning two graphs consists of identifying pairs of corresponding
nodes. We do not require, however, this to be a 1-to-1 correspondence,  as
a node of one graph may have a number of possible matches in the
other. This allows us to model uncertainty of the
correspondence between nodes; and even when the
correspondence is free of uncertainty, we can represent
redundancy in graphs such as the equivalent blank nodes $\B_2$
and $\B_3$ in the graph in Figure~\ref{fig:rdf-alignment}.
Given a partition $\lambda$ we can simply define an alignment of $G_1$ and $G_2$ 
as 
\[
\Align(\lambda) = 
\{ 
(n,m) \in N_1\times N_2 
\mid 
\lambda(n) = \lambda(m)
\}.
\]
Such alignments are precisely those binary relations that have the
{\em crossover property}.  An alignment $A$ of $G_1$ and $G_2$ has
this property if whenever $(n,m)\in A$, $(n,m')\in A$, and $(n',m)\in
A$, then also $(n',m') \in A$.

\eat{Because we
use mainly binary equivalence relations on the combined node set to
define alignment, we identify the consequence of the transitivity of
the equivalence relation.
\begin{definition}
  \label{def:alignment}
  An \emph{alignment} of $G_1=(N_1,E_1,\ell_1)$ and
  $G_2=(N_2,E_2,\ell_2)$ is a binary relation $A\subseteq N_1\times
  N_2$. $A$ has the \emph{crossover} property if for any $n,n'\in N_1$
  and any $m,m'\in N_2$ if $(n,m)\in A$, $(n,m')\in A$, and $(n',m)\in
  A$, then also $(n',m')\in A$. \qed
\end{definition}
Since we represent equivalence relations on the nodes set of $G$ with
partitions of $G$, the alignment defined by a partition $\lambda$ of
$G$ is $\Align(\lambda)=R_\lambda\cap(N_1\times N_2)$, or equivalently,
\[
\Align(\lambda) = 
\{ 
(n,m) \in N_1\times N_2 
\mid 
\lambda(n) = \lambda(m)
\}.
\]
We point out that alignments defined with partitions are precisely
those that have the crossover property.}

An example of an alignment defined with a partition is the
\emph{trivial alignment} that uses label equality on non-blank nodes,
defined with the following partition of $G$ ($n\in N_G$):
\[
\lambda_\Trivial(n) =
\begin{cases}
  \ell_G(n) & \text{if $n$ is a non-blank node,}\\ 
  n & \text{if $n$ is a blank node.}
\end{cases}
\]
Indeed, $\lambda_\Trivial$ aligns only non-blank nodes with the same
label as illustrated in Figure~\ref{fig:rdf-alignment}.

The alignment methods we propose in this paper work progressively,
aligning previously unaligned nodes. The {\em unaligned} nodes in
$G_1$ are those which $\lambda$ does not associate with a node of
$G_2$: $\Unaligned_1(\lambda) = \{n\in N_1 \mid \nexists m\in N_2.\
\lambda(n)=\lambda(m)\}$. $\Unaligned_2(\lambda)$ is defined
similarly, and $\Unaligned(\lambda) = \Unaligned_1(\lambda)\cup
\Unaligned_2(\lambda)$.

\subsection{Partition refinement}
As a first step we employ \emph{partition refinement} technique to
improve on trivial alignment with bisimulation.
\begin{definition}
  \label{def:partition-refinement}
  A (\emph{one-step}) \emph{partition refinement} is a function
  $\Lambda$ that maps one partition of $G$ to another partition of $G$ 
   such that $\Lambda(\lambda)$ is finer than $\lambda$ and
  $\Lambda(\lambda_1)\equiv\Lambda(\lambda_2)$ whenever
  $\lambda_1\equiv\lambda_2$. \qed
\end{definition}
The first condition is natural and requires the process to be monotone;
the second condition allows only those refinement functions that
are independent of the representation of the partition. The refinement function is applied
iteratively to a given initial partition until the process stabilizes
i.e., further applications of the function yield an equivalent partition.
Taken together, these  two conditions guarantee termination.
\begin{definition}
  \label{def:stable-refinement}
  The \emph{refinement} $\Lambda^*(\lambda)$ of $\lambda$
  w.r.t.\ $\Lambda$ is $\Lambda^n(\lambda)$ where $n$ is minimal 
  such that $\Lambda^n(\lambda)\equiv\Lambda^{n+1}(\lambda)$   
  \qed
\end{definition}
$\Lambda^*(\lambda)$ is -- to within recoloring --  a fixpoint
$\Lambda$ on $\lambda$. We incorporate bisimulation by coloring a node with the
combined colors of its outbound node pairs:
\begin{equation}
  \label{eq:recollor}
  \recolor_\lambda(n) = 
  (\lambda(n), \{(\lambda(p), \lambda(o)) \mid (p,o)\in \out_G(n)\}),
\end{equation}
where $\lambda$ is the current partition. The inclusion of the
original color of $n$ is to ensure that the procedure yields
progressively finer partitions. 
We use this function on a selected subset of nodes
without changing the color of the other nodes. Formally, the
(one step) \emph{bisimulation partition refinement}
$\BisimRefine_X(\lambda)$ on $X\subseteq N_G$ is the partition defined
as
\begin{equation}
  \label{eq:partition-refinement}
\lambda'(n)=
\begin{cases}
  \recolor_\lambda(n) &\text{if $n\in X$,}\\
  \lambda(n) & \text{otherwise.}
\end{cases}
\end{equation}
This partition refinement captures bisimulation when applied
to all nodes with the node labeling function $\ell_G$ defining the
initial partition.
\begin{proposition}
  \label{proposition:bisim-refinement}
  For any graph $G=(N_G,E_G,\ell_G)$, the partition
  $\lambda_\Bisim=\BisimRefine_{N_G}^*(\ell_G)$ captures the
  maximal bisimulation on $G$ i.e., $\Align(\lambda_\Bisim) =
  \Bisim(G)$.
\end{proposition}
The color assigned to a node is essentially a
derivation tree rooted at the node, and because of the recursive nature
of the bisimulation process, it can be compactly presented as a DAG and
implemented with a simple hashing technique.
\begin{example}
  \label{ex:2}
  Figure~\ref{fig:bisimulation} shows the fixpoint
  computation of $\lambda_\Bisim$ on the graph in
  Figure~\ref{fig:rdf-graph}, where colors are presented using derivation
  trees.  Note that a node with no outgoing edges, in
  particular a literal or a URI used solely as a predicate, essentially
  maintains the same color through all iterations of the process.  For
  clarity we use only the original color of such nodes and illustrate
  the refinement process only on nodes whose  color changes.
\begin{figure}[htb]
\centering
  \begin{tikzpicture}[scale=0.85]
    \path[use as bounding box] (-.5,0.25) rectangle (9.25,-6.25);
    \begin{scope}[yshift=0cm]
      \node at (1, 0) {\sf w};
      \draw (2.125, -6.25) -- (2.125, .25);
      \node at (3.5, 0) {\sf u};
      \draw (5, -6.25) -- (5, .25);
      \node at (6.5, 0) {$\B_1$};
      \draw (8, -.5) -- (8, .25);
      \draw (8, -6.25) -- (8, -1.5);
      \node at (8.5, 0) {$\B_2$};
      \node at (9, 0) {$\B_3$};
    \end{scope}

    %
    %
    \begin{scope}[yshift=-1cm]
      \draw (0,0.5) -- (9.25,0.5);
      \node[left] at (0,0) (l0) {$\lambda_0$:};
      \begin{scope}[xshift=1cm]
        \korzen{w}{0}

        \token{w}{Cw}{\textsf{w}}
      \end{scope}
      \begin{scope}[xshift=3.5cm]
        \korzen{u}{0}

        \token{u}{Cu}{\textsf{u}}
      \end{scope}
      \begin{scope}[xshift=7.5cm]
        \korzen{B0}{0}

        \token{B0}{CB}{\color{white}$\B$}
      \end{scope}
    \end{scope}

    %
    %
    \begin{scope}[yshift=-2cm]
      \draw (0,0.5) -- (9.25,0.5);
      \node[left] at (0,0) (l1) {$\lambda_1$:};
      \begin{scope}[xshift=1cm]
        \korzen{w}{0}
        \child{B0}{w}{0}{0}
        
        \HullA
        
        \token{w}{Cw}{\textsf{w}}
        \token{B0}{CB}{\color{white}$\B$}
        
        \batton{B0}{Cp}{\textsf{p}}
      \end{scope}
      \begin{scope}[xshift=3.5cm]
        \korzen{u}{0}
        \child{a1}{u}{-60}{0}
        \child{w}{u}{-20}{0}
        \child{B2}{u}{20}{0}
        \child{b1}{u}{60}{0}
        
        \HullB
        
        \token{u}{Cu}{\textsf{u}}
        \token{a1}{Ca}{\textit{a}}
        \token{w}{Cw}{\textsf{w}}
        \token{B2}{CB}{\color{white}$\B$}
        \token{b1}{Cb}{\textit{b}}
        
        \batton{a1}{Cq}{\textsf{q}}
        \batton{w}{Cr}{\textsf{r}}
        \batton{B2}{Cp}{\textsf{p}}
        \batton{b1}{Cq}{\textsf{q}}
      \end{scope}
      \begin{scope}[xshift=6.5cm]
        \korzen{B1}{0}
        \child{b2}{B1}{-20}{0}
        \child{u}{B1}{20}{0}
        
        \HullC

        \token{B1}{CB}{\color{white}$\B$}
        \token{b2}{Cb}{\textit{b}}
        \token{u}{Cu}{\textsf{u}}
        
        \batton{b2}{Cr}{\textsf{r}}
        \batton{u}{Cq}{\textsf{q}}
      \end{scope}
      \begin{scope}[xshift=8.75cm]
        \korzen{B2}{0}
        \child{a2}{B2}{0}{0}
        
        \HullD
        
        \token{B2}{CB}{\color{white}$\B$}
        \token{a2}{Ca}{\textit{a}}
        
        \batton{a2}{Cq}{\textsf{q}}
      \end{scope}
    \end{scope}

    %
    %
    \begin{scope}[yshift=-4cm]
      \draw (0,0.5) -- (9.25,0.5);
      \node[left] at (0,0) (l2) {$\lambda_2$:};
      \begin{scope}[xshift=1cm]
        \korzen{w}{0}
        \child{B1}{w}{-20}{20}
        \child{b2}{B1}{-20}{0}
        \child{u}{B1}{20}{0}
        \child{B2}{w}{20}{0}
        \child{a2}{B2}{20}{0}
        
        \HullE
        \HullAA
        \HullC
        \HullD
        
        \token{w}{Cw}{\textsf{w}}
        \token{B1}{CB}{\color{white}$\B$}
        \token{u}{Cu}{\textsf{u}}
        \token{b2}{Cb}{\textit{b}}
        \token{B2}{CB}{\color{white}$\B$}
        \token{a2}{Ca}{\textit{a}}
        
        \batton{B1}{Cp}{\textsf{p}}
        \batton{u}{Cp}{\textsf{p}}
        \batton{b2}{Cr}{\textsf{r}}
        \batton{B2}{Cp}{\textsf{p}}
        \batton{a2}{Cp}{\textsf{p}}
      \end{scope}
      \begin{scope}[xshift=3.5cm]
        \korzen{u}{0}
        \child{a1}{u}{-60}{0}
        \child{w}{u}{-20}{5}
        \child{B2}{u}{20}{-5}
        \child{b1}{u}{60}{0}
        \child{B0}{w}{-15}{0}
        \child{a2}{B2}{15}{0}
        
        \HullF
        \HullB
        \HullA
        \HullD
        
        \token{u}{Cu}{\textsf{u}}
        \token{a1}{Ca}{\textit{a}}
        \token{w}{Cw}{\textsf{w}}
        \token{B0}{CB}{\color{white}$\B$}
        \token{b1}{Cb}{\textit{b}}
        \token{B2}{CB}{\color{white}$\B$}
        \token{a2}{Ca}{\textit{a}}
        
        \batton{a1}{Cq}{\textsf{q}}
        \batton{w}{Cr}{\textsf{r}}
        \batton{B2}{Cp}{\textsf{p}}
        \batton{b1}{Cq}{\textsf{q}}
        \batton{B0}{Cp}{\textsf{p}}
        \batton{a2}{Cq}{\textsf{q}}
      \end{scope}
      \begin{scope}[xshift=6.35cm]
        \korzen{B1}{-15}
        \child{b2}{B1}{-40}{0}
        \child{u}{B1}{10}{-10}
        \child{a1}{u}{-60}{0}
        \child{w}{u}{-20}{0}
        \child{B2}{u}{20}{0}
        \child{b1}{u}{60}{0}
        
        \HullG
        \HullCTilt
        \HullB
        
        \token{B1}{CB}{\color{white}$\B$}
        \token{b2}{Cb}{\textit{b}}
        \token{u}{Cu}{\textsf{u}}
        \token{a1}{Ca}{\textit{a}}
        \token{w}{Cw}{\textsf{w}}
        \token{B2}{CB}{\color{white}$\B$}
        \token{b1}{Cb}{\textit{b}}
        
        \batton{b2}{Cr}{\textsf{r}}
        \batton{u}{Cq}{\textsf{q}}
        \batton{a1}{Cq}{\textsf{q}}
        \batton{w}{Cr}{\textsf{r}}
        \batton{B2}{Cp}{\textsf{p}}
        \batton{b1}{Cq}{\textsf{q}}
      \end{scope}
      \begin{scope}[xshift=8.75cm]
        \korzen{B2}{0}
        \child{a2}{B2}{0}{0}
        
        \HullDT
        \HullD
        
        \token{B2}{CB}{\color{white}$\B$}
        \token{a2}{Ca}{\textit{a}}
        
        \batton{a2}{Cq}{\textsf{q}}
      \end{scope}
    \end{scope}

    \draw[bend angle=15] (l0) edge[->, bend right] (l1);
    \draw[bend angle=15] (l1) edge[->, bend right] (l2);

  \end{tikzpicture}
  \caption{\label{fig:bisimulation}Fixpoint color computation in $\Bisim$. 
  }
  \end{figure}
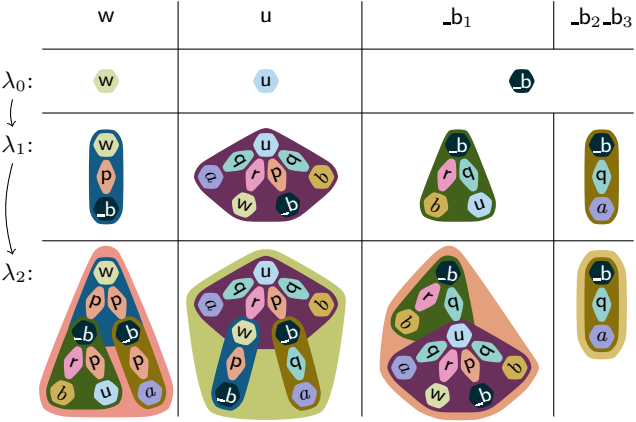
  We use (derivation) trees to represent the colors and point out that
  every iteration essentially unfolds by one level the tree from the
  previous iteration. Depending on the node the derivation tree is
  rooted at the unfolding may yield different results, and
  consequently, different derivation trees may be assigned to nodes that
  previously had the same derivation tree. For instance, the nodes
  $\B_1$, $\B_2$, and $\B_3$ all have initially ($\lambda_0$) the same
  tree, but after the first iteration they are split into two separate
  classes.  Since the partition $\lambda_2$ is the same as the
  previous partition $\lambda_1$, the end result is $\lambda_1$. \qed
\end{example}

\subsection{Deblanking alignment}
Deblanking alignment improves on trivial alignment
(Section~\ref{sec:alignment-by-partition}) by using bisimulation on
blank nodes:
\[
\lambda_\Deblank = \BisimRefine_{\Blanks(G)}^*(\ell_G).
\]
Intuitively, the bisimulation partition refinement assigns to every
blank node a color that characterizes it by its contents (the URIs and
data values reachable from the node). Two blank nodes are aligned if
they have the same contents.  The deblanking partition defines an
equivalence relation that is similar to bisimulation and captures the
essence of the deblanking process (described in the appendix).
\begin{example}
  \label{ex:deblanking}
  Figure~\ref{fig:deblanking-colors} shows the final colors of blank
  nodes of the graphs in Figure~\ref{fig:rdf-alignment} obtained
  during the iterative refinement of deblanking
  alignment. Derivation trees are used for colors; in the case of deblanking alignment the trees are unfolded
  only at blank nodes. The unfolding halts at URIs and literals, and
  in particular, the derivation trees of URIs and literals consist of a root node
  alone. 
  \begin{figure}[htb]
    \centering
    \begin{tikzpicture}[scale=0.85]
      \path[use as bounding box] (-1.5,0.25) rectangle (4.5,-2.125);
      \begin{scope}
        \node at (-.5,0) {$\B_1$};
        \node at (1,0) {$\B_2$};
        \node at (1.5,0) {$\B_3$};
        \node at (2,0) {$\B_4$};
        \node at (3.5,0) {$\B_5$};        
      \end{scope}
      \draw (-1.5,-0.5) -- (4.5,-0.5);
      \draw (2.5, 0.25) -- (2.5,-2.125);
      \draw (0.5, 0.25) -- (0.5,-2.125);

      \begin{scope}[xshift=-.5cm, yshift=-1cm]
        \korzen{B1}{0}
        \child{b2}{B1}{-20}{0}
        \child{u}{B1}{20}{0}

        \HullC
        
        \token{B1}{CB}{\color{white}$\B$}
        \token{b2}{Cb}{\textit{b}}
        \token{u}{Cu}{\textsf{u}}
        
        \batton{b2}{Cr}{\textsf{r}}
        \batton{u}{Cq}{\textsf{q}}
      \end{scope}

      \begin{scope}[xshift=1.5cm, yshift=-1cm]
        \korzen{B2}{0}
        \child{a2}{B2}{0}{0}
        
        \HullD
        
        \token{B2}{CB}{\color{white}$\B$}
        \token{a2}{Ca}{\textit{a}}
        
        \batton{a2}{Cq}{\textsf{q}}
      \end{scope}
      \begin{scope}[xshift=3.5cm, yshift=-1cm]
        \korzen{B1}{0}
        \child{b2}{B1}{-20}{0}
        \child{u}{B1}{20}{0}
        
        \HullH
        
        \token{B1}{CB}{\color{white}$\B$}
        \token{b2}{Cb}{\textit{b}}
        \token{u}{Cv}{\textsf{v}}
        
        \batton{b2}{Cr}{\textsf{r}}
        \batton{u}{Cq}{\textsf{q}}
      \end{scope}

    \end{tikzpicture}
    \caption{\label{fig:deblanking-colors}Colors of blank nodes in $\Deblank$.}
  \end{figure}
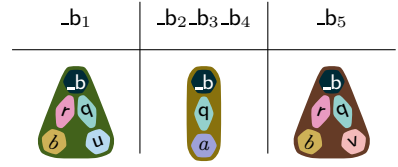

  As a result both the nodes $\B_2$ and $\B_3$ are aligned to
  $\B_4$. On the other hand, the node $\B_1$ is not aligned to $\B_5$
  because their colors differ. \qed
\end{example}

The use of $\BisimRefine$ determines the identity of
a blank node solely on its contents i.e., the identity of nodes
reachable with the outgoing edges. In general, however, the proposed
framework could easily accommodate approaches that consider the
incoming edges or only a selected subset of edges, such as those
determined by the type of a node. 

\subsection{Hybrid alignment}
Hybrid alignment improves on deblanking alignment by applying
bisimulation to unaligned URI nodes. Deblanking
alignment colors those nodes with their URI label; however, the
bisimulation refinement function incorporates this label in every
iteration, and consequently, an unaligned URI cannot be aligned in
this fashion to another unaligned node with a different URI label. We
also note that aligning URI nodes with different labels could permit
aligning previously unaligned blank nodes whose color in the
deblanking alignment might incorporate the different URI
labels. Therefore, we begin by modifying the deblanking partition by
resetting the color of unaligned URI and blank nodes to the neutral
blank color: essentially, we place all unaligned non-literal nodes in
the same cluster and then use bisimulation refinement to define their
identity. Formally, for a set of nodes $X\subseteq N_G$ we define an
auxiliary function that blanks their colors in the given partition:
$\Blank(\lambda, X)=\lambda'$, where
\begin{equation}
\label{eq:blank}
\lambda'(n)=
\begin{cases}
  \B & \text{if $n\in X$}, \\
  \lambda(n) & \text{otherwise.}
\end{cases}\end{equation}
We also identify unaligned non-literal nodes:
\begin{equation}
  \label{eq:unalined-non-literals}
  \mathit{UN}(\lambda) = \Unaligned(\lambda)\minus\Literals(G)
\end{equation}
We define the hybrid partitioning as follows:
\[
\lambda_\Hybrid =
\BisimRefine^*_{\mathit{UN}(\lambda_\Deblank)}(\Blank(\lambda_\Deblank,\mathit{UN}(\lambda_\Deblank))).
\]
Using $\lambda_\Trivial$ instead of $\lambda_\Deblank$ above yields the same result. 
\begin{example}
  \label{ex:hybrid}
  In Figure~\ref{fig:hybrid} we present the final colors of unaligned
  nodes (by $\Deblank$) of the graphs in
  Figure~\ref{fig:rdf-alignment} obtained during the iterative
  refinement procedure of the hybrid alignment. Again, we represent
  the colors as trees but note that technically speaking they are no
  longer derivation trees because for unaligned nodes we use the blank
  label rather than the label of the node (cf.~colors of \textsf{u}
  and \textsf{v}).
  \begin{figure}[htb]
    \centering
    \begin{tikzpicture}[scale=0.85]
      \path[use as bounding box] (-2,0.125) rectangle (6,-4.125);
      \begin{scope}
        \node at (-.25,0) {\sf u};
        \node at (.25,0) {\sf v};
        \node at (3.75,0) {$\B_1$};        
        \node at (4.35,0) {$\B_5$};        
      \end{scope}
      \draw (-2,-0.5) -- (6,-0.5);
      \draw (2, 0.25) -- (2,-4.25);

      \begin{scope}[xshift=0cm, yshift=-1.125cm]
        \korzen{u}{0}
        \child{a1}{u}{-60}{0}
        \child{w}{u}{-20}{0}
        \child{B2}{u}{20}{-20}
        \child{b1}{u}{60}{0}
        \child{a2}{B2}{0}{0}

        \HullIT
        \HullI
        \HullBB
        \HullD

        \token{u}{CB}{\color{white}$\B$}
        \token{a1}{Ca}{\textit{a}}
        \token{w}{Cw}{\textsf{w}}
        \token{B2}{CB}{\color{white}$\B$}
        \token{b1}{Cb}{\textit{b}}
        \token{a2}{Ca}{\textit{a}}
        
        \batton{a1}{Cq}{\textsf{q}}
        \batton{w}{Cr}{\textsf{r}}
        \batton{B2}{Cp}{\textsf{p}}
        \batton{b1}{Cq}{\textsf{q}}
        \batton{a2}{Cq}{\textsf{q}}
      \end{scope}

      \begin{scope}[xshift=4cm, yshift=-1.125cm]
        \korzen{B1}{-15}
        \child{b2}{B1}{-40}{0}
        \child{u}{B1}{10}{-10}
        \child{a1}{u}{-60}{0}
        \child{w}{u}{-20}{0}
        \child{B2}{u}{20}{-20}
        \child{b1}{u}{60}{0}
        \child{a2}{B2}{0}{0}
        
        \HullJ
        \HullGG
        \HullCCTilt
        \HullI
        \HullBB
        \HullD
        
        \token{B1}{CB}{\color{white}$\B$}
        \token{b2}{Cb}{\textit{b}}
        \token{u}{CB}{\color{white}$\B$}
        \token{a1}{Ca}{\textit{a}}
        \token{w}{Cw}{\textsf{w}}
        \token{B2}{CB}{\color{white}$\B$}
        \token{b1}{Cb}{\textit{b}}
        \token{a2}{Ca}{\textit{a}}
        
        \batton{b2}{Cr}{\textsf{r}}
        \batton{u}{Cq}{\textsf{q}}
        \batton{a1}{Cq}{\textsf{q}}
        \batton{w}{Cr}{\textsf{r}}
        \batton{B2}{Cp}{\textsf{p}}
        \batton{b1}{Cq}{\textsf{q}}
        \batton{a2}{Cq}{\textsf{q}}
      \end{scope}

    \end{tikzpicture}
    \caption{\label{fig:hybrid}Colors of selected nodes in $\Hybrid$.}
  \end{figure}
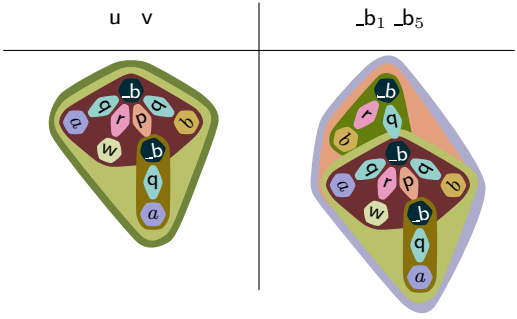
  Also, the depth of the trees may be greater than the number of
  iterations of the refinement process because for aligned nodes
  colors from the deblanking alignments are used, as it is the case
  with the colors of nodes $\B_2$, $\B_3$, and $\B_4$. Naturally, the
  final colors of nodes \textsf{u} and \textsf{v} coincide and
  therefore these two nodes are aligned by $\Hybrid$. Similarly,
  $\Hybrid$ aligns the blank nodes $\B_1$ and $\B_5$.\qed
\end{example}
Finally, we point out that because the constructed partitions have
been defined by improving one on another, the corresponding alignments
create a (proper) hierarchy:
\[
\Align(\lambda_\Trivial) \subseteq \Align(\lambda_\Deblank) \subseteq
\Align(\lambda_\Hybrid).
\]

\section{Similarity alignment}
\label{sec:dissimilarity}
In this section we outline a method of further refining the
bisimulation-based $\Hybrid$ alignment with pairs of similar nodes as
identified with a distance function. More precisely, we define two
distance functions on nodes: $\sigma_\Edit$, which defines an
alignment robust under editing operations but is computationally
expensive, and $\sigma_\Overlap$, which approximates $\sigma_\Edit$
and scales well in practice. We continue to work with a single
combined graph $G=(N_G,E_G,\ell_G)$ that is the disjoint union of the
source graph $G_1=(N_1,E_1,\ell_1)$ and the target graph $G_2=(N_2,
E_2, \ell_2)$.

\subsection{Node distance functions}
We investigate a natural manner of aligning nodes using the standard
notion of a distance function
$\sigma:N_1\times{}N_2\rightarrow{}[0,1]$ in which similar nodes have
a low value of $\sigma$.  Although we do not require $\sigma$ to
satisfy the triangle equality, we will only work with metrics as they
are sometimes used to represent distances in a graph.  
When we wish to combine (add) distance values, in order 
that the result is again in $[0,1]$, we use an infix addition
operator $\oplus:[0,1]\times [0,1]\rightarrow [0,1]$. This operator
can have a number of natural definitions, the only requirement being
compatibility with the triangle inequality: 
$\sigma(n,z)\oplus\sigma(z,m)\leq \sigma(n,m)$ for
any nodes $n,m,z$. We shall use a 
rudimentary definition of this operator: $x\oplus y = \min\{x+y, 1\}$
for $x,y\in{}[0,1]$.

The alignment defined by a node distance function $\sigma$ is
additionally parameterized by a threshold value $\theta\in{}[0,1]$:
\[
\Align_\theta(\sigma)=\{(n,m)\in{}N_1\times{}N_2\mid\sigma(n,m)\leq\theta\}.
\]
Alignments captured with distance functions do not necessarily have
the cross-over property, and are more expressive than alignments
captured with partitions; but for every partition there exists a
distance function and threshold that defines the same alignment.

\subsection{Edit distance alignment}
We define a natural node distance function $\sigma_\Edit$ that
attempts to address two important aspects of evolving RDF data sets:
1) editing changes in the literal values, 2) editing changes in the
structure of the graph. In essence, $\sigma_\Edit$ attempts to refine
the $\Hybrid$ alignment by incorporating string edit distance on
literal values and graph edit distance on non-literal nodes while
iteratively propagating the distances throughout the graph. We omit
its formal definition and illustrate its workings on an example below.
\begin{example}
\label{ex:3}
Consider the two RDF graphs $G_1$ and $G_2$ presented in
Figure~\ref{fig:edit-distance}, where we present the distance between
pairs of nodes. For clarity, we indicate the distances between closest
pairs of nodes.
\begin{figure}[htb]
\centering
\begin{tikzpicture}[>=latex, xscale=0.85]
  \begin{scope}[xshift=-1cm]
    \node at (1.25, 2.5) {$G_1$};
    \node[right] at (1,1) (w0) {$\mathsf{w}$};
    \node[right] at (2.5, 1.75) (u1) {$\mathsf{u}$};
    \node[right] at (2.5,0.5) (v2) {$\mathsf{v}$};
    \node[right] at (4,0) (abc6) {\textit{``abc''}};
    \node[right] at (4,1) (c7) {\textit{``c''}};
    \node[right] at (4,1.65) (b8) {\textit{``b''}};
    \node[right] at (4,2.5) (a9) {\textit{``a''}};
    \draw (w0) edge[semithick, ->] node[above, sloped] {$\mathsf{r}$} (u1);
    \draw (w0) edge[semithick, ->] node[below, sloped] {$\mathsf{q}$} (v2);
    \draw (u1) edge[semithick, ->] node[above, sloped] {$\mathsf{p}$} (a9);
    \draw (u1) edge[semithick, ->] node[above=-3pt, sloped] {$\mathsf{p}$} (b8);
    \draw (u1) edge[semithick, ->] node[below, sloped] {$\mathsf{p}$} (c7);
    \draw (v2) edge[semithick, ->] node[below=-1pt, sloped] {$\mathsf{p}$} (c7);
    \draw (v2) edge[semithick, ->] node[below, sloped] {$\mathsf{q}$} (abc6);
  \end{scope}

  \begin{scope}[xshift=11cm]
    \node at (-1.25, 2.5) {$G_2$};
    \node[left] at (-1,1) (w10) {$\mathsf{w}'$};
    \node[left] at (-2.5, 1.75) (u11) {$\mathsf{u}'$};
    \node[left] at (-2.5,0.5) (v12) {$\mathsf{v}'$};
    \node[left] at (-4,0) (bcd16) {\textit{``ac''}};
    \node[left] at (-4,1.25) (c17) {\textit{``c''}};
    \node[left] at (-4,2.25) (a18) {\textit{``a''}};
    \draw (w10) edge[semithick, ->] node[above, sloped] {$\mathsf{r}$} (u11);
    \draw (w10) edge[semithick, ->] node[below, sloped] {$\mathsf{q}$} (v12);
    \draw (u11) edge[semithick, ->] node[above, sloped] {$\mathsf{p}$} (a18);
    \draw (u11) edge[semithick, ->] node[below, sloped] {$\mathsf{p}$} (c17);
    \draw (v12) edge[semithick, ->] node[below, sloped] {$\mathsf{p}$} (c17);
    \draw (v12) edge[semithick, ->] node[below, sloped] {$\mathsf{q}$} (bcd16);
  \end{scope}

  \begin{scope}[thick,o-o,Fhybrid]
      
    \draw (a9) edge node[above, sloped] {$0$} (a18);
    \draw (c7) edge node[above, sloped] {$0$} (c17);
    \draw (0.75,-1.45) -- (3.25,-1.45);
  \end{scope}
  \node[below, text width=2cm, text centered] at (2,-1.45) {$\Hybrid$ alignment};

  \begin{scope}[thick,dashed,o-o,Fedit]
    \draw (abc6) edge node[above, sloped] {$\frac{1}{3}$} (bcd16);
    
    \draw (3.75,-1.45) -- (6.25,-1.45);
  \end{scope}
  \node[below, text width=2cm, text centered] at (5,-1.45) {string edit distance};

  \begin{scope}[thick,densely dotted,o-o,Fpropagation]
    \path (u1) node[above=2pt] (above_u1) {};
    \path (u11) node[above=2pt] (above_u11) {};
    
    \draw plot[smooth] coordinates {
      (above_u1)
      (2.5,3)
      (4,3.25)
      (6,3.25)
      (7.5,3)
      (above_u11)
    };
    \node[above] at (5,3.25) {$\frac{1}{3}$};

    \path (v2) node[below=2pt] (below_v2) {};
    \path (v12) node[below=2pt] (below_v12) {};
    
    \draw plot[smooth] coordinates {
      (below_v2)
      (2.5,-0.6)
      (4,-0.95)
      (6,-0.95)
      (7.5,-0.6)
      (below_v12)
    };
    \node[above] at (5,-0.95) {$\frac{1}{6}$};

    \path (w0) node[above=2pt] (above_w0) {};
    \path (w10) node[above=2pt] (above_w10) {};
    
    \draw plot[smooth] coordinates {
      (above_w0)
      (1.5,3.5)
      (3,4)
      (7,4)
      (8.5,3.5)
      (above_w10)
    };
    \node[above] at (5,4) {$\frac{1}{4}$};

    \draw (6.75,-1.45) -- (9.25,-1.45);
  \end{scope}
  \node[below, text width=2cm, text centered] at (8,-1.45) {distance propagation};
  
\end{tikzpicture}
\caption{\label{fig:edit-distance}Alignment with the distance function $\sigma_\Edit$.}

\end{figure}
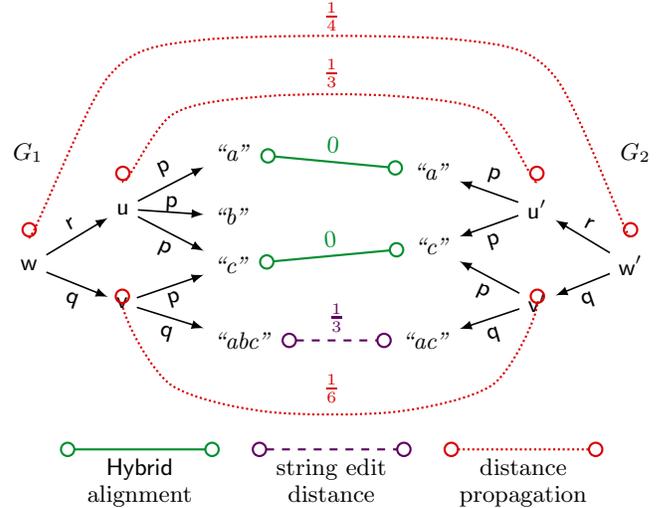
Because $\sigma_\Edit$ refines the $\Hybrid$ alignment, the distance
between any pair of aligned nodes aligned by the $\Hybrid$ partition
is $0$, as it is the case with the trivially aligned literal
nodes \textit{``c''} or with the trivially aligned URIs used as edge
labels. On pairs of unaligned literal nodes we use a string edit
distance. For instance, the distance between the nodes
\textit{``abc''} and \textit{``ac''} is $\frac{1}{3}$ because they
differ by the presence of \textit{b} and the length of both is bounded
by $3$. Note that $\sigma_\Edit$ is $1$ on any other pair
that involves at least one node aligned by the $\Hybrid$
partitioning. For example, the distance between \textit{``a''} and
\textit{``ac''} is $1$ even though the normalized edit distance is
$\frac{1}{2}$.

On unaligned non-literal nodes we propagate the distance established
on other nodes. For example, the distance between $\mathsf{v}$ and
$\mathsf{v}'$ is the average of the distances between the pairs of
nodes on the edges with corresponding labels. Because a node may have
more than one edge with a given label, this process consists of
finding a matching that maximizes this average. Furthermore, when the
numbers of edges with a given label are different the matching
accounts for the differences in a manner consistent with graph edit
distance. The distance between $\mathsf{u}$ and $\mathsf{u}'$ is
$\frac{1}{3}$ because the main difference is in the presence of the
outgoing edge to a node labeled with \textit{``b''} and the size of
their outbound neighborhood is bounded by $3$. Finally, an optimal
matching is found using the Hungarian algorithm~\cite{Ku05}. \qed
\end{example}
The main obstacle to the practical use of $\sigma_\Edit$ is the
significant computational cost of constructing $\sigma_\Edit$: we need
to materialize a matrix whose size is quadratic in the size of the
input graphs, which makes this approach scale poorly. Furthermore,
lower bounds on computing edit distance on strings~\cite{WoCh76} and
graphs~\cite{JuHe06} suggest that the limitations on practical use of
$\sigma_\Edit$ may be fundamental. These observations motivate us to
investigate a heuristic approach that approximates $\sigma_\Edit$ and
has better computational properties.

\subsection{Weighted partitions}
\label{sec:weighted-partitions}
We begin with a simple intuition: for aligning nodes with
$\sigma_\Edit$ we do not necessarily need to know the distance between
every pair of nodes but only wish to find pairs of nodes that are
close to each other. In the context of alignment of evolving RDF, it
is natural to expect the number of such pairs to be relatively low and
more manageable. Ideally, the alignment is a one-to-one correspondence
between the source and the target nodes, which translates to a linear
number of pairs of close nodes. We contrast it with the unlikely case
of a complete bipartite alignment, where every source node is aligned
to every target node, which yields a quadratic number of pairs of
close nodes and has space requirements on a par with materializing
$\sigma_\Edit$. Encouraged by the very good performance of the basic
refinement algorithm, we investigate a generalization of bisimulation
geared towards clustering nodes that are in close proximity.

We begin by generalizing partitions by assigning to every node of a
cluster the distance from the center of the cluster. By using the
triangle inequality, the distance from the center allows us to
estimate the relative distance between any pair of nodes in the same
cluster. Formally, a \emph{weighted partition} of a graph $G$ is a
pair $\xi=(\lambda,\omega)$, where $\lambda:N_G\rightarrow\C$ is a
partition of $G$ and $\omega:N_G\rightarrow[0,1]$ is a \emph{weight}
function. A weighted partition $\xi=(\lambda,\omega)$ defines the
following distance function (for $n,m\in N_G$)
\begin{equation}
\label{eq:distance-weighted-partition}
\sigma_{\xi}(n, m) = 
\begin{cases}
  \omega(n)\oplus\omega(m) &\text{if $\lambda(n)=\lambda(m)$,}\\
  1&\text{otherwise.}
\end{cases}
\end{equation}
Naturally, the alignment defined by the weighted partition $\xi$ w.r.t.\
the threshold value $\theta\in [0,1]$ is
\[
\Align_\theta(\xi) = \{
(n,m) \in N_1 \times N_2 \mid 
\lambda(n)=\lambda(m),\ 
\omega(n)\oplus\omega(m)<\theta
\}.
\]
We propose a method for constructing a weighted
partition that approximates $\sigma_\Edit$ and
thus produces an alignment that approximates
$\Align_\theta(\sigma_\Edit)$.
\begin{example}
\label{ex:4}
Figure~\ref{fig:weighted-partition} presents a weighted partition of
the graph from Figure~\ref{fig:edit-distance} that captures the
essence of the alignment defined with $\sigma_\Edit$.
\begin{figure}[htb]
\centering
\begin{tikzpicture}[>=latex,xscale=0.85]
  \begin{scope}[xshift=-1cm]
    \node at (1.25, 2.75) {$G_1$};
    \node[right] at (1,1) (w0) {$\mathsf{w}$};
    \node[right] at (2.5, 1.75) (u1) {$\mathsf{u}$};
    \node[right] at (2.5,0.5) (v2) {$\mathsf{v}$};
    \node[right] at (4,0) (abc6) {\textit{``abc''}};
    \node[right] at (4,1) (c7) {\textit{``c''}};
    \node[right] at (4,1.75) (b8) {\textit{``b''}};
    \node[right] at (4,2.5) (a9) {\textit{``a''}};
    \draw (w0) edge[semithick, ->] node[above, sloped] {$\mathsf{r}$} (u1);
    \draw (w0) edge[semithick, ->] node[below, sloped] {$\mathsf{q}$} (v2);
    \draw (u1) edge[semithick, ->] node[above, sloped] {$\mathsf{p}$} (a9);
    \draw (u1) edge[semithick, ->] node[above=-4pt, sloped] {$\mathsf{p}$} (b8);
    \draw (u1) edge[semithick, ->] node[below, sloped] {$\mathsf{p}$} (c7);
    \draw (v2) edge[semithick, ->] node[below=-1pt, sloped] {$\mathsf{p}$} (c7);
    \draw (v2) edge[semithick, ->] node[below, sloped] {$\mathsf{q}$} (abc6);
  \end{scope}

  \begin{scope}[xshift=11cm]
    \node at (-1.25, 2.75) {$G_2$};
    \node[left] at (-1,1) (w10) {$\mathsf{w}'$};
    \node[left] at (-2.5, 1.75) (u11) {$\mathsf{u}'$};
    \node[left] at (-2.5,0.5) (v12) {$\mathsf{v}'$};
    \node[left] at (-4,0) (ac16) {\textit{``ac''}};
    \node[left] at (-4,1.25) (c17) {\textit{``c''}};
    \node[left] at (-4,2.25) (a18) {\textit{``a''}};
    \draw (w10) edge[semithick, ->] node[above, sloped] {$\mathsf{r}$} (u11);
    \draw (w10) edge[semithick, ->] node[below, sloped] {$\mathsf{q}$} (v12);
    \draw (u11) edge[semithick, ->] node[above, sloped] {$\mathsf{p}$} (a18);
    \draw (u11) edge[semithick, ->] node[below, sloped] {$\mathsf{p}$} (c17);
    \draw (v12) edge[semithick, ->] node[below, sloped] {$\mathsf{p}$} (c17);
    \draw (v12) edge[semithick, ->] node[below, sloped] {$\mathsf{q}$} (ac16);
  \end{scope}

  \poz{Nabc1}{(abc6.west)}
  \poz{Nabc2}{(abc6.north)}
  \poz{Nabc3}{(ac16.north)}
  \poz{Nabc4}{(ac16.east)}
  \poz{Nabc5}{(ac16.south)}
  \poz{Nabc6}{(abc6.south)}
  \begin{scope}[Wabc, semithick]
    \draw[fill=Wabc, fill opacity=0.2] plot[smooth cycle, tension=0.45] coordinates {
      (Nabc1)
      (Nabc2)
      (Nabc3)
      (Nabc4)
      (Nabc5)
      (Nabc6)
    };
    \path[draw, thick] (abc6.east) circle(1.25pt) node[right, black] {$\frac{2}{9}$};
    \path[draw, thick] (ac16.west) circle(1.25pt) node[left, black] {$\frac{1}{9}$};
  \end{scope}

  \poz{Nc1}{(c7.west)}
  \poz{Nc2}{(c7.north)}
  \poz{Nc3}{(c17.north)}
  \poz{Nc4}{(c17.east)}
  \poz{Nc5}{(c17.south)}
  \poz{Nc6}{(c7.south)}
  \begin{scope}[Wc, semithick]
    \draw[fill=Wc, fill opacity=0.2] plot[smooth cycle, tension=0.4] coordinates {
      (Nc1)
      (Nc2)
      (Nc3)
      (Nc4)
      (Nc5)
      (Nc6)
    };
    \path[draw, thick] (c7.east) circle(1.25pt) node[right, black] {$0$};
    \path[draw, thick] (c17.west) circle(1.25pt) node[left, black] {$0$};
  \end{scope}

  \poz{Na1}{(a9.west)}
  \poz{Na2}{(a9.north)}
  \poz{Na3}{(a18.north)}
  \poz{Na4}{(a18.east)}
  \poz{Na5}{(a18.south)}
  \poz{Na6}{(a9.south)}
  \begin{scope}[Wa, semithick]
    \draw[fill=Wa, fill opacity=0.2] plot[smooth cycle, tension=0.45] coordinates {
      (Na1)
      (Na2)
      (Na3)
      (Na4)
      (Na5)
      (Na6)
    };
    \path[draw, thick] (a9.east) circle(1.25pt) node[right, black] {$0$};
    \path[draw, thick] (a18.west) circle(1.25pt) node[left, black] {$0$};
  \end{scope}

  \poz{Nb1}{(b8.west)}
  \poz{Nb2}{(b8.north)}
  \poz{Nb3}{(4.15, 1.95)}
  \poz{Nb4}{(4.15, 1.55)}
  \poz{Nb5}{(b8.south)}
  \begin{scope}[Wb, semithick]
    \draw[fill=Wb, fill opacity=0.2] plot[smooth cycle, tension=0.75] coordinates {
      (Nb1)
      (Nb2)
      (Nb3)
      (Nb4)
      (Nb5)
    };

    \path[draw, thick] (b8.east) circle(1.25pt) node[right, black] {$0$};
  \end{scope}

  \poz{Nu1}{(1.8, 2.75)}
  \poz{Nu2}{(u1.west)}
  \poz{Nu3}{(u1.south)+(0.0,-0.02)}
  \poz{Nu4}{(u1.east)}
  \poz{Nu5}{(2.2, 2.55)}
  \poz{Nu6r}{(4, 2.9)}
  \poz{Nu6l}{(6, 2.9)}
  \poz{Nu7}{(7.75, 2.55)}
  \poz{Nu8}{(u11.west)+(0.025,-0.1)}
  \poz{Nu9}{(u11.south)+(-0.05,-0.075)}
  \poz{Nu10}{(u11.east)+(-0.05,-0.1)}
  \poz{Nu11}{(8.2, 2.75)}
  \poz{Nu12r}{(5.9,3.175)}
  \poz{Nu12l}{(4.1,3.175)}
  \begin{scope}[Wu, semithick]
    \draw[fill=Wu, fill opacity=0.2] plot[smooth cycle, tension=0.4] coordinates {
      (Nu1)
      (Nu2)
      (Nu3)
      (Nu4)
      (Nu5)
      (Nu6r)
      (Nu6l)
      (Nu7)
      (Nu8)
      (Nu9)
      (Nu10)
      (Nu11)
      (Nu12r)
      (Nu12l)
    };
    \path[draw, thick] (u1.north) circle(1.25pt) node[above, black] {$\frac{1}{3}$};
    \path[draw, thick] (u11.north) circle(1.25pt) node[above, black] {$0$};
  \end{scope}

  \poz{Nv1}{(1.8, -.45)}
  \poz{Nv2}{(v2.west)}
  \poz{Nv3}{(v2.north)+(0.025,0.025)}
  \poz{Nv4}{(v2.east)+(0.05,0.05)}
  \poz{Nv5}{(2.25, -.25)}
  \poz{Nv6l}{(4, -0.55)}
  \poz{Nv6r}{(6, -0.55)}
  \poz{Nv7}{(7.8, -.25)}
  \poz{Nv8}{(v12.west)+(0.075,0.05)}
  \poz{Nv9}{(v12.north)+(0.05,0.0)}
  \poz{Nv10}{(v12.east)+(0.0,-0.01)}
  \poz{Nv11}{(8.2, -.45)}
  \poz{Nv12r}{(5.9,-.85)}
  \poz{Nv12l}{(4.1,-.85)}
  \begin{scope}[Wv, semithick]
    \draw[fill=Wv, fill opacity=0.2] plot[smooth cycle] coordinates {
      (Nv1)
      (Nv2)
      (Nv3)
      (Nv4)
      (Nv5)
      (Nv6l)
      (Nv6r)
      (Nv7)
      (Nv8)
      (Nv9)
      (Nv10)
      (Nv11)
      (Nv12r)
      (Nv12l)
    };
    \path[draw, thick] (v2.south) circle(1.25pt) node[below, black] {$\frac{1}{9}$};
    \path[draw, thick] (v12.south) circle(1.25pt) node[below, black] {$\frac{1}{18}$};
  \end{scope}

  \poz{Nw1}{(.85, 3)}
  \poz{Nw1'}{(.1, 2)}
  \poz{Nw2}{(w0.west)}
  \poz{Nw3}{(w0.south)+(0,-0.05)}
  \poz{Nw4}{(w0.east)}
  \poz{Nw4a}{(0.6,2)}
  \poz{Nw5}{(1.05, 2.75)}
  \poz{Nw6r}{(4, 3.25)}
  \poz{Nw6l}{(6, 3.25)}
  \poz{Nw7}{(8.95, 2.75)}
  \poz{Nw8z}{(9.3,2)}
  \poz{Nw8}{(w10.west)}
  \poz{Nw9}{(w10.south)+(0,-0.0125)}
  \poz{Nw10}{(w10.east)+(-0.05,0)}
  \poz{Nw10a}{(9.85,2)}
  \poz{Nw11}{(9.15, 3)}
  \poz{Nw12l}{(5.9,3.55)}
  \poz{Nw12r}{(4.1,3.55)}
  \begin{scope}[Ww, semithick]
    \draw[fill=Ww, fill opacity=0.2] plot[smooth cycle, tension=0.4] coordinates {
      (Nw1)
      (Nw1')
      (Nw2)
      (Nw3)
      (Nw4)
      (Nw4a)
      (Nw5)
      (Nw6r)
      (Nw6l)
      (Nw7)
      (Nw8z)
      (Nw8)
      (Nw9)
      (Nw10)
      (Nw10a)
      (Nw11)
      (Nw12l)
      (Nw12r)
    };
    \path[draw, thick] (w0.north) circle(1.25pt) node[above, black] {$\frac{2}{9}$};
    \path[draw, thick] (w10.north) circle(1.25pt) node[above, black] {$\frac{1}{36}$};
  \end{scope}
\end{tikzpicture}
\caption{\label{fig:weighted-partition}Weighted partition approximating $\sigma_\Edit$.}
\end{figure}
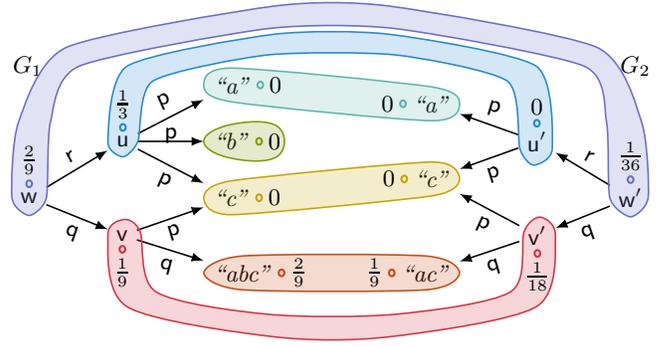
For instance, the distance between the nodes \textit{``abc''} and
\textit{``ac''} is $\frac{2}{9}\oplus\frac{1}{9}=\frac{1}{3}$ and the
distance between nodes the nodes $\mathsf{w}$ and $\mathsf{w}'$ is
$\frac{2}{9}\oplus\frac{1}{36}=\frac{1}{4}$.  However,
the two node distance functions are not equal: for the nodes $u$
and $v'$ the weighted partition defines distance  $1$ because
those nodes are in different clusters while $\sigma_\Edit$ gives this
pair a lower value of $\frac{1}{3}$ because one outgoing edge can be
matched. \qed
\end{example}
We view the weight function $\omega$ of a weighted partition
$\xi=(\lambda, \omega)$ only as a measure of uncertainty that a given
node $n$ has been correctly assigned to its the cluster (labeled)
$\lambda(n)$. The node $n$ belongs to the cluster $\lambda(n)$ even
for the extreme weight value $\omega(n)=1$ and the weight value is
only taken under consideration during the construction of the
alignment. While the precise threshold value $\theta$ identifies the
sets of nodes unaligned by $\Align_\theta(\xi)$, for simplicity of
notations we use the same definitions of unaligned nodes as for
standard partitions: a node of one graph is unaligned if it belongs to
a class with no nodes of the opposite graph. We remark that our
methods can be easily extended to incorporate the threshold value in
identifying unaligned nodes.

\subsection{Enrichment}
Our approach can be described as a simple iterative procedure: start
with an initial weighted partition and while there exist previously
unaligned but close and easily identifiable pairs of nodes, enrich the
partition correspondingly and propagate this information to other
unaligned nodes. The exact method used to identify pairs of close
nodes typically depends on the precise nature of data and later on we
propose one generic method. Here, we present a general approach of
enriching a given weighted partition with newly discovered pairs of
close nodes.

We assume a weighted partition $\xi=(\lambda,\omega)$ of $G$, and the
newly discovered pairs of close nodes are given in the form of a
weighted bipartite graph $H=(A,B,M,d)$, where
$A\subseteq\Unaligned_1(\xi)$ is a subset of unaligned source nodes,
$B\subseteq\Unaligned_2(\xi)$ is a subset of unaligned target nodes,
$M\subseteq A\times B$ is the set of newly discovered close pairs of
nodes, and $d:M\rightarrow [0,1]$ is a distance function on those
pairs of nodes. We view $H$ as an undirected graph and w.l.o.g. assume
that no node in $H$ is isolated i.e., every node in $H$ is connected
to at least one node (isolated nodes can be removed from
consideration). For two arbitrary nodes $v$ and $w$ of $H$, we use
$d^*(v,w)$ the length of the shortest path connecting $v$ and $w$
calculated using $\oplus$, and $1$ if $v$ and $w$ are not connected.
A number of ways of enriching $\xi$ with $H$ can be envisioned, and we
use a rather simple one because in practice $H$ will have a very
sparse structure (close to a one-to-one correspondence).

In the first step we decompose $H$ into a maximal set of disjoint
connected components $\mathcal{X}=\{X_1,\ldots,X_k\}$, where two nodes
belong to the same component if and only if they are
connected. Because we work with $H$ with no isolated nodes, every
component $X_i$ contains both nodes from $A$ and $B$. To incorporate
these components into the weighted partition we need to assign to
every element of each component a weight value that is consistent with
the distances in $H$ i.e., we need to define a function
$w:\bigcup\mathcal{X}\rightarrow [0,1]$ such that for any $X_i$, any
$a\in A\cap X_i$, and any $b\in B\cap X_i$ we have $d^*(a,b) \leq
w(a)\oplus w(b)$. We propose a simple approach where for every source
node in a component we take the half of the maximum distance to any
target node in the same component and {\em vice versa}.  Now, the
\emph{enrichment} of $\xi$ by $H$ is a weighted coloring
$\Enrich(\xi,H) = (\lambda',\omega')$, where
\begin{align*}
&  \lambda'(n)=
  \begin{cases}
    X_i & \text{if $n\in A\cup B$ and $n\in X_i$,}\\
    \lambda(n) & \text{otherwise}
  \end{cases}\\
& \omega'(n)=
  \begin{cases}
    w(n) & \text{if $n\in A\cup B$,}\\
    \omega(n) & \text{otherwise.}
  \end{cases}
\end{align*}
\subsection{Propagation}
Once the newly discovered pairs of close nodes have been incorporated
into the weighted coloring, we propagate this new information in a
manner inspired by the coloring refinement procedure that allows to
identify further close nodes.

The weight of the new color will be an average of the weights of the
colors of outbound nodes that constitute the new color:
\[
\reweight_\omega(n) = 
\bigoplus\left.\left\{
  \frac{\omega(p)\oplus\omega(o)}{|\out_G(n)|} 
  \right|
  (p,o)\in \out_G(n)
  \right\},
\]
where $\omega$ is the current weight function of a weighted
partition. This function is defined only for
nodes having one or more outgoing edges; for a node $n\in N_G$ with no
outgoing edges we set $\reweight_\omega(n)=\omega(n)$. Analogously to
the refinement procedure, we recolor only a selected subset of
(previously unaligned) nodes. We define one-step refinement of a
weighted coloring $\xi=(\lambda,\omega)$ on a set of nodes $X\subseteq
N_G$ as $\BisimRefine_X(\xi)=(\lambda',\omega')$ where $\lambda'$ is
defined as for non-weighted partitions in
\eqref{eq:partition-refinement}, and
\begin{align*}
  & \omega'(n) =
  \begin{cases}
    \reweight_\omega(n), &\text{if $n\in X$},\\
    \omega(n) & \text{otherwise.}
  \end{cases}
\end{align*}
We shall apply this refinement operation iteratively until the
partition no longer changes and the weights stabilize i.e., the weight
assigned to any node changes by less than some fixed small value
$\epsilon>0$. The property that ensures stabilization is that the
initial weights of the nodes in $X$ will all be $0$, and will only
increase during the refinement process We use $\BisimRefine_X^*(\xi)$
to denote the weighted partition obtained with sufficient iterations
of $\BisimRefine$ to ensure a fix-point partition
(cf. Definition~\ref{def:stable-refinement}) and stabilization of the
weight value function. The exact definition is presented in appendix
of the complete version of the paper.

Because we use propagation extensively we introduce a convenient
shorthand $\Propagate(\xi)$ that propagates the alignment information
in a weighted partition $\xi=(\lambda,\omega)$ to unaligned nodes.
The set of unaligned non-literal nodes $\mathit{UN}(\xi)$ is defined
as for non-weighted partitions~\eqref{eq:unalined-non-literals} and we
extend the blank-out operation to weighted partitions: for a set of
(unaligned) nodes $X\subseteq N_G$ let $\Blank(\xi,X) =
(\lambda',\omega')$, where $\lambda'$ is defined as for non-weighted
partitions \eqref{eq:blank} and (for $n\in N_G$)
\begin{align*}
  \omega'(n)=
  \begin{cases}
    0 &\text{if $n\in X$,}\\
    \omega(n) & \text{otherwise.}
  \end{cases}
\end{align*}
Finally, we define
\[
\Propagate(\xi) =
\BisimRefine_{\mathit{UN}(\xi)}^*(\Blank(\xi,\mathit{UN}(\xi))).
\]
There is a natural relationship between the propagation and the
hybrid partition obtained with the coloring refinement algorithm:
$
\Propagate((\lambda_\Trivial,\mathbb0)) =
\Propagate((\lambda_\Deblank,\mathbb0)) =
(\lambda_\Hybrid,\mathbb0)$, where $\mathbb{0}$ is a constant weight
function that assigns $0$ to every node.

\subsection{Overlap heuristic}
\label{sec:overlap}
Similar nodes are identified with a heuristic based on the
overlap measure combined with inverted indexes and identification of
least frequent elements as outlined in
Algorithm~\ref{alg:overlap-heuristic}. Recall that the overlap measure
between two sets of objects $O_1$ and $O_2$ is defined as the fraction
of elements in common over the number of all elements:
\[
\overlap(O_1, O_2) = \frac{|O_1\cap O_2|}{|O_1 \cup O_2|},
\]
This similarity measure has a natural distance counterpart that
measures the fraction of elements present in exactly one of the sets
($\div$ is the symmetric difference operator):
\[
\diff(O_1,O_2) = \frac{|O_1\div O_2|}{|O_1\cup O_2|} = 1 -
\overlap(O_1, O_2),
\]
Note that $\diff(X,X)=0$ but since the above formula is valid only if
one of the sets $O_1$ and $O_2$ is nonempty, we set $\diff(\emptyset,
\emptyset)=0$ and $\overlap(\emptyset,\emptyset)=1$.

\begin{algorithm}
  \caption{\label{alg:overlap-heuristic}Overlap heuristic}
  \FUNCTION $\OverlapMatch(A,B,\theta,\mathit{char},\dist)$\\
  {\bf Input:} $A, B$ -- two (disjoint) sets of nodes\\
  \phantom{\bf Input:} $\theta\in  {[0,1]}$ -- similarity threshold value\\
  \phantom{\bf Input:} $\mathit{char}:A\cup{}B\rightarrow\mathcal{P}(\mathcal{O})$ -- node characterizing function\\
  \phantom{\bf Input:} $\sigma:A\times B\rightarrow {[0,1]}$ -- similarity measure \\
  {\bf Output:} $(A,B,M,w)$ -- weighted bipartite graph\\
  \LN $O\colonequals\bigcup\{\mathit{char}(n) \mid n \in B\}$\\
  \LN $\Inv\colonequals\mathbf{hashtable}()$\\
  \LN $\freq\colonequals\mathbf{hashtable}()$\\
  \LN \FOR $o\in O$ \DO\\
  \LN \TAB $\Inv[o]\colonequals\{n\in B \mid o\in \mathit{char}(n)\}$\\
  \LN \TAB $\freq[o]\colonequals |\Inv[o]|$\\
  \LN $M\colonequals\emptyset$\\
  \LN $w\colonequals\mathbf{hashtable}()$\\
  \LN \FOR $n\in A$ \DO\\
  \LN \TAB $C \colonequals \emptyset$\\
  \LN \TAB $(o_1,\ldots,o_k) \colonequals
  \mathsf{sort}(\mathit{char}(n), \freq)$\hspace{10pt}\makebox[10pt][l]{\fontshape{it}\fontsize{6}{10}\selectfont //$\freq[o_i]\leq\freq[o_{i+1}]$}\\
  \LN \TAB \FOR $i\in\{1,\ldots, \lceil k*\theta \rceil\}$ \DO\\
  \LN \TAB \TAB \FOR $m\in \Inv[o_i]$ \DO\\
  \LN \TAB \TAB \TAB \IF $\overlap(\mathit{char}(n),\mathit{char}(m)) \geq \theta$ \THEN\\
  \LN \TAB \TAB \TAB \TAB $C\colonequals C\cup\{m\}$\\
  \LN \TAB \FOR $m\in C$ \DO\\
  \LN \TAB \TAB \IF $\sigma(n,m)< \theta$ \THEN\\
  \LN \TAB \TAB \TAB $M\colonequals M\cup\{(n,m)\}$\\
  \LN \TAB \TAB \TAB $w(n,m) \colonequals \sigma(n,m)$\\
  \LN \RETURN $(A,B,M,w)$
\end{algorithm}

Our approach (Algorithm~\ref{alg:overlap-heuristic}) identifies
candidate pairs of nodes by representing a node $n$ with a set of
objects $\mathit{char}(n)$ that characterize $n$ in a manner that
exhibits a high coincidence between $\sigma^\dist(n, m) <\theta$ and
$\diff(\mathit{char}(n), \mathit{char}(m)) <\theta$. Intuitively, the
more similar two nodes are the more objects they have in common. We
use inverted indexes to identify pairs of nodes that have the same
object in common and we use frequency counts to use the less frequent,
and thus more discriminating, objects when identifying the set $C$ of
potentially close nodes. Additionally, we use the threshold value
$\theta$ to inspect only a fraction of all objects $char(n)$
characterizing the node $n$ since this fraction must contain objects
of any node $m$ that has overlap above $\theta$. Every candidate pair
is then tested with a distance function $\sigma$ that filters out the
wrong candidates (this function needs not have the same definition as
$\sigma^\dist$).

\subsection{Overlap alignment}
We use the overlap heuristic to construct a weighted partition
$\xi_\Overlap$ (Algorithm~\ref{alg:overlap-partition}) and the
corresponding \emph{overlap} alignment. It defines a distance measure
$\sigma_\Overlap$ that for the purposes of alignment closely captures
the edit distance $\sigma_\Edit$. First, literal nodes are
characterized with the function $\mathit{split}$ that takes the label
of the literal node and splits it into a set of words and the
similarity measure $\sigma_\Literals$ is defined in the same way as
$\sigma_\Dist$ on literal nodes. Then, for a given weighted partition
$\xi=(\lambda,\omega)$, non-literal nodes are characterized with the
set of colors of their outgoing edges
\[
\outcolor_\xi(n) = \{(\lambda(p),\lambda(o)) \mid (p,o) \in \out_G(n)\}.
\]
The distance function on non-literals $\sigma_{\mathit{NL}}^\xi$ is
defined in a manner that captures the weight of the optimal (Hungarian
algorithm) matching among the outgoing edges of two nodes given that
only the weight function is at our disposal. For $n\in N_1$ and $m\in
N_2$ the value of $\sigma^{\mathit{NL}}_\xi(n,m)$ is defined as
\[
\bigoplus
\left.\left\{
\frac{\sigma_\xi(p_1,p_2) \oplus \sigma_\xi(o_1, o_2)
}{
f
}
\right|
((p_1,o_1), (p_2, o_2)) \in M
\right\} 
\oplus \frac{R}{f},
\]
where $\sigma_\xi$ is the distance on nodes induced by $\xi$ as
defined in~\eqref{eq:distance-weighted-partition},
$f=\max\{|\outcolor_\xi(n)|,|\outcolor_\xi(m)|\}$, $M$ is a binary
relation coupling the outgoing edges of $n$ and $m$ with the same
color and having the same position in the list of outgoing edges with
the same colors ordered by their weight, and $R$ is the number of
outgoing edges of $n$ and $m$ that are not coupled in $M$ (when one
node has more outgoing edges of a given color than the other
node). Interestingly, computing $M$ and $R$ can be easily done without
the use of the Hungarian algorithm. The overlap heuristic is applied
iteratively until no further close pair of non-literal nodes can be
found.
\begin{algorithm}
  \caption{\label{alg:overlap-partition}Overlap weighted partition}
  \FUNCTION $\Overlap(G,\theta)$\\
  {\bf Input:} $G=G_1\uplus G_2$ -- combined graph\\
  {\bf Parameter:} $\theta\in {[0,1]}$ -- similarity threshold value\\
  \LN $\xi_0 \colonequals (\lambda_\Hybrid, \mathbb{0})$\\
  \LN $A_0\colonequals \Unaligned_1(\xi_0)\cap \Literals(G_1)$\\
  \LN $B_0\colonequals \Unaligned_2(\xi_0)\cap \Literals(G_2)$\\
  \LN $H_0 \colonequals \OverlapMatch(A_0,B_0,\theta,\mathit{split},\sigma_\Literals)$\\
  \LN $i\colonequals 0$\\
  \LN \DO \\
  \LN \TAB $i \colonequals i + 1$\\
  \LN \TAB $\xi_i\colonequals\Propagate(\Enrich(\xi_{i-1},H_{i-1}))$\\
  \LN \TAB $A_i\colonequals \Unaligned_1(\xi_i)\minus \Literals(G_1)$\\
  \LN \TAB $B_i\colonequals \Unaligned_2(\xi_i)\minus \Literals(G_2)$\\
  \LN \TAB $H_i \colonequals \OverlapMatch(A_i, B_i, \theta, \outcolor_{\xi_i},\sigma^{\mathit{NL}}_{\xi_i})$\\
  \LN \UNTIL $H_i$ has no edges\\
  \LN \RETURN $\xi_i$
\end{algorithm}
The fundamental result is that the overlap alignment only aligns pairs
of nodes that are similar and is stated below. 
\begin{theorem}
  \label{thm:1}
  Let $\xi_\Overlap=(\lambda_\Overlap, \omega_\Overlap)$ be the
  overlap alignment of $G_1=(N_1, E_1,\ell_1)$ and $G_2=(N_2,
  E_2,\ell_2)$. Then, for any $n\in N_1$ and $m\in N_2$, if
  $\lambda_\Overlap(n)=\lambda_\Overlap(m)$, then $\sigma_\Edit(n,
  m)\leq \omega_\Overlap(n)*\omega_\Overlap(m)$.
\end{theorem}
\section{Experimental results}
\label{sec:experiments}
In this section we report on experimental evaluation of the proposed
solutions on three practical data sets: EFO -- Experimental Factor
Ontology~\cite{EFO} supported by the European
Bioinformatics Institute, GtoPdb -- The Guide to Pharmacology
database~\cite{IUPHAR} supported by the
International Union of Pharmacologists and the British Pharmacological
Society, and a subset of DBpedia with Wikipedia category information.  

A brief word about the first two databases.  EFO provides a systematic
description of many experimental variables available in other
databases and combines parts of several biological ontologies.  It is
expressed in OWL, which is in turn reasonably directly expressed in
RDF.  GtoPdb is a relational database that contains curated
information from hundreds of experts about drugs in clinical use and
some experimental drugs, together with information on the cellular
targets of the drugs and their mechanisms of action in the body. We
converted GtoPdb into RDF using a standard (W3C recommended) approach
\cite{DMRDF12}. Both databases are evolving; new versions are released
every few months.  Both databases are usually viewed through a Web
interface and despite their internal representions have a similar
general nature consisting of classification hierarchies along with a
rich annotation.  However their representation in RDF is very
different.  For example, in EFO the notion of a subclass is directly
represented, while in GtoPdb it is inferred from the relational
database.

\subsection{Experimental Factor Ontology (EFO)}
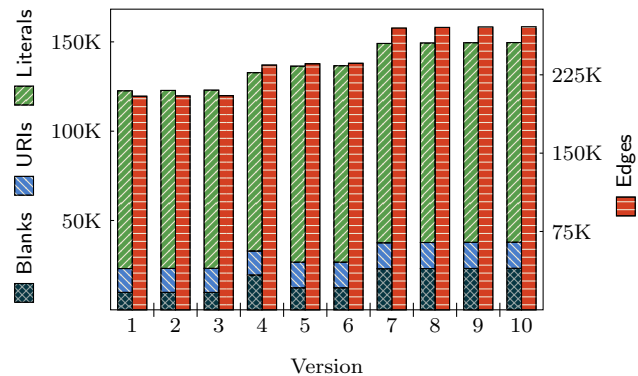
\begin{figure}[htb]
  \centering
\begin{tikzpicture}[>=latex]
\path[use as bounding box] (-1.25,-0.75) rectangle ( 7.0 , 4 );
\draw[fill=white] (0,0) -- ( 5.75 ,0) -- ( 5.75 , 4 ) -- (0, 4 ) -- cycle;
\draw ( 0.575 ,  -0.09375 ) -- ( 0.575 , 0.09375 );
\draw ( 1.15 ,  -0.09375 ) -- ( 1.15 , 0.09375 );
\draw ( 1.725 ,  -0.09375 ) -- ( 1.725 , 0.09375 );
\draw ( 2.3 ,  -0.09375 ) -- ( 2.3 , 0.09375 );
\draw ( 2.875 ,  -0.09375 ) -- ( 2.875 , 0.09375 );
\draw ( 3.45 ,  -0.09375 ) -- ( 3.45 , 0.09375 );
\draw ( 4.025 ,  -0.09375 ) -- ( 4.025 , 0.09375 );
\draw ( 4.6 ,  -0.09375 ) -- ( 4.6 , 0.09375 );
\draw ( 5.175 ,  -0.09375 ) -- ( 5.175 , 0.09375 );
\path ( 0.2875 ,0) node[below] {\small  $ 1 $};
\path ( 0.8625 ,0) node[below] {\small  $ 2 $};
\path ( 1.4375 ,0) node[below] {\small  $ 3 $};
\path ( 2.0125 ,0) node[below] {\small  $ 4 $};
\path ( 2.5875 ,0) node[below] {\small  $ 5 $};
\path ( 3.1625 ,0) node[below] {\small  $ 6 $};
\path ( 3.7375 ,0) node[below] {\small  $ 7 $};
\path ( 4.3125 ,0) node[below] {\small  $ 8 $};
\path ( 4.8875 ,0) node[below] {\small  $ 9 $};
\path ( 5.4625 ,0) node[below] {\small  $ 10 $};
\path ( 2.875 , -0.75) node {\small  Version};
\draw ( 0.046875 , 1.18831882594  ) -- (0.0, 1.18831882594 );
\path (0.0, 1.18831882594  ) node[left] {\small $ 50 $K };
\draw ( 0.046875 , 2.37663765188  ) -- (0.0, 2.37663765188 );
\path (0.0, 2.37663765188  ) node[left] {\small $ 100 $K };
\draw ( 0.046875 , 3.56495647782  ) -- (0.0, 3.56495647782 );
\path (0.0, 3.56495647782  ) node[left] {\small $ 150 $K };
\draw ( 5.703125 , 1.04218051956  ) -- ( 5.75 , 1.04218051956 );
\path ( 5.75 , 1.04218051956  ) node[right] {\small $ 75 $K };
\draw ( 5.703125 , 2.08436103912  ) -- ( 5.75 , 2.08436103912 );
\path ( 5.75 , 2.08436103912  ) node[right] {\small $ 150 $K };
\draw ( 5.703125 , 3.12654155869  ) -- ( 5.75 , 3.12654155869 );
\path ( 5.75 , 3.12654155869  ) node[right] {\small $ 225 $K };
\begin{scope}[yshift= 1.125 cm,xshift= 6.6 cm]
\draw[Eedges] (0.125,0.125) rectangle (0.375,0.375);
\draw[EedgesP] (0.125,0.125) rectangle (0.375,0.375);
\path (0.25,0.375) node[right,rotate=90] {\small  \sf Edges};
\end{scope}
\begin{scope}[yshift= -1.15 cm,xshift= -1.4 cm]
\begin{scope}[yshift=1.125cm,xshift=0.0cm]
\draw[Eblanks] (0.125,0.125) rectangle (0.375,0.375);
\draw[EblanksP] (0.125,0.125) rectangle (0.375,0.375);
\path (0.25,0.375) node[right,rotate=90] {\small  \sf Blanks};
\end{scope}
\begin{scope}[yshift=2.55cm,xshift=0.0cm]
\draw[Euris] (0.125,0.125) rectangle (0.375,0.375);
\draw[EurisP] (0.125,0.125) rectangle (0.375,0.375);
\path (0.25,0.375) node[right,rotate=90] {\small  \sf URIs};
\end{scope}
\begin{scope}[yshift=3.75cm,xshift=0.0cm]
\draw[Eliterals] (0.125,0.125) rectangle (0.375,0.375);
\draw[EliteralsP] (0.125,0.125) rectangle (0.375,0.375);
\path (0.25,0.375) node[right,rotate=90] {\small  \sf Literals};
\end{scope}
\end{scope}
\draw[ Eedges ] ( 0.2875 ,0.0) rectangle ( 0.48125 , 2.84192900666 );
\draw[EedgesP] ( 0.2875 ,0.0) rectangle ( 0.48125 , 2.84192900666 );
\draw[ Eblanks ] ( 0.09375 ,0.0) rectangle ( 0.2875 , 0.231722171058 );
\draw[EblanksP] ( 0.09375 ,0.0) rectangle ( 0.2875 , 0.231722171058 );
\draw[ Euris ] ( 0.09375 , 0.231722171058 ) rectangle ( 0.2875 , 0.549431092362 );
\draw[EurisP] ( 0.09375 , 0.231722171058 ) rectangle ( 0.2875 , 0.549431092362 );
\draw[ Eliterals ] ( 0.09375 , 0.549431092362 ) rectangle ( 0.2875 , 2.91418555598 );
\draw[EliteralsP] ( 0.09375 , 0.549431092362 ) rectangle ( 0.2875 , 2.91418555598 );
\draw[ Eedges ] ( 0.8625 ,0.0) rectangle ( 1.05625 , 2.84616720744 );
\draw[EedgesP] ( 0.8625 ,0.0) rectangle ( 1.05625 , 2.84616720744 );
\draw[ Eblanks ] ( 0.66875 ,0.0) rectangle ( 0.8625 , 0.231722171058 );
\draw[EblanksP] ( 0.66875 ,0.0) rectangle ( 0.8625 , 0.231722171058 );
\draw[ Euris ] ( 0.66875 , 0.231722171058 ) rectangle ( 0.8625 , 0.550690710318 );
\draw[EurisP] ( 0.66875 , 0.231722171058 ) rectangle ( 0.8625 , 0.550690710318 );
\draw[ Eliterals ] ( 0.66875 , 0.550690710318 ) rectangle ( 0.8625 , 2.91865363477 );
\draw[EliteralsP] ( 0.66875 , 0.550690710318 ) rectangle ( 0.8625 , 2.91865363477 );
\draw[ Eedges ] ( 1.4375 ,0.0) rectangle ( 1.63125 , 2.84868233643 );
\draw[EedgesP] ( 1.4375 ,0.0) rectangle ( 1.63125 , 2.84868233643 );
\draw[ Eblanks ] ( 1.24375 ,0.0) rectangle ( 1.4375 , 0.231460740917 );
\draw[EblanksP] ( 1.24375 ,0.0) rectangle ( 1.4375 , 0.231460740917 );
\draw[ Euris ] ( 1.24375 , 0.231460740917 ) rectangle ( 1.4375 , 0.55142746799 );
\draw[EurisP] ( 1.24375 , 0.231460740917 ) rectangle ( 1.4375 , 0.55142746799 );
\draw[ Eliterals ] ( 1.24375 , 0.55142746799 ) rectangle ( 1.4375 , 2.92221859125 );
\draw[EliteralsP] ( 1.24375 , 0.55142746799 ) rectangle ( 1.4375 , 2.92221859125 );
\draw[ Eedges ] ( 2.0125 ,0.0) rectangle ( 2.20625 , 3.25617491958 );
\draw[EedgesP] ( 2.0125 ,0.0) rectangle ( 2.20625 , 3.25617491958 );
\draw[ Eblanks ] ( 1.81875 ,0.0) rectangle ( 2.0125 , 0.462992780963 );
\draw[EblanksP] ( 1.81875 ,0.0) rectangle ( 2.0125 , 0.462992780963 );
\draw[ Euris ] ( 1.81875 , 0.462992780963 ) rectangle ( 2.0125 , 0.783529901072 );
\draw[EurisP] ( 1.81875 , 0.462992780963 ) rectangle ( 2.0125 , 0.783529901072 );
\draw[ Eliterals ] ( 1.81875 , 0.783529901072 ) rectangle ( 2.0125 , 3.15491518374 );
\draw[EliteralsP] ( 1.81875 , 0.783529901072 ) rectangle ( 2.0125 , 3.15491518374 );
\draw[ Eedges ] ( 2.5875 ,0.0) rectangle ( 2.78125 , 3.27322499288 );
\draw[EedgesP] ( 2.5875 ,0.0) rectangle ( 2.78125 , 3.27322499288 );
\draw[ Eblanks ] ( 2.39375 ,0.0) rectangle ( 2.5875 , 0.293752413773 );
\draw[EblanksP] ( 2.39375 ,0.0) rectangle ( 2.5875 , 0.293752413773 );
\draw[ Euris ] ( 2.39375 , 0.293752413773 ) rectangle ( 2.5875 , 0.633873028133 );
\draw[EurisP] ( 2.39375 , 0.293752413773 ) rectangle ( 2.5875 , 0.633873028133 );
\draw[ Eliterals ] ( 2.39375 , 0.633873028133 ) rectangle ( 2.5875 , 3.24154362615 );
\draw[EliteralsP] ( 2.39375 , 0.633873028133 ) rectangle ( 2.5875 , 3.24154362615 );
\draw[ Eedges ] ( 3.1625 ,0.0) rectangle ( 3.35625 , 3.27995053116 );
\draw[EedgesP] ( 3.1625 ,0.0) rectangle ( 3.35625 , 3.27995053116 );
\draw[ Eblanks ] ( 2.96875 ,0.0) rectangle ( 3.1625 , 0.293752413773 );
\draw[EblanksP] ( 2.96875 ,0.0) rectangle ( 3.1625 , 0.293752413773 );
\draw[ Euris ] ( 2.96875 , 0.293752413773 ) rectangle ( 3.1625 , 0.63382549538 );
\draw[EurisP] ( 2.96875 , 0.293752413773 ) rectangle ( 3.1625 , 0.63382549538 );
\draw[ Eliterals ] ( 2.96875 , 0.63382549538 ) rectangle ( 3.1625 , 3.24710495826 );
\draw[EliteralsP] ( 2.96875 , 0.63382549538 ) rectangle ( 3.1625 , 3.24710495826 );
\draw[ Eedges ] ( 3.7375 ,0.0) rectangle ( 3.93125 , 3.75022406881 );
\draw[EedgesP] ( 3.7375 ,0.0) rectangle ( 3.93125 , 3.75022406881 );
\draw[ Eblanks ] ( 3.54375 ,0.0) rectangle ( 3.7375 , 0.546983155581 );
\draw[EblanksP] ( 3.54375 ,0.0) rectangle ( 3.7375 , 0.546983155581 );
\draw[ Euris ] ( 3.54375 , 0.546983155581 ) rectangle ( 3.7375 , 0.892617569294 );
\draw[EurisP] ( 3.54375 , 0.546983155581 ) rectangle ( 3.7375 , 0.892617569294 );
\draw[ Eliterals ] ( 3.54375 , 0.892617569294 ) rectangle ( 3.7375 , 3.54370933722 );
\draw[EliteralsP] ( 3.54375 , 0.892617569294 ) rectangle ( 3.7375 , 3.54370933722 );
\draw[ Eedges ] ( 4.3125 ,0.0) rectangle ( 4.50625 , 3.75693571136 );
\draw[EedgesP] ( 4.3125 ,0.0) rectangle ( 4.50625 , 3.75693571136 );
\draw[ Eblanks ] ( 4.11875 ,0.0) rectangle ( 4.3125 , 0.54924096135 );
\draw[EblanksP] ( 4.11875 ,0.0) rectangle ( 4.3125 , 0.54924096135 );
\draw[ Euris ] ( 4.11875 , 0.54924096135 ) rectangle ( 4.3125 , 0.895540833606 );
\draw[EurisP] ( 4.11875 , 0.54924096135 ) rectangle ( 4.3125 , 0.895540833606 );
\draw[ Eliterals ] ( 4.11875 , 0.895540833606 ) rectangle ( 4.3125 , 3.54953209946 );
\draw[EliteralsP] ( 4.11875 , 0.895540833606 ) rectangle ( 4.3125 , 3.54953209946 );
\draw[ Eedges ] ( 4.8875 ,0.0) rectangle ( 5.08125 , 3.76191038637 );
\draw[EedgesP] ( 4.8875 ,0.0) rectangle ( 5.08125 , 3.76191038637 );
\draw[ Eblanks ] ( 4.69375 ,0.0) rectangle ( 4.8875 , 0.55026291554 );
\draw[EblanksP] ( 4.69375 ,0.0) rectangle ( 4.8875 , 0.55026291554 );
\draw[ Euris ] ( 4.69375 , 0.55026291554 ) rectangle ( 4.8875 , 0.897537209233 );
\draw[EurisP] ( 4.69375 , 0.55026291554 ) rectangle ( 4.8875 , 0.897537209233 );
\draw[ Eliterals ] ( 4.69375 , 0.897537209233 ) rectangle ( 4.8875 , 3.55373874811 );
\draw[EliteralsP] ( 4.69375 , 0.897537209233 ) rectangle ( 4.8875 , 3.55373874811 );
\draw[ Eedges ] ( 5.4625 ,0.0) rectangle ( 5.65625 , 3.7647173259 );
\draw[EedgesP] ( 5.4625 ,0.0) rectangle ( 5.65625 , 3.7647173259 );
\draw[ Eblanks ] ( 5.26875 ,0.0) rectangle ( 5.4625 , 0.551783963637 );
\draw[EblanksP] ( 5.26875 ,0.0) rectangle ( 5.4625 , 0.551783963637 );
\draw[ Euris ] ( 5.26875 , 0.551783963637 ) rectangle ( 5.4625 , 0.899367220225 );
\draw[EurisP] ( 5.26875 , 0.551783963637 ) rectangle ( 5.4625 , 0.899367220225 );
\draw[ Eliterals ] ( 5.26875 , 0.899367220225 ) rectangle ( 5.4625 , 3.5555687591 );
\draw[EliteralsP] ( 5.26875 , 0.899367220225 ) rectangle ( 5.4625 , 3.5555687591 );
\end{tikzpicture}

  \caption{\label{fig:efo-versions}EFO dataset versions.}
\end{figure}

In Figure~\ref{fig:efo-versions} we present node and edge counts of
ten versions of the Experimental Factor Ontology (versions 2.34
through 2.44; 2.40 not accessible).
We point out that literals comprise over $75$\% of the contents of
every version. While the number of URIs is generally proportional to
the total number of nodes (approx. $10$\%), the number of blank nodes
fluctuates quite significantly $7$--$15$\%. After a closer inspection
we found that the fluctuations are due to duplication (bisimilar blank
nodes) and normalized counts of blank nodes do not fluctuate but grow
steadily.

\begin{figure}[htb]
  \centering
\begin{tikzpicture}
\path[use as bounding box] (0,-0.85) rectangle ( 3.75 , 3.3 );
\path (1.5,3.0) node[above=5pt] {\footnotesize $\Trivial$};
\draw[gray, very thin, step=0.3cm] (0.0,0.0) grid (3.0,3.0);
\path (0.15,0.0)
node[below] {\footnotesize $1$};
\path (0.45,0.0)
node[below] {\footnotesize $2$};
\path (0.75,0.0)
node[below] {\footnotesize $3$};
\path (1.05,0.0)
node[below] {\footnotesize $4$};
\path (1.35,0.0)
node[below] {\footnotesize $5$};
\path (1.65,0.0)
node[below] {\footnotesize $6$};
\path (1.95,0.0)
node[below] {\footnotesize $7$};
\path (2.25,0.0)
node[below] {\footnotesize $8$};
\path (2.55,0.0)
node[below] {\footnotesize $9$};
\path (2.85,0.0)
node[below] {\footnotesize $10$};
\node at ( 1.5 , -0.75) {\footnotesize Source version};
\path (0.0,0.15)
node[left] {\footnotesize $1$};
\path (0.0,0.45)
node[left] {\footnotesize $2$};
\path (0.0,0.75)
node[left] {\footnotesize $3$};
\path (0.0,1.05)
node[left] {\footnotesize $4$};
\path (0.0,1.35)
node[left] {\footnotesize $5$};
\path (0.0,1.65)
node[left] {\footnotesize $6$};
\path (0.0,1.95)
node[left] {\footnotesize $7$};
\path (0.0,2.25)
node[left] {\footnotesize $8$};
\path (0.0,2.55)
node[left] {\footnotesize $9$};
\path (0.0,2.85)
node[left] {\footnotesize $10$};
\node[rotate=90] at (-0.75, 1.5 ) {\footnotesize Target version};
\begin{scope}[xshift=0.125cm]
\draw[top color=EmaxR, bottom color=EminR]
(3.25,0) rectangle (3.5,3) ;
\path (3.375,0) node[below] {$0.4$};
\path (3.375,3) node[above] {$1.0$};
\end{scope}
\draw[black, fill=EmaxR!60!EminR]
(0.0,0.0) rectangle (0.3,0.3) ;
\draw[black, fill=EmaxR!35!EminR]
(0.0,0.3) rectangle (0.3,0.6) ;
\draw[black, fill=EmaxR!33!EminR]
(0.0,0.6) rectangle (0.3,0.9) ;
\draw[black, fill=EmaxR!24!EminR]
(0.0,0.9) rectangle (0.3,1.2) ;
\draw[black, fill=EmaxR!23!EminR]
(0.0,1.2) rectangle (0.3,1.5) ;
\draw[black, fill=EmaxR!23!EminR]
(0.0,1.5) rectangle (0.3,1.8) ;
\draw[black, fill=EmaxR!15!EminR]
(0.0,1.8) rectangle (0.3,2.1) ;
\draw[black, fill=EmaxR!11!EminR]
(0.0,2.1) rectangle (0.3,2.4) ;
\draw[black, fill=EmaxR!11!EminR]
(0.0,2.4) rectangle (0.3,2.7) ;
\draw[black, fill=EmaxR!11!EminR]
(0.0,2.7) rectangle (0.3,3.0) ;
\draw[black, fill=EmaxR!60!EminR]
(0.3,0.3) rectangle (0.6,0.6) ;
\draw[black, fill=EmaxR!33!EminR]
(0.3,0.6) rectangle (0.6,0.9) ;
\draw[black, fill=EmaxR!24!EminR]
(0.3,0.9) rectangle (0.6,1.2) ;
\draw[black, fill=EmaxR!23!EminR]
(0.3,1.2) rectangle (0.6,1.5) ;
\draw[black, fill=EmaxR!23!EminR]
(0.3,1.5) rectangle (0.6,1.8) ;
\draw[black, fill=EmaxR!15!EminR]
(0.3,1.8) rectangle (0.6,2.1) ;
\draw[black, fill=EmaxR!12!EminR]
(0.3,2.1) rectangle (0.6,2.4) ;
\draw[black, fill=EmaxR!12!EminR]
(0.3,2.4) rectangle (0.6,2.7) ;
\draw[black, fill=EmaxR!11!EminR]
(0.3,2.7) rectangle (0.6,3.0) ;
\draw[black, fill=EmaxR!60!EminR]
(0.6,0.6) rectangle (0.9,0.9) ;
\draw[black, fill=EmaxR!26!EminR]
(0.6,0.9) rectangle (0.9,1.2) ;
\draw[black, fill=EmaxR!25!EminR]
(0.6,1.2) rectangle (0.9,1.5) ;
\draw[black, fill=EmaxR!24!EminR]
(0.6,1.5) rectangle (0.9,1.8) ;
\draw[black, fill=EmaxR!15!EminR]
(0.6,1.8) rectangle (0.9,2.1) ;
\draw[black, fill=EmaxR!12!EminR]
(0.6,2.1) rectangle (0.9,2.4) ;
\draw[black, fill=EmaxR!12!EminR]
(0.6,2.4) rectangle (0.9,2.7) ;
\draw[black, fill=EmaxR!12!EminR]
(0.6,2.7) rectangle (0.9,3.0) ;
\draw[black, fill=EmaxR!45!EminR]
(0.9,0.9) rectangle (1.2,1.2) ;
\draw[black, fill=EmaxR!17!EminR]
(0.9,1.2) rectangle (1.2,1.5) ;
\draw[black, fill=EmaxR!17!EminR]
(0.9,1.5) rectangle (1.2,1.8) ;
\draw[black, fill=EmaxR!9!EminR]
(0.9,1.8) rectangle (1.2,2.1) ;
\draw[black, fill=EmaxR!6!EminR]
(0.9,2.1) rectangle (1.2,2.4) ;
\draw[black, fill=EmaxR!6!EminR]
(0.9,2.4) rectangle (1.2,2.7) ;
\draw[black, fill=EmaxR!6!EminR]
(0.9,2.7) rectangle (1.2,3.0) ;
\draw[black, fill=EmaxR!54!EminR]
(1.2,1.2) rectangle (1.5,1.5) ;
\draw[black, fill=EmaxR!28!EminR]
(1.2,1.5) rectangle (1.5,1.8) ;
\draw[black, fill=EmaxR!18!EminR]
(1.2,1.8) rectangle (1.5,2.1) ;
\draw[black, fill=EmaxR!15!EminR]
(1.2,2.1) rectangle (1.5,2.4) ;
\draw[black, fill=EmaxR!15!EminR]
(1.2,2.4) rectangle (1.5,2.7) ;
\draw[black, fill=EmaxR!15!EminR]
(1.2,2.7) rectangle (1.5,3.0) ;
\draw[black, fill=EmaxR!54!EminR]
(1.5,1.5) rectangle (1.8,1.8) ;
\draw[black, fill=EmaxR!19!EminR]
(1.5,1.8) rectangle (1.8,2.1) ;
\draw[black, fill=EmaxR!16!EminR]
(1.5,2.1) rectangle (1.8,2.4) ;
\draw[black, fill=EmaxR!15!EminR]
(1.5,2.4) rectangle (1.8,2.7) ;
\draw[black, fill=EmaxR!15!EminR]
(1.5,2.7) rectangle (1.8,3.0) ;
\draw[black, fill=EmaxR!41!EminR]
(1.8,1.8) rectangle (2.1,2.1) ;
\draw[black, fill=EmaxR!10!EminR]
(1.8,2.1) rectangle (2.1,2.4) ;
\draw[black, fill=EmaxR!10!EminR]
(1.8,2.4) rectangle (2.1,2.7) ;
\draw[black, fill=EmaxR!10!EminR]
(1.8,2.7) rectangle (2.1,3.0) ;
\draw[black, fill=EmaxR!41!EminR]
(2.1,2.1) rectangle (2.4,2.4) ;
\draw[black, fill=EmaxR!12!EminR]
(2.1,2.4) rectangle (2.4,2.7) ;
\draw[black, fill=EmaxR!12!EminR]
(2.1,2.7) rectangle (2.4,3.0) ;
\draw[black, fill=EmaxR!41!EminR]
(2.4,2.4) rectangle (2.7,2.7) ;
\draw[black, fill=EmaxR!12!EminR]
(2.4,2.7) rectangle (2.7,3.0) ;
\draw[black, fill=EmaxR!41!EminR]
(2.7,2.7) rectangle (3.0,3.0) ;
\draw[black, fill=EmaxR!35!EminR]
(0.0,0.3) rectangle (0.3,0.6) ;
\draw[black, fill=EmaxR!33!EminR]
(0.3,0.6) rectangle (0.6,0.9) ;
\draw[black, fill=EmaxR!26!EminR]
(0.6,0.9) rectangle (0.9,1.2) ;
\draw[black, fill=EmaxR!17!EminR]
(0.9,1.2) rectangle (1.2,1.5) ;
\draw[black, fill=EmaxR!28!EminR]
(1.2,1.5) rectangle (1.5,1.8) ;
\draw[black, fill=EmaxR!19!EminR]
(1.5,1.8) rectangle (1.8,2.1) ;
\draw[black, fill=EmaxR!10!EminR]
(1.8,2.1) rectangle (2.1,2.4) ;
\draw[black, fill=EmaxR!12!EminR]
(2.1,2.4) rectangle (2.4,2.7) ;
\draw[black, fill=EmaxR!12!EminR]
(2.4,2.7) rectangle (2.7,3.0) ;
\end{tikzpicture}
\begin{tikzpicture}
\path[use as bounding box] (0,-0.85) rectangle ( 3 , 3.3 );
\path (1.5,3.0) node[above=5pt] {\footnotesize $\Deblank$};
\draw[gray, very thin, step=0.3cm] (0.0,0.0) grid (3.0,3.0);
\path (0.15,0.0)
node[below] {\footnotesize $1$};
\path (0.45,0.0)
node[below] {\footnotesize $2$};
\path (0.75,0.0)
node[below] {\footnotesize $3$};
\path (1.05,0.0)
node[below] {\footnotesize $4$};
\path (1.35,0.0)
node[below] {\footnotesize $5$};
\path (1.65,0.0)
node[below] {\footnotesize $6$};
\path (1.95,0.0)
node[below] {\footnotesize $7$};
\path (2.25,0.0)
node[below] {\footnotesize $8$};
\path (2.55,0.0)
node[below] {\footnotesize $9$};
\path (2.85,0.0)
node[below] {\footnotesize $10$};
\node at ( 1.5 , -0.75) {\footnotesize Source version};
\path (3.0,0.15)
node[right] {\footnotesize $1$};
\path (3.0,0.45)
node[right] {\footnotesize $2$};
\path (3.0,0.75)
node[right] {\footnotesize $3$};
\path (3.0,1.05)
node[right] {\footnotesize $4$};
\path (3.0,1.35)
node[right] {\footnotesize $5$};
\path (3.0,1.65)
node[right] {\footnotesize $6$};
\path (3.0,1.95)
node[right] {\footnotesize $7$};
\path (3.0,2.25)
node[right] {\footnotesize $8$};
\path (3.0,2.55)
node[right] {\footnotesize $9$};
\path (3.0,2.85)
node[right] {\footnotesize $10$};
\node[rotate=90] at ( 3.75 , 1.5 ) {\footnotesize Target version};
\draw[black, fill=EmaxR!100!EminR]
(0.0,0.0) rectangle (0.3,0.3) ;
\draw[black, fill=EmaxR!99!EminR]
(0.0,0.3) rectangle (0.3,0.6) ;
\draw[black, fill=EmaxR!73!EminR]
(0.0,0.6) rectangle (0.3,0.9) ;
\draw[black, fill=EmaxR!60!EminR]
(0.0,0.9) rectangle (0.3,1.2) ;
\draw[black, fill=EmaxR!59!EminR]
(0.0,1.2) rectangle (0.3,1.5) ;
\draw[black, fill=EmaxR!58!EminR]
(0.0,1.5) rectangle (0.3,1.8) ;
\draw[black, fill=EmaxR!46!EminR]
(0.0,1.8) rectangle (0.3,2.1) ;
\draw[black, fill=EmaxR!43!EminR]
(0.0,2.1) rectangle (0.3,2.4) ;
\draw[black, fill=EmaxR!43!EminR]
(0.0,2.4) rectangle (0.3,2.7) ;
\draw[black, fill=EmaxR!43!EminR]
(0.0,2.7) rectangle (0.3,3.0) ;
\draw[black, fill=EmaxR!100!EminR]
(0.3,0.3) rectangle (0.6,0.6) ;
\draw[black, fill=EmaxR!73!EminR]
(0.3,0.6) rectangle (0.6,0.9) ;
\draw[black, fill=EmaxR!60!EminR]
(0.3,0.9) rectangle (0.6,1.2) ;
\draw[black, fill=EmaxR!59!EminR]
(0.3,1.2) rectangle (0.6,1.5) ;
\draw[black, fill=EmaxR!59!EminR]
(0.3,1.5) rectangle (0.6,1.8) ;
\draw[black, fill=EmaxR!47!EminR]
(0.3,1.8) rectangle (0.6,2.1) ;
\draw[black, fill=EmaxR!43!EminR]
(0.3,2.1) rectangle (0.6,2.4) ;
\draw[black, fill=EmaxR!43!EminR]
(0.3,2.4) rectangle (0.6,2.7) ;
\draw[black, fill=EmaxR!43!EminR]
(0.3,2.7) rectangle (0.6,3.0) ;
\draw[black, fill=EmaxR!100!EminR]
(0.6,0.6) rectangle (0.9,0.9) ;
\draw[black, fill=EmaxR!83!EminR]
(0.6,0.9) rectangle (0.9,1.2) ;
\draw[black, fill=EmaxR!81!EminR]
(0.6,1.2) rectangle (0.9,1.5) ;
\draw[black, fill=EmaxR!81!EminR]
(0.6,1.5) rectangle (0.9,1.8) ;
\draw[black, fill=EmaxR!65!EminR]
(0.6,1.8) rectangle (0.9,2.1) ;
\draw[black, fill=EmaxR!61!EminR]
(0.6,2.1) rectangle (0.9,2.4) ;
\draw[black, fill=EmaxR!61!EminR]
(0.6,2.4) rectangle (0.9,2.7) ;
\draw[black, fill=EmaxR!61!EminR]
(0.6,2.7) rectangle (0.9,3.0) ;
\draw[black, fill=EmaxR!100!EminR]
(0.9,0.9) rectangle (1.2,1.2) ;
\draw[black, fill=EmaxR!68!EminR]
(0.9,1.2) rectangle (1.2,1.5) ;
\draw[black, fill=EmaxR!68!EminR]
(0.9,1.5) rectangle (1.2,1.8) ;
\draw[black, fill=EmaxR!75!EminR]
(0.9,1.8) rectangle (1.2,2.1) ;
\draw[black, fill=EmaxR!69!EminR]
(0.9,2.1) rectangle (1.2,2.4) ;
\draw[black, fill=EmaxR!68!EminR]
(0.9,2.4) rectangle (1.2,2.7) ;
\draw[black, fill=EmaxR!68!EminR]
(0.9,2.7) rectangle (1.2,3.0) ;
\draw[black, fill=EmaxR!100!EminR]
(1.2,1.2) rectangle (1.5,1.5) ;
\draw[black, fill=EmaxR!99!EminR]
(1.2,1.5) rectangle (1.5,1.8) ;
\draw[black, fill=EmaxR!81!EminR]
(1.2,1.8) rectangle (1.5,2.1) ;
\draw[black, fill=EmaxR!77!EminR]
(1.2,2.1) rectangle (1.5,2.4) ;
\draw[black, fill=EmaxR!77!EminR]
(1.2,2.4) rectangle (1.5,2.7) ;
\draw[black, fill=EmaxR!77!EminR]
(1.2,2.7) rectangle (1.5,3.0) ;
\draw[black, fill=EmaxR!100!EminR]
(1.5,1.5) rectangle (1.8,1.8) ;
\draw[black, fill=EmaxR!82!EminR]
(1.5,1.8) rectangle (1.8,2.1) ;
\draw[black, fill=EmaxR!78!EminR]
(1.5,2.1) rectangle (1.8,2.4) ;
\draw[black, fill=EmaxR!78!EminR]
(1.5,2.4) rectangle (1.8,2.7) ;
\draw[black, fill=EmaxR!78!EminR]
(1.5,2.7) rectangle (1.8,3.0) ;
\draw[black, fill=EmaxR!100!EminR]
(1.8,1.8) rectangle (2.1,2.1) ;
\draw[black, fill=EmaxR!93!EminR]
(1.8,2.1) rectangle (2.1,2.4) ;
\draw[black, fill=EmaxR!93!EminR]
(1.8,2.4) rectangle (2.1,2.7) ;
\draw[black, fill=EmaxR!93!EminR]
(1.8,2.7) rectangle (2.1,3.0) ;
\draw[black, fill=EmaxR!100!EminR]
(2.1,2.1) rectangle (2.4,2.4) ;
\draw[black, fill=EmaxR!99!EminR]
(2.1,2.4) rectangle (2.4,2.7) ;
\draw[black, fill=EmaxR!99!EminR]
(2.1,2.7) rectangle (2.4,3.0) ;
\draw[black, fill=EmaxR!100!EminR]
(2.4,2.4) rectangle (2.7,2.7) ;
\draw[black, fill=EmaxR!99!EminR]
(2.4,2.7) rectangle (2.7,3.0) ;
\draw[black, fill=EmaxR!100!EminR]
(2.7,2.7) rectangle (3.0,3.0) ;
\draw[black, fill=EmaxR!99!EminR]
(0.0,0.3) rectangle (0.3,0.6) ;
\draw[black, fill=EmaxR!73!EminR]
(0.3,0.6) rectangle (0.6,0.9) ;
\draw[black, fill=EmaxR!83!EminR]
(0.6,0.9) rectangle (0.9,1.2) ;
\draw[black, fill=EmaxR!68!EminR]
(0.9,1.2) rectangle (1.2,1.5) ;
\draw[black, fill=EmaxR!99!EminR]
(1.2,1.5) rectangle (1.5,1.8) ;
\draw[black, fill=EmaxR!82!EminR]
(1.5,1.8) rectangle (1.8,2.1) ;
\draw[black, fill=EmaxR!93!EminR]
(1.8,2.1) rectangle (2.1,2.4) ;
\draw[black, fill=EmaxR!99!EminR]
(2.1,2.4) rectangle (2.4,2.7) ;
\draw[black, fill=EmaxR!99!EminR]
(2.4,2.7) rectangle (2.7,3.0) ;
\end{tikzpicture}
  \caption{\label{fig:efo-trivial-and-deblank-alignment-matrix}$\Trivial$ and $\Deblank$ alignments (EFO).}
\end{figure}
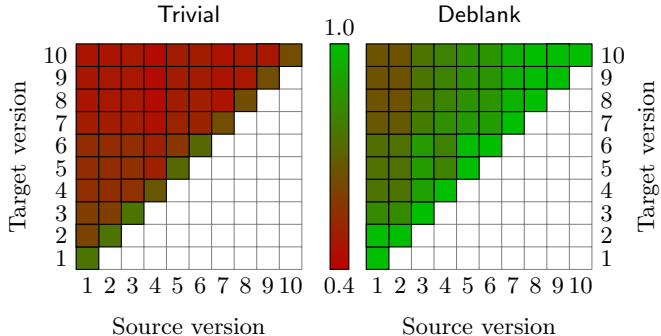

We analyzed the alignments obtained with the presented methods
between any pair of versions of EFO. We also measured the number of
aligned edges -- the results are virtually the same if we measure
the number of aligned nodes. For the trivial and deblanking alignment
in Figure~\ref{fig:efo-trivial-and-deblank-alignment-matrix} we report
the ratio of the number of aligned edges to the total number of edges
in both graphs (edges using precisely the same identifiers are counted
precisely once). 
The diagonal of the matrix is the result of self-alignment, the
alignment of a version with itself, and ideally we wish it to be a
complete alignment with ratio equal to $1$, as it is for the
deblanking alignment. The ratios for trivial alignment are
significantly worse because of the impact of blank nodes that are not
aligned. Overall, we observe an expected descending gradient from the diagonal
towards the upper left point of the matrix, except for version $3$ due
to fluctuations in the number of blanks. This gradient has a natural
explanation: the further apart the two aligned versions are, the more
significant changes they have undergone, and consequently, less edges
can be aligned.

\begin{figure}[htb]
  \centering
\begin{tikzpicture}
\path[use as bounding box] (0,-0.85) rectangle ( 3.75 , 3.3 );
\path (1.5,3.0) node[above=5pt] {\footnotesize $\Hybrid$ vs $\Deblank$};
\draw[gray, very thin, step=0.3cm] (0.0,0.0) grid (3.0,3.0);
\path (0.15,0.0)
node[below] {\footnotesize $1$};
\path (0.45,0.0)
node[below] {\footnotesize $2$};
\path (0.75,0.0)
node[below] {\footnotesize $3$};
\path (1.05,0.0)
node[below] {\footnotesize $4$};
\path (1.35,0.0)
node[below] {\footnotesize $5$};
\path (1.65,0.0)
node[below] {\footnotesize $6$};
\path (1.95,0.0)
node[below] {\footnotesize $7$};
\path (2.25,0.0)
node[below] {\footnotesize $8$};
\path (2.55,0.0)
node[below] {\footnotesize $9$};
\path (2.85,0.0)
node[below] {\footnotesize $10$};
\node at ( 1.5 , -0.75) {\footnotesize Source version};
\path (0.0,0.15)
node[left] {\footnotesize $1$};
\path (0.0,0.45)
node[left] {\footnotesize $2$};
\path (0.0,0.75)
node[left] {\footnotesize $3$};
\path (0.0,1.05)
node[left] {\footnotesize $4$};
\path (0.0,1.35)
node[left] {\footnotesize $5$};
\path (0.0,1.65)
node[left] {\footnotesize $6$};
\path (0.0,1.95)
node[left] {\footnotesize $7$};
\path (0.0,2.25)
node[left] {\footnotesize $8$};
\path (0.0,2.55)
node[left] {\footnotesize $9$};
\path (0.0,2.85)
node[left] {\footnotesize $10$};
\node[rotate=90] at (-0.75, 1.5 ) {\footnotesize Target version};
\begin{scope}[xshift=0.125cm]
\draw[top color=EmaxA, bottom color=EminA]
(3.25,0) rectangle (3.5,3) ;
\path (3.375,0) node[below] {$0$};
\path (3.375,3) node[above] {$12$K};
\end{scope}
\draw[black, fill=EmaxA!0!EminA]
(0.0,0.0) rectangle (0.3,0.3) ;
\draw[black, fill=EmaxA!1!EminA]
(0.0,0.3) rectangle (0.3,0.6) ;
\draw[black, fill=EmaxA!1!EminA]
(0.0,0.6) rectangle (0.3,0.9) ;
\draw[black, fill=EmaxA!1!EminA]
(0.0,0.9) rectangle (0.3,1.2) ;
\draw[black, fill=EmaxA!45!EminA]
(0.0,1.2) rectangle (0.3,1.5) ;
\draw[black, fill=EmaxA!45!EminA]
(0.0,1.5) rectangle (0.3,1.8) ;
\draw[black, fill=EmaxA!27!EminA]
(0.0,1.8) rectangle (0.3,2.1) ;
\draw[black, fill=EmaxA!46!EminA]
(0.0,2.1) rectangle (0.3,2.4) ;
\draw[black, fill=EmaxA!46!EminA]
(0.0,2.4) rectangle (0.3,2.7) ;
\draw[black, fill=EmaxA!46!EminA]
(0.0,2.7) rectangle (0.3,3.0) ;
\draw[black, fill=EmaxA!0!EminA]
(0.3,0.3) rectangle (0.6,0.6) ;
\draw[black, fill=EmaxA!0!EminA]
(0.3,0.6) rectangle (0.6,0.9) ;
\draw[black, fill=EmaxA!0!EminA]
(0.3,0.9) rectangle (0.6,1.2) ;
\draw[black, fill=EmaxA!44!EminA]
(0.3,1.2) rectangle (0.6,1.5) ;
\draw[black, fill=EmaxA!44!EminA]
(0.3,1.5) rectangle (0.6,1.8) ;
\draw[black, fill=EmaxA!26!EminA]
(0.3,1.8) rectangle (0.6,2.1) ;
\draw[black, fill=EmaxA!46!EminA]
(0.3,2.1) rectangle (0.6,2.4) ;
\draw[black, fill=EmaxA!46!EminA]
(0.3,2.4) rectangle (0.6,2.7) ;
\draw[black, fill=EmaxA!46!EminA]
(0.3,2.7) rectangle (0.6,3.0) ;
\draw[black, fill=EmaxA!0!EminA]
(0.6,0.6) rectangle (0.9,0.9) ;
\draw[black, fill=EmaxA!0!EminA]
(0.6,0.9) rectangle (0.9,1.2) ;
\draw[black, fill=EmaxA!0!EminA]
(0.6,1.2) rectangle (0.9,1.5) ;
\draw[black, fill=EmaxA!0!EminA]
(0.6,1.5) rectangle (0.9,1.8) ;
\draw[black, fill=EmaxA!0!EminA]
(0.6,1.8) rectangle (0.9,2.1) ;
\draw[black, fill=EmaxA!20!EminA]
(0.6,2.1) rectangle (0.9,2.4) ;
\draw[black, fill=EmaxA!20!EminA]
(0.6,2.4) rectangle (0.9,2.7) ;
\draw[black, fill=EmaxA!20!EminA]
(0.6,2.7) rectangle (0.9,3.0) ;
\draw[black, fill=EmaxA!0!EminA]
(0.9,0.9) rectangle (1.2,1.2) ;
\draw[black, fill=EmaxA!0!EminA]
(0.9,1.2) rectangle (1.2,1.5) ;
\draw[black, fill=EmaxA!0!EminA]
(0.9,1.5) rectangle (1.2,1.8) ;
\draw[black, fill=EmaxA!0!EminA]
(0.9,1.8) rectangle (1.2,2.1) ;
\draw[black, fill=EmaxA!30!EminA]
(0.9,2.1) rectangle (1.2,2.4) ;
\draw[black, fill=EmaxA!29!EminA]
(0.9,2.4) rectangle (1.2,2.7) ;
\draw[black, fill=EmaxA!29!EminA]
(0.9,2.7) rectangle (1.2,3.0) ;
\draw[black, fill=EmaxA!0!EminA]
(1.2,1.2) rectangle (1.5,1.5) ;
\draw[black, fill=EmaxA!0!EminA]
(1.2,1.5) rectangle (1.5,1.8) ;
\draw[black, fill=EmaxA!1!EminA]
(1.2,1.8) rectangle (1.5,2.1) ;
\draw[black, fill=EmaxA!22!EminA]
(1.2,2.1) rectangle (1.5,2.4) ;
\draw[black, fill=EmaxA!22!EminA]
(1.2,2.4) rectangle (1.5,2.7) ;
\draw[black, fill=EmaxA!22!EminA]
(1.2,2.7) rectangle (1.5,3.0) ;
\draw[black, fill=EmaxA!0!EminA]
(1.5,1.5) rectangle (1.8,1.8) ;
\draw[black, fill=EmaxA!1!EminA]
(1.5,1.8) rectangle (1.8,2.1) ;
\draw[black, fill=EmaxA!22!EminA]
(1.5,2.1) rectangle (1.8,2.4) ;
\draw[black, fill=EmaxA!22!EminA]
(1.5,2.4) rectangle (1.8,2.7) ;
\draw[black, fill=EmaxA!22!EminA]
(1.5,2.7) rectangle (1.8,3.0) ;
\draw[black, fill=EmaxA!0!EminA]
(1.8,1.8) rectangle (2.1,2.1) ;
\draw[black, fill=EmaxA!93!EminA]
(1.8,2.1) rectangle (2.1,2.4) ;
\draw[black, fill=EmaxA!88!EminA]
(1.8,2.4) rectangle (2.1,2.7) ;
\draw[black, fill=EmaxA!88!EminA]
(1.8,2.7) rectangle (2.1,3.0) ;
\draw[black, fill=EmaxA!0!EminA]
(2.1,2.1) rectangle (2.4,2.4) ;
\draw[black, fill=EmaxA!0!EminA]
(2.1,2.4) rectangle (2.4,2.7) ;
\draw[black, fill=EmaxA!0!EminA]
(2.1,2.7) rectangle (2.4,3.0) ;
\draw[black, fill=EmaxA!0!EminA]
(2.4,2.4) rectangle (2.7,2.7) ;
\draw[black, fill=EmaxA!0!EminA]
(2.4,2.7) rectangle (2.7,3.0) ;
\draw[black, fill=EmaxA!0!EminA]
(2.7,2.7) rectangle (3.0,3.0) ;
\draw[black, fill=EmaxA!1!EminA]
(0.0,0.3) rectangle (0.3,0.6) ;
\draw[black, fill=EmaxA!0!EminA]
(0.3,0.6) rectangle (0.6,0.9) ;
\draw[black, fill=EmaxA!0!EminA]
(0.6,0.9) rectangle (0.9,1.2) ;
\draw[black, fill=EmaxA!0!EminA]
(0.9,1.2) rectangle (1.2,1.5) ;
\draw[black, fill=EmaxA!0!EminA]
(1.2,1.5) rectangle (1.5,1.8) ;
\draw[black, fill=EmaxA!1!EminA]
(1.5,1.8) rectangle (1.8,2.1) ;
\draw[black, fill=EmaxA!93!EminA]
(1.8,2.1) rectangle (2.1,2.4) ;
\draw[black, fill=EmaxA!0!EminA]
(2.1,2.4) rectangle (2.4,2.7) ;
\draw[black, fill=EmaxA!0!EminA]
(2.4,2.7) rectangle (2.7,3.0) ;
\end{tikzpicture}
\begin{tikzpicture}
\path[use as bounding box] (0,-0.85) rectangle ( 3 , 3.3 );
\path (1.5,3.0) node[above=5pt] {\footnotesize $\Overlap$ vs $\Hybrid$};
\draw[gray, very thin, step=0.3cm] (0.0,0.0) grid (3.0,3.0);
\path (0.15,0.0)
node[below] {\footnotesize $1$};
\path (0.45,0.0)
node[below] {\footnotesize $2$};
\path (0.75,0.0)
node[below] {\footnotesize $3$};
\path (1.05,0.0)
node[below] {\footnotesize $4$};
\path (1.35,0.0)
node[below] {\footnotesize $5$};
\path (1.65,0.0)
node[below] {\footnotesize $6$};
\path (1.95,0.0)
node[below] {\footnotesize $7$};
\path (2.25,0.0)
node[below] {\footnotesize $8$};
\path (2.55,0.0)
node[below] {\footnotesize $9$};
\path (2.85,0.0)
node[below] {\footnotesize $10$};
\node at ( 1.5 , -0.75) {\footnotesize Source version};
\path (3.0,0.15)
node[right] {\footnotesize $1$};
\path (3.0,0.45)
node[right] {\footnotesize $2$};
\path (3.0,0.75)
node[right] {\footnotesize $3$};
\path (3.0,1.05)
node[right] {\footnotesize $4$};
\path (3.0,1.35)
node[right] {\footnotesize $5$};
\path (3.0,1.65)
node[right] {\footnotesize $6$};
\path (3.0,1.95)
node[right] {\footnotesize $7$};
\path (3.0,2.25)
node[right] {\footnotesize $8$};
\path (3.0,2.55)
node[right] {\footnotesize $9$};
\path (3.0,2.85)
node[right] {\footnotesize $10$};
\node[rotate=90] at ( 3.75 , 1.5 ) {\footnotesize Target version};
\draw[black, fill=EmaxA!0!EminA]
(0.0,0.0) rectangle (0.3,0.3) ;
\draw[black, fill=EmaxA!0!EminA]
(0.0,0.3) rectangle (0.3,0.6) ;
\draw[black, fill=EmaxA!0!EminA]
(0.0,0.6) rectangle (0.3,0.9) ;
\draw[black, fill=EmaxA!0!EminA]
(0.0,0.9) rectangle (0.3,1.2) ;
\draw[black, fill=EmaxA!0!EminA]
(0.0,1.2) rectangle (0.3,1.5) ;
\draw[black, fill=EmaxA!0!EminA]
(0.0,1.5) rectangle (0.3,1.8) ;
\draw[black, fill=EmaxA!0!EminA]
(0.0,1.8) rectangle (0.3,2.1) ;
\draw[black, fill=EmaxA!43!EminA]
(0.0,2.1) rectangle (0.3,2.4) ;
\draw[black, fill=EmaxA!43!EminA]
(0.0,2.4) rectangle (0.3,2.7) ;
\draw[black, fill=EmaxA!43!EminA]
(0.0,2.7) rectangle (0.3,3.0) ;
\draw[black, fill=EmaxA!0!EminA]
(0.3,0.3) rectangle (0.6,0.6) ;
\draw[black, fill=EmaxA!0!EminA]
(0.3,0.6) rectangle (0.6,0.9) ;
\draw[black, fill=EmaxA!0!EminA]
(0.3,0.9) rectangle (0.6,1.2) ;
\draw[black, fill=EmaxA!0!EminA]
(0.3,1.2) rectangle (0.6,1.5) ;
\draw[black, fill=EmaxA!0!EminA]
(0.3,1.5) rectangle (0.6,1.8) ;
\draw[black, fill=EmaxA!0!EminA]
(0.3,1.8) rectangle (0.6,2.1) ;
\draw[black, fill=EmaxA!43!EminA]
(0.3,2.1) rectangle (0.6,2.4) ;
\draw[black, fill=EmaxA!43!EminA]
(0.3,2.4) rectangle (0.6,2.7) ;
\draw[black, fill=EmaxA!43!EminA]
(0.3,2.7) rectangle (0.6,3.0) ;
\draw[black, fill=EmaxA!0!EminA]
(0.6,0.6) rectangle (0.9,0.9) ;
\draw[black, fill=EmaxA!0!EminA]
(0.6,0.9) rectangle (0.9,1.2) ;
\draw[black, fill=EmaxA!0!EminA]
(0.6,1.2) rectangle (0.9,1.5) ;
\draw[black, fill=EmaxA!0!EminA]
(0.6,1.5) rectangle (0.9,1.8) ;
\draw[black, fill=EmaxA!0!EminA]
(0.6,1.8) rectangle (0.9,2.1) ;
\draw[black, fill=EmaxA!43!EminA]
(0.6,2.1) rectangle (0.9,2.4) ;
\draw[black, fill=EmaxA!43!EminA]
(0.6,2.4) rectangle (0.9,2.7) ;
\draw[black, fill=EmaxA!43!EminA]
(0.6,2.7) rectangle (0.9,3.0) ;
\draw[black, fill=EmaxA!0!EminA]
(0.9,0.9) rectangle (1.2,1.2) ;
\draw[black, fill=EmaxA!0!EminA]
(0.9,1.2) rectangle (1.2,1.5) ;
\draw[black, fill=EmaxA!0!EminA]
(0.9,1.5) rectangle (1.2,1.8) ;
\draw[black, fill=EmaxA!0!EminA]
(0.9,1.8) rectangle (1.2,2.1) ;
\draw[black, fill=EmaxA!61!EminA]
(0.9,2.1) rectangle (1.2,2.4) ;
\draw[black, fill=EmaxA!62!EminA]
(0.9,2.4) rectangle (1.2,2.7) ;
\draw[black, fill=EmaxA!62!EminA]
(0.9,2.7) rectangle (1.2,3.0) ;
\draw[black, fill=EmaxA!0!EminA]
(1.2,1.2) rectangle (1.5,1.5) ;
\draw[black, fill=EmaxA!3!EminA]
(1.2,1.5) rectangle (1.5,1.8) ;
\draw[black, fill=EmaxA!3!EminA]
(1.2,1.8) rectangle (1.5,2.1) ;
\draw[black, fill=EmaxA!45!EminA]
(1.2,2.1) rectangle (1.5,2.4) ;
\draw[black, fill=EmaxA!45!EminA]
(1.2,2.4) rectangle (1.5,2.7) ;
\draw[black, fill=EmaxA!46!EminA]
(1.2,2.7) rectangle (1.5,3.0) ;
\draw[black, fill=EmaxA!0!EminA]
(1.5,1.5) rectangle (1.8,1.8) ;
\draw[black, fill=EmaxA!0!EminA]
(1.5,1.8) rectangle (1.8,2.1) ;
\draw[black, fill=EmaxA!42!EminA]
(1.5,2.1) rectangle (1.8,2.4) ;
\draw[black, fill=EmaxA!42!EminA]
(1.5,2.4) rectangle (1.8,2.7) ;
\draw[black, fill=EmaxA!42!EminA]
(1.5,2.7) rectangle (1.8,3.0) ;
\draw[black, fill=EmaxA!0!EminA]
(1.8,1.8) rectangle (2.1,2.1) ;
\draw[black, fill=EmaxA!9!EminA]
(1.8,2.1) rectangle (2.1,2.4) ;
\draw[black, fill=EmaxA!14!EminA]
(1.8,2.4) rectangle (2.1,2.7) ;
\draw[black, fill=EmaxA!14!EminA]
(1.8,2.7) rectangle (2.1,3.0) ;
\draw[black, fill=EmaxA!0!EminA]
(2.1,2.1) rectangle (2.4,2.4) ;
\draw[black, fill=EmaxA!0!EminA]
(2.1,2.4) rectangle (2.4,2.7) ;
\draw[black, fill=EmaxA!0!EminA]
(2.1,2.7) rectangle (2.4,3.0) ;
\draw[black, fill=EmaxA!0!EminA]
(2.4,2.4) rectangle (2.7,2.7) ;
\draw[black, fill=EmaxA!0!EminA]
(2.4,2.7) rectangle (2.7,3.0) ;
\draw[black, fill=EmaxA!0!EminA]
(2.7,2.7) rectangle (3.0,3.0) ;
\draw[black, fill=EmaxA!0!EminA]
(0.0,0.3) rectangle (0.3,0.6) ;
\draw[black, fill=EmaxA!0!EminA]
(0.3,0.6) rectangle (0.6,0.9) ;
\draw[black, fill=EmaxA!0!EminA]
(0.6,0.9) rectangle (0.9,1.2) ;
\draw[black, fill=EmaxA!0!EminA]
(0.9,1.2) rectangle (1.2,1.5) ;
\draw[black, fill=EmaxA!3!EminA]
(1.2,1.5) rectangle (1.5,1.8) ;
\draw[black, fill=EmaxA!0!EminA]
(1.5,1.8) rectangle (1.8,2.1) ;
\draw[black, fill=EmaxA!9!EminA]
(1.8,2.1) rectangle (2.1,2.4) ;
\draw[black, fill=EmaxA!0!EminA]
(2.1,2.4) rectangle (2.4,2.7) ;
\draw[black, fill=EmaxA!0!EminA]
(2.4,2.7) rectangle (2.7,3.0) ;
\end{tikzpicture}
  \caption{\label{fig:efo-hybrid-and-overlap-alignment-matrix}$\Hybrid$ and $\Overlap$ alignments (EFO).}
\end{figure}
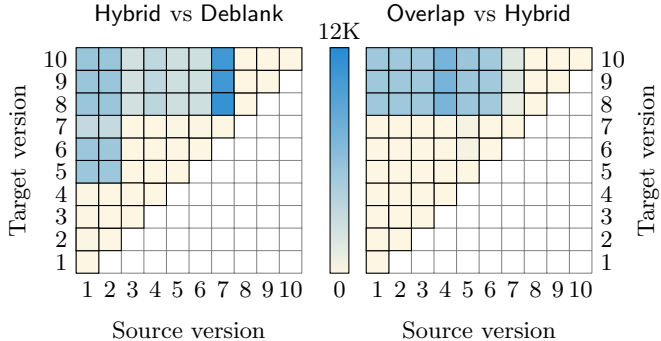

The relative improvement offered by the hybrid and overlap alignments
is subtle, and to highlight it in
Figure~\ref{fig:efo-hybrid-and-overlap-alignment-matrix} we show the
absolute number of edges that are additionally aligned by the hybrid
alignment (compared with the deblanking alignment) and the overlap
alignment (compared with the hybrid alignment).
In both cases, the improvements come mainly from ontology changes
manifested by change of URI prefix e.g.,
\texttt{http://purl.org/obo/owl/} to
\texttt{http://purl.obolibrary.org/obo/}. This process can be quite
straightforward e.g., a large number of URIs using old prefix in
version $7$ is replaced by URIs with new prefix in version $8$. This
change also involves changes in the contents of the affected nodes,
which are captured with the overlap alignment.  Ontology change may
take more time with URIs disappearing in between: a number of URIs
using the old prefix in the first two versions are removed in version
$3$, and then reappear in version $5$ with the new prefix.

Because our methods focus on the outgoing neighborhood of a node, they
make errors by incorrectly aligning URIs that are used as predicates
only: these URIs typically are present as subject in one triple that
declares the type of the URI (and uses \texttt{rdf:type} as
predicate). The number of such incorrectly aligned predicates is
relatively small ($<15$). A better solution would identify
URIs that are predominantly used as predicates and use a different
refinement process, for instance, one that incorporates the colors of
the subject and the object in any triple that uses the given predicate.

Finally, we found the quality of the hybrid and overlap alignments to
be overall satisfying: very few URIs undergoing changes are missed and
no URIs are aligned in error. Unfortunately we cannot precisely
evaluate it because we lack the appropriate ground truth for the EFO
dataset and we present a more detailed discussion in the appendix of
the complete paper. In the following subsection, we run experiments on
a dataset for which the ground truth is easily obtained.

\subsection{Guide to Pharmacology database (GtoPdb)}
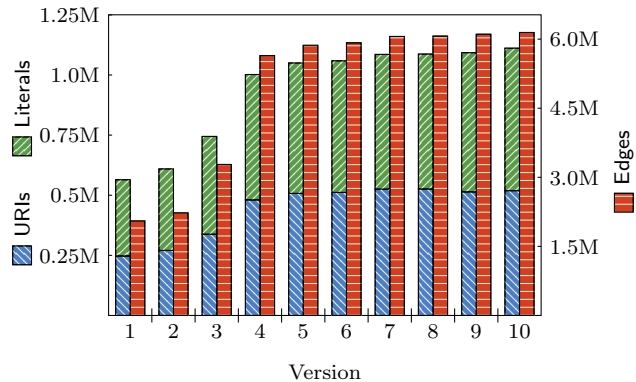
\begin{figure}[htb]
  \centering
\begin{tikzpicture}[>=latex]
\path[use as bounding box] (-1.25,-0.75) rectangle ( 7.0 , 4 );
\draw[fill=white] (0,0) -- ( 5.75 ,0) -- ( 5.75 , 4 ) -- (0, 4 ) -- cycle;
\draw ( 0.575 ,  -0.09375 ) -- ( 0.575 , 0.09375 );
\draw ( 1.15 ,  -0.09375 ) -- ( 1.15 , 0.09375 );
\draw ( 1.725 ,  -0.09375 ) -- ( 1.725 , 0.09375 );
\draw ( 2.3 ,  -0.09375 ) -- ( 2.3 , 0.09375 );
\draw ( 2.875 ,  -0.09375 ) -- ( 2.875 , 0.09375 );
\draw ( 3.45 ,  -0.09375 ) -- ( 3.45 , 0.09375 );
\draw ( 4.025 ,  -0.09375 ) -- ( 4.025 , 0.09375 );
\draw ( 4.6 ,  -0.09375 ) -- ( 4.6 , 0.09375 );
\draw ( 5.175 ,  -0.09375 ) -- ( 5.175 , 0.09375 );
\path ( 0.2875 ,0) node[below] {\small  $ 1 $};
\path ( 0.8625 ,0) node[below] {\small  $ 2 $};
\path ( 1.4375 ,0) node[below] {\small  $ 3 $};
\path ( 2.0125 ,0) node[below] {\small  $ 4 $};
\path ( 2.5875 ,0) node[below] {\small  $ 5 $};
\path ( 3.1625 ,0) node[below] {\small  $ 6 $};
\path ( 3.7375 ,0) node[below] {\small  $ 7 $};
\path ( 4.3125 ,0) node[below] {\small  $ 8 $};
\path ( 4.8875 ,0) node[below] {\small  $ 9 $};
\path ( 5.4625 ,0) node[below] {\small  $ 10 $};
\path ( 2.875 , -0.75) node {\small  Version};
\draw ( 0.046875 , 0.799797171437  ) -- (0.0, 0.799797171437 );
\path (0.0, 0.799797171437  ) node[left] {\small $ 0.25 $M };
\draw ( 0.046875 , 1.59959434287  ) -- (0.0, 1.59959434287 );
\path (0.0, 1.59959434287  ) node[left] {\small $ 0.5 $M };
\draw ( 0.046875 , 2.39939151431  ) -- (0.0, 2.39939151431 );
\path (0.0, 2.39939151431  ) node[left] {\small $ 0.75 $M };
\draw ( 0.046875 , 3.19918868575  ) -- (0.0, 3.19918868575 );
\path (0.0, 3.19918868575  ) node[left] {\small $ 1.0 $M };
\draw ( 0.046875 , 3.99898585719  ) -- (0.0, 3.99898585719 );
\path (0.0, 3.99898585719  ) node[left] {\small $ 1.25 $M };
\draw ( 5.703125 , 0.919086250577  ) -- ( 5.75 , 0.919086250577 );
\path ( 5.75 , 0.919086250577  ) node[right] {\small $ 1.5 $M };
\draw ( 5.703125 , 1.83817250115  ) -- ( 5.75 , 1.83817250115 );
\path ( 5.75 , 1.83817250115  ) node[right] {\small $ 3.0 $M };
\draw ( 5.703125 , 2.75725875173  ) -- ( 5.75 , 2.75725875173 );
\path ( 5.75 , 2.75725875173  ) node[right] {\small $ 4.5 $M };
\draw ( 5.703125 , 3.67634500231  ) -- ( 5.75 , 3.67634500231 );
\path ( 5.75 , 3.67634500231  ) node[right] {\small $ 6.0 $M };
\begin{scope}[yshift= 1.25 cm,xshift= 6.6 cm]
\draw[Eedges] (0.125,0.125) rectangle (0.375,0.375);
\draw[EedgesP] (0.125,0.125) rectangle (0.375,0.375);
\path (0.25,0.375) node[right,rotate=90] {\small  \sf Edges};
\end{scope}
\begin{scope}[yshift= -2 cm,xshift= -1.4 cm]
\begin{scope}[yshift=2.55cm,xshift=0.0cm]
\draw[Euris] (0.125,0.125) rectangle (0.375,0.375);
\draw[EurisP] (0.125,0.125) rectangle (0.375,0.375);
\path (0.25,0.375) node[right,rotate=90] {\small  \sf URIs};
\end{scope}
\begin{scope}[yshift=4cm,xshift=0.0cm]
\draw[Eliterals] (0.125,0.125) rectangle (0.375,0.375);
\draw[EliteralsP] (0.125,0.125) rectangle (0.375,0.375);
\path (0.25,0.375) node[right,rotate=90] {\small  \sf Literals};
\end{scope}
\end{scope}
\draw[ Eedges ] ( 0.2875 ,0.0) rectangle ( 0.48125 , 1.25888836824 );
\draw[EedgesP] ( 0.2875 ,0.0) rectangle ( 0.48125 , 1.25888836824 );
\draw[ Eblanks ] ( 0.09375 ,0.0) rectangle ( 0.2875 , 0.0 );
\draw[EblanksP] ( 0.09375 ,0.0) rectangle ( 0.2875 , 0.0 );
\draw[ Euris ] ( 0.09375 , 0.0 ) rectangle ( 0.2875 , 0.792205496686 );
\draw[EurisP] ( 0.09375 , 0.0 ) rectangle ( 0.2875 , 0.792205496686 );
\draw[ Eliterals ] ( 0.09375 , 0.792205496686 ) rectangle ( 0.2875 , 1.80577565529 );
\draw[EliteralsP] ( 0.09375 , 0.792205496686 ) rectangle ( 0.2875 , 1.80577565529 );
\draw[ Eedges ] ( 0.8625 ,0.0) rectangle ( 1.05625 , 1.36631974735 );
\draw[EedgesP] ( 0.8625 ,0.0) rectangle ( 1.05625 , 1.36631974735 );
\draw[ Eblanks ] ( 0.66875 ,0.0) rectangle ( 0.8625 , 0.0 );
\draw[EblanksP] ( 0.66875 ,0.0) rectangle ( 0.8625 , 0.0 );
\draw[ Euris ] ( 0.66875 , 0.0 ) rectangle ( 0.8625 , 0.864993437664 );
\draw[EurisP] ( 0.66875 , 0.0 ) rectangle ( 0.8625 , 0.864993437664 );
\draw[ Eliterals ] ( 0.66875 , 0.864993437664 ) rectangle ( 0.8625 , 1.9494160281 );
\draw[EliteralsP] ( 0.66875 , 0.864993437664 ) rectangle ( 0.8625 , 1.9494160281 );
\draw[ Eedges ] ( 1.4375 ,0.0) rectangle ( 1.63125 , 2.00879105999 );
\draw[EedgesP] ( 1.4375 ,0.0) rectangle ( 1.63125 , 2.00879105999 );
\draw[ Eblanks ] ( 1.24375 ,0.0) rectangle ( 1.4375 , 0.0 );
\draw[EblanksP] ( 1.24375 ,0.0) rectangle ( 1.4375 , 0.0 );
\draw[ Euris ] ( 1.24375 , 0.0 ) rectangle ( 1.4375 , 1.08068273886 );
\draw[EurisP] ( 1.24375 , 0.0 ) rectangle ( 1.4375 , 1.08068273886 );
\draw[ Eliterals ] ( 1.24375 , 1.08068273886 ) rectangle ( 1.4375 , 2.38168080575 );
\draw[EliteralsP] ( 1.24375 , 1.08068273886 ) rectangle ( 1.4375 , 2.38168080575 );
\draw[ Eedges ] ( 2.0125 ,0.0) rectangle ( 2.20625 , 3.45798481455 );
\draw[EedgesP] ( 2.0125 ,0.0) rectangle ( 2.20625 , 3.45798481455 );
\draw[ Eblanks ] ( 1.81875 ,0.0) rectangle ( 2.0125 , 0.0 );
\draw[EblanksP] ( 1.81875 ,0.0) rectangle ( 2.0125 , 0.0 );
\draw[ Euris ] ( 1.81875 , 0.0 ) rectangle ( 2.0125 , 1.53846104628 );
\draw[EurisP] ( 1.81875 , 0.0 ) rectangle ( 2.0125 , 1.53846104628 );
\draw[ Eliterals ] ( 1.81875 , 1.53846104628 ) rectangle ( 2.0125 , 3.20561905501 );
\draw[EliteralsP] ( 1.81875 , 1.53846104628 ) rectangle ( 2.0125 , 3.20561905501 );
\draw[ Eedges ] ( 2.5875 ,0.0) rectangle ( 2.78125 , 3.59381534608 );
\draw[EedgesP] ( 2.5875 ,0.0) rectangle ( 2.78125 , 3.59381534608 );
\draw[ Eblanks ] ( 2.39375 ,0.0) rectangle ( 2.5875 , 0.0 );
\draw[EblanksP] ( 2.39375 ,0.0) rectangle ( 2.5875 , 0.0 );
\draw[ Euris ] ( 2.39375 , 0.0 ) rectangle ( 2.5875 , 1.62380580285 );
\draw[EurisP] ( 2.39375 , 0.0 ) rectangle ( 2.5875 , 1.62380580285 );
\draw[ Eliterals ] ( 2.39375 , 1.62380580285 ) rectangle ( 2.5875 , 3.35906814032 );
\draw[EliteralsP] ( 2.39375 , 1.62380580285 ) rectangle ( 2.5875 , 3.35906814032 );
\draw[ Eedges ] ( 3.1625 ,0.0) rectangle ( 3.35625 , 3.62722290584 );
\draw[EedgesP] ( 3.1625 ,0.0) rectangle ( 3.35625 , 3.62722290584 );
\draw[ Eblanks ] ( 2.96875 ,0.0) rectangle ( 3.1625 , 0.0 );
\draw[EblanksP] ( 2.96875 ,0.0) rectangle ( 3.1625 , 0.0 );
\draw[ Euris ] ( 2.96875 , 0.0 ) rectangle ( 3.1625 , 1.6365705657 );
\draw[EurisP] ( 2.96875 , 0.0 ) rectangle ( 3.1625 , 1.6365705657 );
\draw[ Eliterals ] ( 2.96875 , 1.6365705657 ) rectangle ( 3.1625 , 3.38788003362 );
\draw[EliteralsP] ( 2.96875 , 1.6365705657 ) rectangle ( 3.1625 , 3.38788003362 );
\draw[ Eedges ] ( 3.7375 ,0.0) rectangle ( 3.93125 , 3.71180641348 );
\draw[EedgesP] ( 3.7375 ,0.0) rectangle ( 3.93125 , 3.71180641348 );
\draw[ Eblanks ] ( 3.54375 ,0.0) rectangle ( 3.7375 , 0.0 );
\draw[EblanksP] ( 3.54375 ,0.0) rectangle ( 3.7375 , 0.0 );
\draw[ Euris ] ( 3.54375 , 0.0 ) rectangle ( 3.7375 , 1.68077375578 );
\draw[EurisP] ( 3.54375 , 0.0 ) rectangle ( 3.7375 , 1.68077375578 );
\draw[ Eliterals ] ( 3.54375 , 1.68077375578 ) rectangle ( 3.7375 , 3.47286968025 );
\draw[EliteralsP] ( 3.54375 , 1.68077375578 ) rectangle ( 3.7375 , 3.47286968025 );
\draw[ Eedges ] ( 4.3125 ,0.0) rectangle ( 4.50625 , 3.71659362739 );
\draw[EedgesP] ( 4.3125 ,0.0) rectangle ( 4.50625 , 3.71659362739 );
\draw[ Eblanks ] ( 4.11875 ,0.0) rectangle ( 4.3125 , 0.0 );
\draw[EblanksP] ( 4.11875 ,0.0) rectangle ( 4.3125 , 0.0 );
\draw[ Euris ] ( 4.11875 , 0.0 ) rectangle ( 4.3125 , 1.68362743208 );
\draw[EurisP] ( 4.11875 , 0.0 ) rectangle ( 4.3125 , 1.68362743208 );
\draw[ Eliterals ] ( 4.11875 , 1.68362743208 ) rectangle ( 4.3125 , 3.47727176388 );
\draw[EliteralsP] ( 4.11875 , 1.68362743208 ) rectangle ( 4.3125 , 3.47727176388 );
\draw[ Eedges ] ( 4.8875 ,0.0) rectangle ( 5.08125 , 3.74134339467 );
\draw[EedgesP] ( 4.8875 ,0.0) rectangle ( 5.08125 , 3.74134339467 );
\draw[ Eblanks ] ( 4.69375 ,0.0) rectangle ( 4.8875 , 0.0 );
\draw[EblanksP] ( 4.69375 ,0.0) rectangle ( 4.8875 , 0.0 );
\draw[ Euris ] ( 4.69375 , 0.0 ) rectangle ( 4.8875 , 1.64559867618 );
\draw[EurisP] ( 4.69375 , 0.0 ) rectangle ( 4.8875 , 1.64559867618 );
\draw[ Eliterals ] ( 4.69375 , 1.64559867618 ) rectangle ( 4.8875 , 3.49617896901 );
\draw[EliteralsP] ( 4.69375 , 1.64559867618 ) rectangle ( 4.8875 , 3.49617896901 );
\draw[ Eedges ] ( 5.4625 ,0.0) rectangle ( 5.65625 , 3.76470595444 );
\draw[EedgesP] ( 5.4625 ,0.0) rectangle ( 5.65625 , 3.76470595444 );
\draw[ Eblanks ] ( 5.26875 ,0.0) rectangle ( 5.4625 , 0.0 );
\draw[EblanksP] ( 5.26875 ,0.0) rectangle ( 5.4625 , 0.0 );
\draw[ Euris ] ( 5.26875 , 0.0 ) rectangle ( 5.4625 , 1.66096757862 );
\draw[EurisP] ( 5.26875 , 0.0 ) rectangle ( 5.4625 , 1.66096757862 );
\draw[ Eliterals ] ( 5.26875 , 1.66096757862 ) rectangle ( 5.4625 , 3.55555591102 );
\draw[EliteralsP] ( 5.26875 , 1.66096757862 ) rectangle ( 5.4625 , 3.55555591102 );
\end{tikzpicture}

  \caption{\label{fig:iuphar-versions}GtoPdb dataset versions.}
\end{figure}

We used 10 versions of the GtoPdb relational database, which we
exported to RDF following the W3C Direct Mapping
recommendation~\cite{DMRDF12} using the D2RQ platform. The mapping
works as follows: 1) every tuple is identified by a URI which is
constituted from a given URI prefix, the table name table, and the
attribute values of the primary key, 2) (non-referential) value 
attributes are translated to edges consisting of the tuple URI, the
attribute name and a literal for the attribute value, 3) referential
attributes are translated to edges pointing to the URI of the
referred tuple. While this experimental setting has been designed to
evaluate the hybrid and overlap alignments, we believe it captures a
common situation in which a relational database is exported to RDF at
different times by different services using similar export schemes
(e.g., the default W3C Direct Mapping configuration). Node and edge counts are shown in
Figure~\ref{fig:iuphar-versions}. These graphs do not have any blank
nodes, and the number of literals is slightly larger than the number
of URIs.

To focus our study on the hybrid and overlap alignments,  we export
every version with a different URI prefix. Because there are no common
URIs and no blank nodes, the trivial and deblanking alignments align no
non-literal nodes. However, since the URI prefixes are known to us and
the key values in the GtoPdb are generally persistent (the same entity does not
change its key over different versions), we are able to identify a
precise alignment between any pair of versions that will serve as
ground truth (\textsf{GtoPdb}).  For example, the calcitonin ligand is
identified in all versions as ligand 685. In version 1 this is given a
URI \url{http://gtopdb.org/ver1/ligand685} and in version 2 
\url{http://gtopdb.org/ver2/ligand685}. 

\begin{figure}[htb]
  \centering
\begin{tikzpicture}[>=latex]
\path[use as bounding box] (-1,-0.6) rectangle (7.125,4);
\draw[fill=white] (0,0) -- ( 7 ,0) -- ( 7 , 4 ) -- (0, 4 ) -- cycle;
\draw ( 0.777777777778 ,  -0.125 ) -- ( 0.777777777778 , 0.125 );
\draw ( 1.55555555556 ,  -0.125 ) -- ( 1.55555555556 , 0.125 );
\draw ( 2.33333333333 ,  -0.125 ) -- ( 2.33333333333 , 0.125 );
\draw ( 3.11111111111 ,  -0.125 ) -- ( 3.11111111111 , 0.125 );
\draw ( 3.88888888889 ,  -0.125 ) -- ( 3.88888888889 , 0.125 );
\draw ( 4.66666666667 ,  -0.125 ) -- ( 4.66666666667 , 0.125 );
\draw ( 5.44444444444 ,  -0.125 ) -- ( 5.44444444444 , 0.125 );
\draw ( 6.22222222222 ,  -0.125 ) -- ( 6.22222222222 , 0.125 );
\path ( 0.0 ,0) node[below] {\small  $ 1 $};
\path ( 0.777777777778 ,0) node[below] {\small  $ 2 $};
\path ( 1.55555555556 ,0) node[below] {\small  $ 3 $};
\path ( 2.33333333333 ,0) node[below] {\small  $ 4 $};
\path ( 3.11111111111 ,0) node[below] {\small  $ 5 $};
\path ( 3.88888888889 ,0) node[below] {\small  $ 6 $};
\path ( 4.66666666667 ,0) node[below] {\small  $ 7 $};
\path ( 5.44444444444 ,0) node[below] {\small  $ 8 $};
\path ( 6.22222222222 ,0) node[below] {\small  $ 9 $};
\path ( 7.0 ,0) node[below] {\small  $ 10 $};
\path ( 3.5 , -0.75) node {\small  Alignment between versions};
\draw ( 0.0625 , 0.845239946716  ) -- (0.0, 0.845239946716 );
\path (0.0, 0.845239946716  ) node[left] {\small $ 0.25 $M};
\draw ( 0.0625 , 1.69047989343  ) -- (0.0, 1.69047989343 );
\path (0.0, 1.69047989343  ) node[left] {\small $ 0.5 $M};
\draw ( 0.0625 , 2.53571984015  ) -- (0.0, 2.53571984015 );
\path (0.0, 2.53571984015  ) node[left] {\small $ 0.75 $M};
\draw ( 0.0625 , 3.38095978686  ) -- (0.0, 3.38095978686 );
\path (0.0, 3.38095978686  ) node[left] {\small $ 1.0 $M};
\begin{scope}[yshift= 2.3 cm,xshift= 0.125 cm]
\begin{scope}[yshift=0cm,xshift=0.0cm]
\draw[Ehybrid] (0.125,0.125) rectangle (0.375,0.375);
\draw[EhybridP] (0.125,0.125) rectangle (0.375,0.375);
\path (0.375,0.25) node[right] {\small  $\mathsf{Hybrid}$};
\end{scope}
\begin{scope}[yshift=0.4cm,xshift=0.0cm]
\draw[Eoverlap] (0.125,0.125) rectangle (0.375,0.375);
\draw[EoverlapP] (0.125,0.125) rectangle (0.375,0.375);
\path (0.375,0.25) node[right] {\small  $\mathsf{Overlap}$};
\end{scope}
\begin{scope}[yshift=0.8cm,xshift=0.0cm]
\draw[EGtoPdb] (0.125,0.125) rectangle (0.375,0.375);
\draw[EGtoPdbP] (0.125,0.125) rectangle (0.375,0.375);
\path (0.375,0.25) node[right] {\small  $\mathsf{GtoPdb}$};
\end{scope}
\begin{scope}[yshift=1.2cm,xshift=0.0cm]
\draw[Etotal] (0.125,0.125) rectangle (0.375,0.375);
\draw[EtotalP] (0.125,0.125) rectangle (0.375,0.375);
\path (0.375,0.25) node[right] {\small  $\mathsf{Total}$};
\end{scope}
\end{scope}
\draw[ Ehybrid ] ( 0.125 , 0.0 ) rectangle ( 0.256944444444 , 0.958055812884 );
\draw[EhybridP] ( 0.125 , 0.0 ) rectangle ( 0.256944444444 , 0.958055812884 );
\draw[ Eoverlap ] ( 0.256944444444 , 0.0 ) rectangle ( 0.388888888889 , 1.51826732573 );
\draw[EoverlapP] ( 0.256944444444 , 0.0 ) rectangle ( 0.388888888889 , 1.51826732573 );
\draw[ EGtoPdb ] ( 0.388888888889 , 0.0 ) rectangle ( 0.520833333333 , 1.63032247594 );
\draw[EGtoPdbP] ( 0.388888888889 , 0.0 ) rectangle ( 0.520833333333 , 1.63032247594 );
\draw[ Etotal ] ( 0.520833333333 , 0.0 ) rectangle ( 0.652777777778 , 1.75135407439 );
\draw[EtotalP] ( 0.520833333333 , 0.0 ) rectangle ( 0.652777777778 , 1.75135407439 );
\draw[ Ehybrid ] ( 0.902777777778 , 0.0 ) rectangle ( 1.03472222222 , 0.658722538154 );
\draw[EhybridP] ( 0.902777777778 , 0.0 ) rectangle ( 1.03472222222 , 0.658722538154 );
\draw[ Eoverlap ] ( 1.03472222222 , 0.0 ) rectangle ( 1.16666666667 , 1.5944843022 );
\draw[EoverlapP] ( 1.03472222222 , 0.0 ) rectangle ( 1.16666666667 , 1.5944843022 );
\draw[ EGtoPdb ] ( 1.16666666667 , 0.0 ) rectangle ( 1.29861111111 , 1.81687707506 );
\draw[EGtoPdbP] ( 1.16666666667 , 0.0 ) rectangle ( 1.29861111111 , 1.81687707506 );
\draw[ Etotal ] ( 1.29861111111 , 0.0 ) rectangle ( 1.43055555556 , 2.0562219803 );
\draw[EtotalP] ( 1.29861111111 , 0.0 ) rectangle ( 1.43055555556 , 2.0562219803 );
\draw[ Ehybrid ] ( 1.68055555556 , 0.0 ) rectangle ( 1.8125 , 0.990878170495 );
\draw[EhybridP] ( 1.68055555556 , 0.0 ) rectangle ( 1.8125 , 0.990878170495 );
\draw[ Eoverlap ] ( 1.8125 , 0.0 ) rectangle ( 1.94444444444 , 2.07321468418 );
\draw[EoverlapP] ( 1.8125 , 0.0 ) rectangle ( 1.94444444444 , 2.07321468418 );
\draw[ EGtoPdb ] ( 1.94444444444 , 0.0 ) rectangle ( 2.07638888889 , 2.05590417008 );
\draw[EGtoPdbP] ( 1.94444444444 , 0.0 ) rectangle ( 2.07638888889 , 2.05590417008 );
\draw[ Etotal ] ( 2.07638888889 , 0.0 ) rectangle ( 2.20833333333 , 2.76795458695 );
\draw[EtotalP] ( 2.07638888889 , 0.0 ) rectangle ( 2.20833333333 , 2.76795458695 );
\draw[ Ehybrid ] ( 2.45833333333 , 0.0 ) rectangle ( 2.59027777778 , 1.32396356678 );
\draw[EhybridP] ( 2.45833333333 , 0.0 ) rectangle ( 2.59027777778 , 1.32396356678 );
\draw[ Eoverlap ] ( 2.59027777778 , 0.0 ) rectangle ( 2.72222222222 , 3.0021773381 );
\draw[EoverlapP] ( 2.59027777778 , 0.0 ) rectangle ( 2.72222222222 , 3.0021773381 );
\draw[ EGtoPdb ] ( 2.72222222222 , 0.0 ) rectangle ( 2.85416666667 , 3.23435798955 );
\draw[EGtoPdbP] ( 2.72222222222 , 0.0 ) rectangle ( 2.85416666667 , 3.23435798955 );
\draw[ Etotal ] ( 2.85416666667 , 0.0 ) rectangle ( 2.98611111111 , 3.341936749 );
\draw[EtotalP] ( 2.85416666667 , 0.0 ) rectangle ( 2.98611111111 , 3.341936749 );
\draw[ Ehybrid ] ( 3.23611111111 , 0.0 ) rectangle ( 3.36805555556 , 1.43802362615 );
\draw[EhybridP] ( 3.23611111111 , 0.0 ) rectangle ( 3.36805555556 , 1.43802362615 );
\draw[ Eoverlap ] ( 3.36805555556 , 0.0 ) rectangle ( 3.5 , 3.20520059234 );
\draw[EoverlapP] ( 3.36805555556 , 0.0 ) rectangle ( 3.5 , 3.20520059234 );
\draw[ EGtoPdb ] ( 3.5 , 0.0 ) rectangle ( 3.63194444444 , 3.40999208855 );
\draw[EGtoPdbP] ( 3.5 , 0.0 ) rectangle ( 3.63194444444 , 3.40999208855 );
\draw[ Etotal ] ( 3.63194444444 , 0.0 ) rectangle ( 3.76388888889 , 3.44562064279 );
\draw[EtotalP] ( 3.63194444444 , 0.0 ) rectangle ( 3.76388888889 , 3.44562064279 );
\draw[ Ehybrid ] ( 4.01388888889 , 0.0 ) rectangle ( 4.14583333333 , 2.05501159669 );
\draw[EhybridP] ( 4.01388888889 , 0.0 ) rectangle ( 4.14583333333 , 2.05501159669 );
\draw[ Eoverlap ] ( 4.14583333333 , 0.0 ) rectangle ( 4.27777777778 , 3.26913116095 );
\draw[EoverlapP] ( 4.14583333333 , 0.0 ) rectangle ( 4.27777777778 , 3.26913116095 );
\draw[ EGtoPdb ] ( 4.27777777778 , 0.0 ) rectangle ( 4.40972222222 , 3.44639826354 );
\draw[EGtoPdbP] ( 4.27777777778 , 0.0 ) rectangle ( 4.40972222222 , 3.44639826354 );
\draw[ Etotal ] ( 4.40972222222 , 0.0 ) rectangle ( 4.54166666667 , 3.50582539371 );
\draw[EtotalP] ( 4.40972222222 , 0.0 ) rectangle ( 4.54166666667 , 3.50582539371 );
\draw[ Ehybrid ] ( 4.79166666667 , 0.0 ) rectangle ( 4.92361111111 , 3.49630799191 );
\draw[EhybridP] ( 4.79166666667 , 0.0 ) rectangle ( 4.92361111111 , 3.49630799191 );
\draw[ Eoverlap ] ( 4.92361111111 , 0.0 ) rectangle ( 5.05555555556 , 3.52204385781 );
\draw[EoverlapP] ( 4.92361111111 , 0.0 ) rectangle ( 5.05555555556 , 3.52204385781 );
\draw[ EGtoPdb ] ( 5.05555555556 , 0.0 ) rectangle ( 5.1875 , 3.55224259063 );
\draw[EGtoPdbP] ( 5.05555555556 , 0.0 ) rectangle ( 5.1875 , 3.55224259063 );
\draw[ Etotal ] ( 5.1875 , 0.0 ) rectangle ( 5.31944444444 , 3.55555593122 );
\draw[EtotalP] ( 5.1875 , 0.0 ) rectangle ( 5.31944444444 , 3.55555593122 );
\draw[ Ehybrid ] ( 5.56944444444 , 0.0 ) rectangle ( 5.70138888889 , 0.917761534144 );
\draw[EhybridP] ( 5.56944444444 , 0.0 ) rectangle ( 5.70138888889 , 0.917761534144 );
\draw[ Eoverlap ] ( 5.70138888889 , 0.0 ) rectangle ( 5.83333333333 , 2.97174870002 );
\draw[EoverlapP] ( 5.70138888889 , 0.0 ) rectangle ( 5.83333333333 , 2.97174870002 );
\draw[ EGtoPdb ] ( 5.83333333333 , 0.0 ) rectangle ( 5.96527777778 , 3.33090467722 );
\draw[EGtoPdbP] ( 5.83333333333 , 0.0 ) rectangle ( 5.96527777778 , 3.33090467722 );
\draw[ Etotal ] ( 5.96527777778 , 0.0 ) rectangle ( 6.09722222222 , 3.51838227836 );
\draw[EtotalP] ( 5.96527777778 , 0.0 ) rectangle ( 6.09722222222 , 3.51838227836 );
\draw[ Ehybrid ] ( 6.34722222222 , 0.0 ) rectangle ( 6.47916666667 , 1.26094585731 );
\draw[EhybridP] ( 6.34722222222 , 0.0 ) rectangle ( 6.47916666667 , 1.26094585731 );
\draw[ Eoverlap ] ( 6.47916666667 , 0.0 ) rectangle ( 6.61111111111 , 3.30920905827 );
\draw[EoverlapP] ( 6.47916666667 , 0.0 ) rectangle ( 6.61111111111 , 3.30920905827 );
\draw[ EGtoPdb ] ( 6.61111111111 , 0.0 ) rectangle ( 6.74305555556 , 3.45504675867 );
\draw[EGtoPdbP] ( 6.61111111111 , 0.0 ) rectangle ( 6.74305555556 , 3.45504675867 );
\draw[ Etotal ] ( 6.74305555556 , 0.0 ) rectangle ( 6.875 , 3.49443494019 );
\draw[EtotalP] ( 6.74305555556 , 0.0 ) rectangle ( 6.875 , 3.49443494019 );
\end{tikzpicture}
  \caption{\label{fig:iuphar-alignment-lines}Alignments (GtoPdb).}
\end{figure}
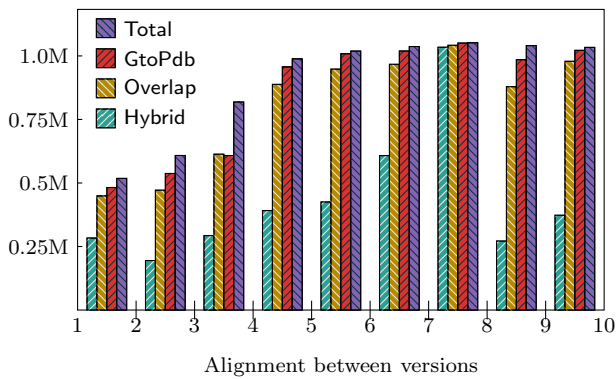

In Figure~\ref{fig:iuphar-alignment-lines} we show the number of
aligned nodes in all pairs of consecutive versions by the hybrid and
overlap alignment together with the number of nodes aligned by ground
truth as well as the total number of nodes (\textsf{Total})
present in both versions. 
All counts are free of duplicates: any two URIs coming from two
versions but representing the same tuple are counted as one.
Comparing the values of \textsf{Total} and \textsf{GtoPdb} allows us
characterize the degree of relative change between versions. In
particular between versions $3$ and $4$ these two values are most
different, which indicates a large number of changes (mainly
insertions of new nodes). On the other hand, the changes are minute
between versions $7$ and $8$. In general, the values of the overlap
alignment are significantly closer to \textsf{GtoPdb} than are those
of the hybrid alignment. This suggest that the hybrid alignment is
sensitive to changes as they propagate throughout the graph, while
overlap may better handle changes.

\begin{figure}[thb]
  \centering
  \begin{tikzpicture}[>=latex]
    \path[use as bounding box] (0,-0.365) rectangle (7.5,2.1);
\draw (0,0) -- ( 7.11 ,0);
\draw ( 0.0 ,  -0.045 ) -- ( 0.0 , 0.045 );
\draw ( 0.79 ,  -0.045 ) -- ( 0.79 , 0.045 );
\draw ( 1.58 ,  -0.045 ) -- ( 1.58 , 0.045 );
\draw ( 2.37 ,  -0.045 ) -- ( 2.37 , 0.045 );
\draw ( 3.16 ,  -0.045 ) -- ( 3.16 , 0.045 );
\draw ( 3.95 ,  -0.045 ) -- ( 3.95 , 0.045 );
\draw ( 4.74 ,  -0.045 ) -- ( 4.74 , 0.045 );
\draw ( 5.53 ,  -0.045 ) -- ( 5.53 , 0.045 );
\draw ( 6.32 ,  -0.045 ) -- ( 6.32 , 0.045 );
\draw ( 7.11 ,  -0.045 ) -- ( 7.11 , 0.045 );
\path ( 0.0 ,0) node[below] {\small  $ 1 $};
\path ( 0.79 ,0) node[below] {\small  $ 2 $};
\path ( 1.58 ,0) node[below] {\small  $ 3 $};
\path ( 2.37 ,0) node[below] {\small  $ 4 $};
\path ( 3.16 ,0) node[below] {\small  $ 5 $};
\path ( 3.95 ,0) node[below] {\small  $ 6 $};
\path ( 4.74 ,0) node[below] {\small  $ 7 $};
\path ( 5.53 ,0) node[below] {\small  $ 8 $};
\path ( 6.32 ,0) node[below] {\small  $ 9 $};
\path ( 7.11 ,0) node[below] {\small  $ 10 $};
\path ( 3.555 , -0.35) node[below] {\small  Version};
\begin{scope}[yshift= 1.83 cm,xshift=0cm]
\begin{scope}[yshift=0cm,xshift= 0.0 cm]
\draw[gray, very thin, Eexact] (0.125,0.125) rectangle (0.375,0.375);
\fill[EexactP ] (0.125,0.125) rectangle (0.375,0.375);
\path (0.375,0.25) node[right] {\small  $\mathsf{ exact }$};
\end{scope}
\begin{scope}[yshift=0cm,xshift= 1.25 cm]
\draw[gray, very thin, Einclusive] (0.125,0.125) rectangle (0.375,0.375);
\fill[EinclusiveP ] (0.125,0.125) rectangle (0.375,0.375);
\path (0.375,0.25) node[right] {\small  $\mathsf{ inclusive }$};
\end{scope}
\begin{scope}[yshift=0cm,xshift= 2.875 cm]
\draw[gray, very thin, Efalse] (0.125,0.125) rectangle (0.375,0.375);
\fill[EfalseP ] (0.125,0.125) rectangle (0.375,0.375);
\path (0.375,0.25) node[right] {\small  $\mathsf{ false }$};
\end{scope}
\begin{scope}[yshift=0cm,xshift= 4.0625 cm]
\draw[gray, very thin, Emissing] (0.125,0.125) rectangle (0.375,0.375);
\fill[EmissingP ] (0.125,0.125) rectangle (0.375,0.375);
\path (0.375,0.25) node[right] {\small  $\mathsf{ missing }$};
\end{scope}
\end{scope}
\path (7.2285, 0.395) node[right] {\small  $\mathsf{ Hybrid }$};
\fill[ Eexact ] (0.395, 0.395) -- +( 90.0 : 0.365 cm) arc ( 90.0 : -124.599533409 : 0.365 cm)-- cycle;
\fill[draw=black,EexactP ] (0.395, 0.395) -- +( 90.0 : 0.365 cm) arc ( 90.0 : -124.599533409 : 0.365 cm)-- cycle;
\fill[ Einclusive ] (0.395, 0.395) -- +( -124.599533409 : 0.365 cm) arc ( -124.599533409 : -140.707759424 : 0.365 cm)-- cycle;
\fill[draw=black,EinclusiveP ] (0.395, 0.395) -- +( -124.599533409 : 0.365 cm) arc ( -124.599533409 : -140.707759424 : 0.365 cm)-- cycle;
\fill[ Efalse ] (0.395, 0.395) -- +( -140.707759424 : 0.365 cm) arc ( -140.707759424 : -140.732461304 : 0.365 cm)-- cycle;
\fill[draw=black,EfalseP ] (0.395, 0.395) -- +( -140.707759424 : 0.365 cm) arc ( -140.707759424 : -140.732461304 : 0.365 cm)-- cycle;
\fill[ Emissing ] (0.395, 0.395) -- +( -140.732461304 : 0.365 cm) arc ( -140.732461304 : -270.0 : 0.365 cm)-- cycle;
\fill[draw=black,EmissingP ] (0.395, 0.395) -- +( -140.732461304 : 0.365 cm) arc ( -140.732461304 : -270.0 : 0.365 cm)-- cycle;
\fill[ Eexact ] (1.185, 0.395) -- +( 90.0 : 0.365 cm) arc ( 90.0 : -86.9655802171 : 0.365 cm)-- cycle;
\fill[draw=black,EexactP ] (1.185, 0.395) -- +( 90.0 : 0.365 cm) arc ( 90.0 : -86.9655802171 : 0.365 cm)-- cycle;
\fill[ Einclusive ] (1.185, 0.395) -- +( -86.9655802171 : 0.365 cm) arc ( -86.9655802171 : -88.2847548493 : 0.365 cm)-- cycle;
\fill[draw=black,EinclusiveP ] (1.185, 0.395) -- +( -86.9655802171 : 0.365 cm) arc ( -86.9655802171 : -88.2847548493 : 0.365 cm)-- cycle;
\fill[ Efalse ] (1.185, 0.395) -- +( -88.2847548493 : 0.365 cm) arc ( -88.2847548493 : -88.3441389163 : 0.365 cm)-- cycle;
\fill[draw=black,EfalseP ] (1.185, 0.395) -- +( -88.2847548493 : 0.365 cm) arc ( -88.2847548493 : -88.3441389163 : 0.365 cm)-- cycle;
\fill[ Emissing ] (1.185, 0.395) -- +( -88.3441389163 : 0.365 cm) arc ( -88.3441389163 : -270.0 : 0.365 cm)-- cycle;
\fill[draw=black,EmissingP ] (1.185, 0.395) -- +( -88.3441389163 : 0.365 cm) arc ( -88.3441389163 : -270.0 : 0.365 cm)-- cycle;
\fill[ Eexact ] (1.975, 0.395) -- +( 90.0 : 0.365 cm) arc ( 90.0 : -148.211593167 : 0.365 cm)-- cycle;
\fill[draw=black,EexactP ] (1.975, 0.395) -- +( 90.0 : 0.365 cm) arc ( 90.0 : -148.211593167 : 0.365 cm)-- cycle;
\fill[ Einclusive ] (1.975, 0.395) -- +( -148.211593167 : 0.365 cm) arc ( -148.211593167 : -159.763255662 : 0.365 cm)-- cycle;
\fill[draw=black,EinclusiveP ] (1.975, 0.395) -- +( -148.211593167 : 0.365 cm) arc ( -148.211593167 : -159.763255662 : 0.365 cm)-- cycle;
\fill[ Efalse ] (1.975, 0.395) -- +( -159.763255662 : 0.365 cm) arc ( -159.763255662 : -159.804526589 : 0.365 cm)-- cycle;
\fill[draw=black,EfalseP ] (1.975, 0.395) -- +( -159.763255662 : 0.365 cm) arc ( -159.763255662 : -159.804526589 : 0.365 cm)-- cycle;
\fill[ Emissing ] (1.975, 0.395) -- +( -159.804526589 : 0.365 cm) arc ( -159.804526589 : -270.0 : 0.365 cm)-- cycle;
\fill[draw=black,EmissingP ] (1.975, 0.395) -- +( -159.804526589 : 0.365 cm) arc ( -159.804526589 : -270.0 : 0.365 cm)-- cycle;
\fill[ Eexact ] (2.765, 0.395) -- +( 90.0 : 0.365 cm) arc ( 90.0 : -62.6492191344 : 0.365 cm)-- cycle;
\fill[draw=black,EexactP ] (2.765, 0.395) -- +( 90.0 : 0.365 cm) arc ( 90.0 : -62.6492191344 : 0.365 cm)-- cycle;
\fill[ Einclusive ] (2.765, 0.395) -- +( -62.6492191344 : 0.365 cm) arc ( -62.6492191344 : -70.554358714 : 0.365 cm)-- cycle;
\fill[draw=black,EinclusiveP ] (2.765, 0.395) -- +( -62.6492191344 : 0.365 cm) arc ( -62.6492191344 : -70.554358714 : 0.365 cm)-- cycle;
\fill[ Efalse ] (2.765, 0.395) -- +( -70.554358714 : 0.365 cm) arc ( -70.554358714 : -70.6489211764 : 0.365 cm)-- cycle;
\fill[draw=black,EfalseP ] (2.765, 0.395) -- +( -70.554358714 : 0.365 cm) arc ( -70.554358714 : -70.6489211764 : 0.365 cm)-- cycle;
\fill[ Emissing ] (2.765, 0.395) -- +( -70.6489211764 : 0.365 cm) arc ( -70.6489211764 : -270.0 : 0.365 cm)-- cycle;
\fill[draw=black,EmissingP ] (2.765, 0.395) -- +( -70.6489211764 : 0.365 cm) arc ( -70.6489211764 : -270.0 : 0.365 cm)-- cycle;
\fill[ Eexact ] (3.555, 0.395) -- +( 90.0 : 0.365 cm) arc ( 90.0 : -58.2460297728 : 0.365 cm)-- cycle;
\fill[draw=black,EexactP ] (3.555, 0.395) -- +( 90.0 : 0.365 cm) arc ( 90.0 : -58.2460297728 : 0.365 cm)-- cycle;
\fill[ Einclusive ] (3.555, 0.395) -- +( -58.2460297728 : 0.365 cm) arc ( -58.2460297728 : -65.9937999192 : 0.365 cm)-- cycle;
\fill[draw=black,EinclusiveP ] (3.555, 0.395) -- +( -58.2460297728 : 0.365 cm) arc ( -58.2460297728 : -65.9937999192 : 0.365 cm)-- cycle;
\fill[ Efalse ] (3.555, 0.395) -- +( -65.9937999192 : 0.365 cm) arc ( -65.9937999192 : -66.0490412406 : 0.365 cm)-- cycle;
\fill[draw=black,EfalseP ] (3.555, 0.395) -- +( -65.9937999192 : 0.365 cm) arc ( -65.9937999192 : -66.0490412406 : 0.365 cm)-- cycle;
\fill[ Emissing ] (3.555, 0.395) -- +( -66.0490412406 : 0.365 cm) arc ( -66.0490412406 : -270.0 : 0.365 cm)-- cycle;
\fill[draw=black,EmissingP ] (3.555, 0.395) -- +( -66.0490412406 : 0.365 cm) arc ( -66.0490412406 : -270.0 : 0.365 cm)-- cycle;
\fill[ Eexact ] (4.345000000000001, 0.395) -- +( 90.0 : 0.365 cm) arc ( 90.0 : -118.069491992 : 0.365 cm)-- cycle;
\fill[draw=black,EexactP ] (4.345000000000001, 0.395) -- +( 90.0 : 0.365 cm) arc ( 90.0 : -118.069491992 : 0.365 cm)-- cycle;
\fill[ Einclusive ] (4.345000000000001, 0.395) -- +( -118.069491992 : 0.365 cm) arc ( -118.069491992 : -129.472627097 : 0.365 cm)-- cycle;
\fill[draw=black,EinclusiveP ] (4.345000000000001, 0.395) -- +( -118.069491992 : 0.365 cm) arc ( -118.069491992 : -129.472627097 : 0.365 cm)-- cycle;
\fill[ Efalse ] (4.345000000000001, 0.395) -- +( -129.472627097 : 0.365 cm) arc ( -129.472627097 : -129.494476107 : 0.365 cm)-- cycle;
\fill[draw=black,EfalseP ] (4.345000000000001, 0.395) -- +( -129.472627097 : 0.365 cm) arc ( -129.472627097 : -129.494476107 : 0.365 cm)-- cycle;
\fill[ Emissing ] (4.345000000000001, 0.395) -- +( -129.494476107 : 0.365 cm) arc ( -129.494476107 : -270.0 : 0.365 cm)-- cycle;
\fill[draw=black,EmissingP ] (4.345000000000001, 0.395) -- +( -129.494476107 : 0.365 cm) arc ( -129.494476107 : -270.0 : 0.365 cm)-- cycle;
\fill[ Eexact ] (5.135, 0.395) -- +( 90.0 : 0.365 cm) arc ( 90.0 : -248.024932027 : 0.365 cm)-- cycle;
\fill[draw=black,EexactP ] (5.135, 0.395) -- +( 90.0 : 0.365 cm) arc ( 90.0 : -248.024932027 : 0.365 cm)-- cycle;
\fill[ Einclusive ] (5.135, 0.395) -- +( -248.024932027 : 0.365 cm) arc ( -248.024932027 : -264.341900511 : 0.365 cm)-- cycle;
\fill[draw=black,EinclusiveP ] (5.135, 0.395) -- +( -248.024932027 : 0.365 cm) arc ( -248.024932027 : -264.341900511 : 0.365 cm)-- cycle;
\fill[ Efalse ] (5.135, 0.395) -- +( -264.341900511 : 0.365 cm) arc ( -264.341900511 : -264.341900511 : 0.365 cm)-- cycle;
\fill[draw=black,EfalseP ] (5.135, 0.395) -- +( -264.341900511 : 0.365 cm) arc ( -264.341900511 : -264.341900511 : 0.365 cm)-- cycle;
\fill[ Emissing ] (5.135, 0.395) -- +( -264.341900511 : 0.365 cm) arc ( -264.341900511 : -270.0 : 0.365 cm)-- cycle;
\fill[draw=black,EmissingP ] (5.135, 0.395) -- +( -264.341900511 : 0.365 cm) arc ( -264.341900511 : -270.0 : 0.365 cm)-- cycle;
\fill[ Eexact ] (5.925000000000001, 0.395) -- +( 90.0 : 0.365 cm) arc ( 90.0 : -27.1169731174 : 0.365 cm)-- cycle;
\fill[draw=black,EexactP ] (5.925000000000001, 0.395) -- +( 90.0 : 0.365 cm) arc ( 90.0 : -27.1169731174 : 0.365 cm)-- cycle;
\fill[ Einclusive ] (5.925000000000001, 0.395) -- +( -27.1169731174 : 0.365 cm) arc ( -27.1169731174 : -35.538117942 : 0.365 cm)-- cycle;
\fill[draw=black,EinclusiveP ] (5.925000000000001, 0.395) -- +( -27.1169731174 : 0.365 cm) arc ( -27.1169731174 : -35.538117942 : 0.365 cm)-- cycle;
\fill[ Efalse ] (5.925000000000001, 0.395) -- +( -35.538117942 : 0.365 cm) arc ( -35.538117942 : -35.5742460984 : 0.365 cm)-- cycle;
\fill[draw=black,EfalseP ] (5.925000000000001, 0.395) -- +( -35.538117942 : 0.365 cm) arc ( -35.538117942 : -35.5742460984 : 0.365 cm)-- cycle;
\fill[ Emissing ] (5.925000000000001, 0.395) -- +( -35.5742460984 : 0.365 cm) arc ( -35.5742460984 : -270.0 : 0.365 cm)-- cycle;
\fill[draw=black,EmissingP ] (5.925000000000001, 0.395) -- +( -35.5742460984 : 0.365 cm) arc ( -35.5742460984 : -270.0 : 0.365 cm)-- cycle;
\fill[ Eexact ] (6.715, 0.395) -- +( 90.0 : 0.365 cm) arc ( 90.0 : -39.4742703408 : 0.365 cm)-- cycle;
\fill[draw=black,EexactP ] (6.715, 0.395) -- +( 90.0 : 0.365 cm) arc ( 90.0 : -39.4742703408 : 0.365 cm)-- cycle;
\fill[ Einclusive ] (6.715, 0.395) -- +( -39.4742703408 : 0.365 cm) arc ( -39.4742703408 : -46.21401933 : 0.365 cm)-- cycle;
\fill[draw=black,EinclusiveP ] (6.715, 0.395) -- +( -39.4742703408 : 0.365 cm) arc ( -39.4742703408 : -46.21401933 : 0.365 cm)-- cycle;
\fill[ Efalse ] (6.715, 0.395) -- +( -46.21401933 : 0.365 cm) arc ( -46.21401933 : -46.2925487029 : 0.365 cm)-- cycle;
\fill[draw=black,EfalseP ] (6.715, 0.395) -- +( -46.21401933 : 0.365 cm) arc ( -46.21401933 : -46.2925487029 : 0.365 cm)-- cycle;
\fill[ Emissing ] (6.715, 0.395) -- +( -46.2925487029 : 0.365 cm) arc ( -46.2925487029 : -270.0 : 0.365 cm)-- cycle;
\fill[draw=black,EmissingP ] (6.715, 0.395) -- +( -46.2925487029 : 0.365 cm) arc ( -46.2925487029 : -270.0 : 0.365 cm)-- cycle;
\path (7.2285, 1.185) node[right] {\small  $\mathsf{ Overlap }$};
\fill[ Eexact ] (0.395, 1.185) -- +( 90.0 : 0.365 cm) arc ( 90.0 : -220.71454883 : 0.365 cm)-- cycle;
\fill[draw=black,EexactP ] (0.395, 1.185) -- +( 90.0 : 0.365 cm) arc ( 90.0 : -220.71454883 : 0.365 cm)-- cycle;
\fill[ Einclusive ] (0.395, 1.185) -- +( -220.71454883 : 0.365 cm) arc ( -220.71454883 : -239.396970769 : 0.365 cm)-- cycle;
\fill[draw=black,EinclusiveP ] (0.395, 1.185) -- +( -220.71454883 : 0.365 cm) arc ( -220.71454883 : -239.396970769 : 0.365 cm)-- cycle;
\fill[ Efalse ] (0.395, 1.185) -- +( -239.396970769 : 0.365 cm) arc ( -239.396970769 : -245.4372305 : 0.365 cm)-- cycle;
\fill[draw=black,EfalseP ] (0.395, 1.185) -- +( -239.396970769 : 0.365 cm) arc ( -239.396970769 : -245.4372305 : 0.365 cm)-- cycle;
\fill[ Emissing ] (0.395, 1.185) -- +( -245.4372305 : 0.365 cm) arc ( -245.4372305 : -270.0 : 0.365 cm)-- cycle;
\fill[draw=black,EmissingP ] (0.395, 1.185) -- +( -245.4372305 : 0.365 cm) arc ( -245.4372305 : -270.0 : 0.365 cm)-- cycle;
\fill[ Eexact ] (1.185, 1.185) -- +( 90.0 : 0.365 cm) arc ( 90.0 : -220.818327761 : 0.365 cm)-- cycle;
\fill[draw=black,EexactP ] (1.185, 1.185) -- +( 90.0 : 0.365 cm) arc ( 90.0 : -220.818327761 : 0.365 cm)-- cycle;
\fill[ Einclusive ] (1.185, 1.185) -- +( -220.818327761 : 0.365 cm) arc ( -220.818327761 : -234.9867299 : 0.365 cm)-- cycle;
\fill[draw=black,EinclusiveP ] (1.185, 1.185) -- +( -220.818327761 : 0.365 cm) arc ( -220.818327761 : -234.9867299 : 0.365 cm)-- cycle;
\fill[ Efalse ] (1.185, 1.185) -- +( -234.9867299 : 0.365 cm) arc ( -234.9867299 : -235.07686643 : 0.365 cm)-- cycle;
\fill[draw=black,EfalseP ] (1.185, 1.185) -- +( -234.9867299 : 0.365 cm) arc ( -234.9867299 : -235.07686643 : 0.365 cm)-- cycle;
\fill[ Emissing ] (1.185, 1.185) -- +( -235.07686643 : 0.365 cm) arc ( -235.07686643 : -270.0 : 0.365 cm)-- cycle;
\fill[draw=black,EmissingP ] (1.185, 1.185) -- +( -235.07686643 : 0.365 cm) arc ( -235.07686643 : -270.0 : 0.365 cm)-- cycle;
\fill[ Eexact ] (1.975, 1.185) -- +( 90.0 : 0.365 cm) arc ( 90.0 : -194.214686543 : 0.365 cm)-- cycle;
\fill[draw=black,EexactP ] (1.975, 1.185) -- +( 90.0 : 0.365 cm) arc ( 90.0 : -194.214686543 : 0.365 cm)-- cycle;
\fill[ Einclusive ] (1.975, 1.185) -- +( -194.214686543 : 0.365 cm) arc ( -194.214686543 : -206.940122181 : 0.365 cm)-- cycle;
\fill[draw=black,EinclusiveP ] (1.975, 1.185) -- +( -194.214686543 : 0.365 cm) arc ( -194.214686543 : -206.940122181 : 0.365 cm)-- cycle;
\fill[ Efalse ] (1.975, 1.185) -- +( -206.940122181 : 0.365 cm) arc ( -206.940122181 : -250.17386641 : 0.365 cm)-- cycle;
\fill[draw=black,EfalseP ] (1.975, 1.185) -- +( -206.940122181 : 0.365 cm) arc ( -206.940122181 : -250.17386641 : 0.365 cm)-- cycle;
\fill[ Emissing ] (1.975, 1.185) -- +( -250.17386641 : 0.365 cm) arc ( -250.17386641 : -270.0 : 0.365 cm)-- cycle;
\fill[draw=black,EmissingP ] (1.975, 1.185) -- +( -250.17386641 : 0.365 cm) arc ( -250.17386641 : -270.0 : 0.365 cm)-- cycle;
\fill[ Eexact ] (2.765, 1.185) -- +( 90.0 : 0.365 cm) arc ( 90.0 : -235.745100845 : 0.365 cm)-- cycle;
\fill[draw=black,EexactP ] (2.765, 1.185) -- +( 90.0 : 0.365 cm) arc ( 90.0 : -235.745100845 : 0.365 cm)-- cycle;
\fill[ Einclusive ] (2.765, 1.185) -- +( -235.745100845 : 0.365 cm) arc ( -235.745100845 : -245.573245723 : 0.365 cm)-- cycle;
\fill[draw=black,EinclusiveP ] (2.765, 1.185) -- +( -235.745100845 : 0.365 cm) arc ( -235.745100845 : -245.573245723 : 0.365 cm)-- cycle;
\fill[ Efalse ] (2.765, 1.185) -- +( -245.573245723 : 0.365 cm) arc ( -245.573245723 : -245.863990011 : 0.365 cm)-- cycle;
\fill[draw=black,EfalseP ] (2.765, 1.185) -- +( -245.573245723 : 0.365 cm) arc ( -245.573245723 : -245.863990011 : 0.365 cm)-- cycle;
\fill[ Emissing ] (2.765, 1.185) -- +( -245.863990011 : 0.365 cm) arc ( -245.863990011 : -270.0 : 0.365 cm)-- cycle;
\fill[draw=black,EmissingP ] (2.765, 1.185) -- +( -245.863990011 : 0.365 cm) arc ( -245.863990011 : -270.0 : 0.365 cm)-- cycle;
\fill[ Eexact ] (3.555, 1.185) -- +( 90.0 : 0.365 cm) arc ( 90.0 : -235.077695248 : 0.365 cm)-- cycle;
\fill[draw=black,EexactP ] (3.555, 1.185) -- +( 90.0 : 0.365 cm) arc ( 90.0 : -235.077695248 : 0.365 cm)-- cycle;
\fill[ Einclusive ] (3.555, 1.185) -- +( -235.077695248 : 0.365 cm) arc ( -235.077695248 : -248.682443982 : 0.365 cm)-- cycle;
\fill[draw=black,EinclusiveP ] (3.555, 1.185) -- +( -235.077695248 : 0.365 cm) arc ( -235.077695248 : -248.682443982 : 0.365 cm)-- cycle;
\fill[ Efalse ] (3.555, 1.185) -- +( -248.682443982 : 0.365 cm) arc ( -248.682443982 : -248.775445194 : 0.365 cm)-- cycle;
\fill[draw=black,EfalseP ] (3.555, 1.185) -- +( -248.682443982 : 0.365 cm) arc ( -248.682443982 : -248.775445194 : 0.365 cm)-- cycle;
\fill[ Emissing ] (3.555, 1.185) -- +( -248.775445194 : 0.365 cm) arc ( -248.775445194 : -270.0 : 0.365 cm)-- cycle;
\fill[draw=black,EmissingP ] (3.555, 1.185) -- +( -248.775445194 : 0.365 cm) arc ( -248.775445194 : -270.0 : 0.365 cm)-- cycle;
\fill[ Eexact ] (4.345000000000001, 1.185) -- +( 90.0 : 0.365 cm) arc ( 90.0 : -238.739528312 : 0.365 cm)-- cycle;
\fill[draw=black,EexactP ] (4.345000000000001, 1.185) -- +( 90.0 : 0.365 cm) arc ( 90.0 : -238.739528312 : 0.365 cm)-- cycle;
\fill[ Einclusive ] (4.345000000000001, 1.185) -- +( -238.739528312 : 0.365 cm) arc ( -238.739528312 : -252.020312752 : 0.365 cm)-- cycle;
\fill[draw=black,EinclusiveP ] (4.345000000000001, 1.185) -- +( -238.739528312 : 0.365 cm) arc ( -238.739528312 : -252.020312752 : 0.365 cm)-- cycle;
\fill[ Efalse ] (4.345000000000001, 1.185) -- +( -252.020312752 : 0.365 cm) arc ( -252.020312752 : -252.073569715 : 0.365 cm)-- cycle;
\fill[draw=black,EfalseP ] (4.345000000000001, 1.185) -- +( -252.020312752 : 0.365 cm) arc ( -252.020312752 : -252.073569715 : 0.365 cm)-- cycle;
\fill[ Emissing ] (4.345000000000001, 1.185) -- +( -252.073569715 : 0.365 cm) arc ( -252.073569715 : -270.0 : 0.365 cm)-- cycle;
\fill[draw=black,EmissingP ] (4.345000000000001, 1.185) -- +( -252.073569715 : 0.365 cm) arc ( -252.073569715 : -270.0 : 0.365 cm)-- cycle;
\fill[ Eexact ] (5.135, 1.185) -- +( 90.0 : 0.365 cm) arc ( 90.0 : -250.489406454 : 0.365 cm)-- cycle;
\fill[draw=black,EexactP ] (5.135, 1.185) -- +( 90.0 : 0.365 cm) arc ( 90.0 : -250.489406454 : 0.365 cm)-- cycle;
\fill[ Einclusive ] (5.135, 1.185) -- +( -250.489406454 : 0.365 cm) arc ( -250.489406454 : -266.945228202 : 0.365 cm)-- cycle;
\fill[draw=black,EinclusiveP ] (5.135, 1.185) -- +( -250.489406454 : 0.365 cm) arc ( -250.489406454 : -266.945228202 : 0.365 cm)-- cycle;
\fill[ Efalse ] (5.135, 1.185) -- +( -266.945228202 : 0.365 cm) arc ( -266.945228202 : -266.945228202 : 0.365 cm)-- cycle;
\fill[draw=black,EfalseP ] (5.135, 1.185) -- +( -266.945228202 : 0.365 cm) arc ( -266.945228202 : -266.945228202 : 0.365 cm)-- cycle;
\fill[ Emissing ] (5.135, 1.185) -- +( -266.945228202 : 0.365 cm) arc ( -266.945228202 : -270.0 : 0.365 cm)-- cycle;
\fill[draw=black,EmissingP ] (5.135, 1.185) -- +( -266.945228202 : 0.365 cm) arc ( -266.945228202 : -270.0 : 0.365 cm)-- cycle;
\fill[ Eexact ] (5.925000000000001, 1.185) -- +( 90.0 : 0.365 cm) arc ( 90.0 : -224.041043775 : 0.365 cm)-- cycle;
\fill[draw=black,EexactP ] (5.925000000000001, 1.185) -- +( 90.0 : 0.365 cm) arc ( 90.0 : -224.041043775 : 0.365 cm)-- cycle;
\fill[ Einclusive ] (5.925000000000001, 1.185) -- +( -224.041043775 : 0.365 cm) arc ( -224.041043775 : -235.029915208 : 0.365 cm)-- cycle;
\fill[draw=black,EinclusiveP ] (5.925000000000001, 1.185) -- +( -224.041043775 : 0.365 cm) arc ( -224.041043775 : -235.029915208 : 0.365 cm)-- cycle;
\fill[ Efalse ] (5.925000000000001, 1.185) -- +( -235.029915208 : 0.365 cm) arc ( -235.029915208 : -235.140270304 : 0.365 cm)-- cycle;
\fill[draw=black,EfalseP ] (5.925000000000001, 1.185) -- +( -235.029915208 : 0.365 cm) arc ( -235.029915208 : -235.140270304 : 0.365 cm)-- cycle;
\fill[ Emissing ] (5.925000000000001, 1.185) -- +( -235.140270304 : 0.365 cm) arc ( -235.140270304 : -270.0 : 0.365 cm)-- cycle;
\fill[draw=black,EmissingP ] (5.925000000000001, 1.185) -- +( -235.140270304 : 0.365 cm) arc ( -235.140270304 : -270.0 : 0.365 cm)-- cycle;
\fill[ Eexact ] (6.715, 1.185) -- +( 90.0 : 0.365 cm) arc ( 90.0 : -239.779968504 : 0.365 cm)-- cycle;
\fill[draw=black,EexactP ] (6.715, 1.185) -- +( 90.0 : 0.365 cm) arc ( 90.0 : -239.779968504 : 0.365 cm)-- cycle;
\fill[ Einclusive ] (6.715, 1.185) -- +( -239.779968504 : 0.365 cm) arc ( -239.779968504 : -254.788033838 : 0.365 cm)-- cycle;
\fill[draw=black,EinclusiveP ] (6.715, 1.185) -- +( -239.779968504 : 0.365 cm) arc ( -239.779968504 : -254.788033838 : 0.365 cm)-- cycle;
\fill[ Efalse ] (6.715, 1.185) -- +( -254.788033838 : 0.365 cm) arc ( -254.788033838 : -254.98435727 : 0.365 cm)-- cycle;
\fill[draw=black,EfalseP ] (6.715, 1.185) -- +( -254.788033838 : 0.365 cm) arc ( -254.788033838 : -254.98435727 : 0.365 cm)-- cycle;
\fill[ Emissing ] (6.715, 1.185) -- +( -254.98435727 : 0.365 cm) arc ( -254.98435727 : -270.0 : 0.365 cm)-- cycle;
\fill[draw=black,EmissingP ] (6.715, 1.185) -- +( -254.98435727 : 0.365 cm) arc ( -254.98435727 : -270.0 : 0.365 cm)-- cycle;
\end{tikzpicture}
  \caption{\label{fig:iuphar-alignment-precision}Alignment precision (GtoPdb).}
\end{figure}
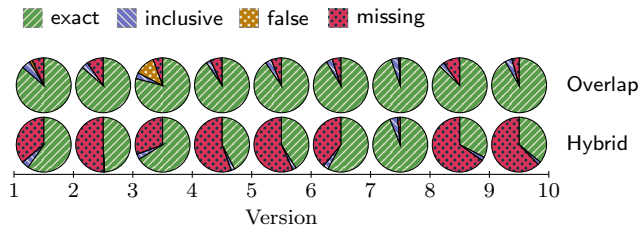

We can now use the ground truth (\textsf{GtoPdb}) to substantiate
these observations and to evaluate the precision of the alignments. In
the ground truth a node is aligned to at most one other node, while
the overlap and hybrid alignment may map a node to multiple
nodes. Consequently, for every alignment we identify the numbers of:
\textsf{exact} matches -- any node that is aligned to the same set of
nodes as the ground truth, \textsf{inclusive} matches -- any node that
is aligned to a set of nodes that properly includes the node indicated
by the ground truth, \textsf{missing} matches -- any node that is
mapped to a set of nodes that does not include the node indicated by
the ground truth, and \textsf{false} matches -- any node that is
aligned to a nonempty set of nodes while the ground truth does not
align the node to any node. We present the results in
Figure~\ref{fig:iuphar-alignment-precision}.
Clearly, the results confirm that the overlap significantly
outperforms the hybrid alignment. We point out that the relative
change between versions (as we read it by comparing the values
\textsf{GtoPdb} and \textsf{Total} in
Figure~\ref{fig:iuphar-alignment-lines}) is not a good indicator of
the performance of the hybrid alignment e.g., the hybrid exhibits
better precision when aligning versions $3$ and $4$, where the
relative change is significant, than it does when aligning versions
$5$ and $6$, where the relative change is smaller. Interestingly, for
the overlap alignment there is a dependence between the relative
change between two versions and the precision of the overlap
alignment. In particular, the overlap alignment between versions $3$
and $4$ has the worst precision overall and even aligns incorrectly a
significant number of nodes.  Our investigations of why nodes are
falsely aligned indicate that it mainly happens to nodes that are
inserted and deleted between the two versions and that the main reason
of false alignment of a node is the number of previously existing
nodes present in its outbound neighbourhood. For instance, out of
$177$K inserted URIs $31$K are falsely aligned, and in case of the
falsely aligned URIs on average only $9$\% of outbound nodes are newly
inserted nodes while in case of inserted nodes that are correctly
unaligned this average is higher and reaches $31$\%.

\begin{figure}[bht]
  \centering
  \begin{tikzpicture}[>=latex]
    \path[use as bounding box] (0.25,-0.85) rectangle (8.625,1.75);
\path ( 0.625 ,-0.125) node[below] {\small  $ 0.35 $};
\path ( 1.875 ,-0.125) node[below] {\small  $ 0.45 $};
\path ( 3.125 ,-0.125) node[below] {\small  $ 0.55 $};
\path ( 4.375 ,-0.125) node[below] {\small  $ 0.65 $};
\path ( 5.625 ,-0.125) node[below] {\small  $ 0.75 $};
\path ( 6.875 ,-0.125) node[below] {\small  $ 0.85 $};
\path ( 8.125 ,-0.125) node[below] {\small  $ 0.95 $};
\path ( 4.375 , -0.75) node[below] {\small  Threshold value $\theta$};
\begin{scope}[yshift= 1.5 cm,xshift=0cm]
\begin{scope}[yshift=0cm,xshift= 0.0 cm]
\draw[gray, very thin,Eexact] (0.125,0.125) rectangle (0.375,0.375);
\fill[EexactP ] (0.125,0.125) rectangle (0.375,0.375);
\path (0.375,0.25) node[right] {\small  $\mathsf{ exact }$};
\end{scope}
\begin{scope}[yshift=0cm,xshift= 1.25 cm]
\draw[gray, very thin,Einclusive] (0.125,0.125) rectangle (0.375,0.375);
\fill[EinclusiveP ] (0.125,0.125) rectangle (0.375,0.375);
\path (0.375,0.25) node[right] {\small  $\mathsf{ inclusive }$};
\end{scope}
\begin{scope}[yshift=0cm,xshift= 2.875 cm]
\draw[gray, very thin,Efalse] (0.125,0.125) rectangle (0.375,0.375);
\fill[EfalseP ] (0.125,0.125) rectangle (0.375,0.375);
\path (0.375,0.25) node[right] {\small  $\mathsf{ false }$};
\end{scope}
\begin{scope}[yshift=0cm,xshift= 4.0625 cm]
\draw[gray, very thin,Emissing] (0.125,0.125) rectangle (0.375,0.375);
\fill[EmissingP ] (0.125,0.125) rectangle (0.375,0.375);
\path (0.375,0.25) node[right] {\small  $\mathsf{ missing }$};
\end{scope}
\end{scope}
\fill[ Eexact ] (0.625, 0.625) -- +( 90.0 : 0.5 cm) arc ( 90.0 : -161.124496743 : 0.5 cm)-- cycle;
\fill[draw=black,EexactP ] (0.625, 0.625) -- +( 90.0 : 0.5 cm) arc ( 90.0 : -161.124496743 : 0.5 cm)-- cycle;
\fill[ Einclusive ] (0.625, 0.625) -- +( -161.124496743 : 0.5 cm) arc ( -161.124496743 : -218.451912764 : 0.5 cm)-- cycle;
\fill[draw=black,EinclusiveP ] (0.625, 0.625) -- +( -161.124496743 : 0.5 cm) arc ( -161.124496743 : -218.451912764 : 0.5 cm)-- cycle;
\fill[ Efalse ] (0.625, 0.625) -- +( -218.451912764 : 0.5 cm) arc ( -218.451912764 : -269.871990176 : 0.5 cm)-- cycle;
\fill[draw=black,EfalseP ] (0.625, 0.625) -- +( -218.451912764 : 0.5 cm) arc ( -218.451912764 : -269.871990176 : 0.5 cm)-- cycle;
\fill[ Emissing ] (0.625, 0.625) -- +( -269.871990176 : 0.5 cm) arc ( -269.871990176 : -270.0 : 0.5 cm)-- cycle;
\fill[draw=black,EmissingP ] (0.625, 0.625) -- +( -269.871990176 : 0.5 cm) arc ( -269.871990176 : -270.0 : 0.5 cm)-- cycle;
\fill[ Eexact ] (1.875, 0.625) -- +( 90.0 : 0.5 cm) arc ( 90.0 : -191.183721689 : 0.5 cm)-- cycle;
\fill[draw=black,EexactP ] (1.875, 0.625) -- +( 90.0 : 0.5 cm) arc ( 90.0 : -191.183721689 : 0.5 cm)-- cycle;
\fill[ Einclusive ] (1.875, 0.625) -- +( -191.183721689 : 0.5 cm) arc ( -191.183721689 : -223.338670314 : 0.5 cm)-- cycle;
\fill[draw=black,EinclusiveP ] (1.875, 0.625) -- +( -191.183721689 : 0.5 cm) arc ( -191.183721689 : -223.338670314 : 0.5 cm)-- cycle;
\fill[ Efalse ] (1.875, 0.625) -- +( -223.338670314 : 0.5 cm) arc ( -223.338670314 : -269.221448446 : 0.5 cm)-- cycle;
\fill[draw=black,EfalseP ] (1.875, 0.625) -- +( -223.338670314 : 0.5 cm) arc ( -223.338670314 : -269.221448446 : 0.5 cm)-- cycle;
\fill[ Emissing ] (1.875, 0.625) -- +( -269.221448446 : 0.5 cm) arc ( -269.221448446 : -270.0 : 0.5 cm)-- cycle;
\fill[draw=black,EmissingP ] (1.875, 0.625) -- +( -269.221448446 : 0.5 cm) arc ( -269.221448446 : -270.0 : 0.5 cm)-- cycle;
\fill[ Eexact ] (3.125, 0.625) -- +( 90.0 : 0.5 cm) arc ( 90.0 : -202.730487634 : 0.5 cm)-- cycle;
\fill[draw=black,EexactP ] (3.125, 0.625) -- +( 90.0 : 0.5 cm) arc ( 90.0 : -202.730487634 : 0.5 cm)-- cycle;
\fill[ Einclusive ] (3.125, 0.625) -- +( -202.730487634 : 0.5 cm) arc ( -202.730487634 : -223.030187623 : 0.5 cm)-- cycle;
\fill[draw=black,EinclusiveP ] (3.125, 0.625) -- +( -202.730487634 : 0.5 cm) arc ( -202.730487634 : -223.030187623 : 0.5 cm)-- cycle;
\fill[ Efalse ] (3.125, 0.625) -- +( -223.030187623 : 0.5 cm) arc ( -223.030187623 : -267.233448882 : 0.5 cm)-- cycle;
\fill[draw=black,EfalseP ] (3.125, 0.625) -- +( -223.030187623 : 0.5 cm) arc ( -223.030187623 : -267.233448882 : 0.5 cm)-- cycle;
\fill[ Emissing ] (3.125, 0.625) -- +( -267.233448882 : 0.5 cm) arc ( -267.233448882 : -270.0 : 0.5 cm)-- cycle;
\fill[draw=black,EmissingP ] (3.125, 0.625) -- +( -267.233448882 : 0.5 cm) arc ( -267.233448882 : -270.0 : 0.5 cm)-- cycle;
\fill[ Eexact ] (4.375, 0.625) -- +( 90.0 : 0.5 cm) arc ( 90.0 : -206.819806936 : 0.5 cm)-- cycle;
\fill[draw=black,EexactP ] (4.375, 0.625) -- +( 90.0 : 0.5 cm) arc ( 90.0 : -206.819806936 : 0.5 cm)-- cycle;
\fill[ Einclusive ] (4.375, 0.625) -- +( -206.819806936 : 0.5 cm) arc ( -206.819806936 : -220.708523107 : 0.5 cm)-- cycle;
\fill[draw=black,EinclusiveP ] (4.375, 0.625) -- +( -206.819806936 : 0.5 cm) arc ( -206.819806936 : -220.708523107 : 0.5 cm)-- cycle;
\fill[ Efalse ] (4.375, 0.625) -- +( -220.708523107 : 0.5 cm) arc ( -220.708523107 : -264.457804169 : 0.5 cm)-- cycle;
\fill[draw=black,EfalseP ] (4.375, 0.625) -- +( -220.708523107 : 0.5 cm) arc ( -220.708523107 : -264.457804169 : 0.5 cm)-- cycle;
\fill[ Emissing ] (4.375, 0.625) -- +( -264.457804169 : 0.5 cm) arc ( -264.457804169 : -270.0 : 0.5 cm)-- cycle;
\fill[draw=black,EmissingP ] (4.375, 0.625) -- +( -264.457804169 : 0.5 cm) arc ( -264.457804169 : -270.0 : 0.5 cm)-- cycle;
\fill[ Eexact ] (5.625, 0.625) -- +( 90.0 : 0.5 cm) arc ( 90.0 : -194.214686543 : 0.5 cm)-- cycle;
\fill[draw=black,EexactP ] (5.625, 0.625) -- +( 90.0 : 0.5 cm) arc ( 90.0 : -194.214686543 : 0.5 cm)-- cycle;
\fill[ Einclusive ] (5.625, 0.625) -- +( -194.214686543 : 0.5 cm) arc ( -194.214686543 : -206.940122181 : 0.5 cm)-- cycle;
\fill[draw=black,EinclusiveP ] (5.625, 0.625) -- +( -194.214686543 : 0.5 cm) arc ( -194.214686543 : -206.940122181 : 0.5 cm)-- cycle;
\fill[ Efalse ] (5.625, 0.625) -- +( -206.940122181 : 0.5 cm) arc ( -206.940122181 : -250.17386641 : 0.5 cm)-- cycle;
\fill[draw=black,EfalseP ] (5.625, 0.625) -- +( -206.940122181 : 0.5 cm) arc ( -206.940122181 : -250.17386641 : 0.5 cm)-- cycle;
\fill[ Emissing ] (5.625, 0.625) -- +( -250.17386641 : 0.5 cm) arc ( -250.17386641 : -270.0 : 0.5 cm)-- cycle;
\fill[draw=black,EmissingP ] (5.625, 0.625) -- +( -250.17386641 : 0.5 cm) arc ( -250.17386641 : -270.0 : 0.5 cm)-- cycle;
\fill[ Eexact ] (6.875, 0.625) -- +( 90.0 : 0.5 cm) arc ( 90.0 : -162.499028462 : 0.5 cm)-- cycle;
\fill[draw=black,EexactP ] (6.875, 0.625) -- +( 90.0 : 0.5 cm) arc ( 90.0 : -162.499028462 : 0.5 cm)-- cycle;
\fill[ Einclusive ] (6.875, 0.625) -- +( -162.499028462 : 0.5 cm) arc ( -162.499028462 : -174.664158804 : 0.5 cm)-- cycle;
\fill[draw=black,EinclusiveP ] (6.875, 0.625) -- +( -162.499028462 : 0.5 cm) arc ( -162.499028462 : -174.664158804 : 0.5 cm)-- cycle;
\fill[ Efalse ] (6.875, 0.625) -- +( -174.664158804 : 0.5 cm) arc ( -174.664158804 : -207.254900437 : 0.5 cm)-- cycle;
\fill[draw=black,EfalseP ] (6.875, 0.625) -- +( -174.664158804 : 0.5 cm) arc ( -174.664158804 : -207.254900437 : 0.5 cm)-- cycle;
\fill[ Emissing ] (6.875, 0.625) -- +( -207.254900437 : 0.5 cm) arc ( -207.254900437 : -270.0 : 0.5 cm)-- cycle;
\fill[draw=black,EmissingP ] (6.875, 0.625) -- +( -207.254900437 : 0.5 cm) arc ( -207.254900437 : -270.0 : 0.5 cm)-- cycle;
\fill[ Eexact ] (8.125, 0.625) -- +( 90.0 : 0.5 cm) arc ( 90.0 : -148.399760613 : 0.5 cm)-- cycle;
\fill[draw=black,EexactP ] (8.125, 0.625) -- +( 90.0 : 0.5 cm) arc ( 90.0 : -148.399760613 : 0.5 cm)-- cycle;
\fill[ Einclusive ] (8.125, 0.625) -- +( -148.399760613 : 0.5 cm) arc ( -148.399760613 : -159.951423109 : 0.5 cm)-- cycle;
\fill[draw=black,EinclusiveP ] (8.125, 0.625) -- +( -148.399760613 : 0.5 cm) arc ( -148.399760613 : -159.951423109 : 0.5 cm)-- cycle;
\fill[ Efalse ] (8.125, 0.625) -- +( -159.951423109 : 0.5 cm) arc ( -159.951423109 : -159.992694036 : 0.5 cm)-- cycle;
\fill[draw=black,EfalseP ] (8.125, 0.625) -- +( -159.951423109 : 0.5 cm) arc ( -159.951423109 : -159.992694036 : 0.5 cm)-- cycle;
\fill[ Emissing ] (8.125, 0.625) -- +( -159.992694036 : 0.5 cm) arc ( -159.992694036 : -270.0 : 0.5 cm)-- cycle;
\fill[draw=black,EmissingP ] (8.125, 0.625) -- +( -159.992694036 : 0.5 cm) arc ( -159.992694036 : -270.0 : 0.5 cm)-- cycle;
\end{tikzpicture}

  \caption{\label{fig:iuphar-threshold}Overlap alignment between versions 3 and 4 (GtoPdb) for different threshold values.}
\end{figure}
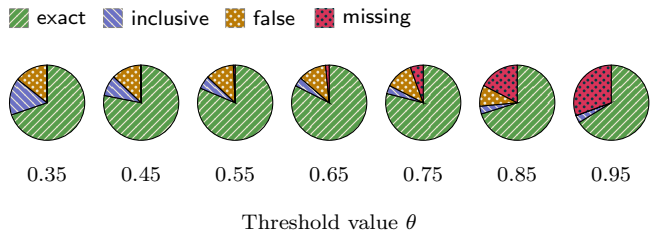

In Figure~\ref{fig:iuphar-threshold} we further investigate how the
precision can be controlled with the threshold value used by the
overlap alignment (between versions $3$ and $4$)
The findings are as expected: the lower the threshold value the lower
the number of \textsf{missing} matches but also the higher number of
\textsf{false} and \textsf{inclusive} matches.  The number of exact
matches reaches maximal value at threshold equal $0.65$.

\subsection{DBpedia}
\begin{figure}[htb]
  \centering
\begin{tikzpicture}[>=latex]
\path[use as bounding box] (-1.25,-0.5) rectangle ( 7.25 , 4 );
\draw[fill=white] (0,0) -- ( 5.75 ,0) -- ( 5.75 , 4 ) -- (0, 4 ) -- cycle;
\path ( 0.479166666667 ,0) node[below] {\small  $ 1 $};
\draw ( 0.479166666667 ,  -0.09375 ) -- ( 0.479166666667 , 0.09375 );
\path ( 1.4375 ,0) node[below] {\small  $ 2 $};
\draw ( 1.4375 ,  -0.09375 ) -- ( 1.4375 , 0.09375 );
\path ( 2.39583333333 ,0) node[below] {\small  $ 3 $};
\draw ( 2.39583333333 ,  -0.09375 ) -- ( 2.39583333333 , 0.09375 );
\path ( 3.35416666667 ,0) node[below] {\small  $ 4 $};
\draw ( 3.35416666667 ,  -0.09375 ) -- ( 3.35416666667 , 0.09375 );
\path ( 4.3125 ,0) node[below] {\small  $ 5 $};
\draw ( 4.3125 ,  -0.09375 ) -- ( 4.3125 , 0.09375 );
\path ( 5.27083333333 ,0) node[below] {\small  $ 6 $};
\draw ( 5.27083333333 ,  -0.09375 ) -- ( 5.27083333333 , 0.09375 );
\path ( 2.875 , -0.65) node {\small  Version};
\draw ( 0.046875 , 1.33333333333  ) -- (0.0, 1.33333333333 );
\path (0.0, 1.33333333333  ) node[left] {\small $ 5 $M};
\draw ( 0.046875 , 2.66666666667  ) -- (0.0, 2.66666666667 );
\path (0.0, 2.66666666667  ) node[left] {\small $ 10 $M};
\draw ( 0.046875 , 4.0  ) -- (0.0, 4.0 );
\path (0.0, 4.0  ) node[left] {\small $ 15 $M};
\draw ( 5.703125 , 0.709219858156  ) -- ( 5.75 , 0.709219858156 );
\path ( 5.75 , 0.709219858156  ) node[right] {\small $ 50 $};
\draw ( 5.703125 , 1.41843971631  ) -- ( 5.75 , 1.41843971631 );
\path ( 5.75 , 1.41843971631  ) node[right] {\small $ 100 $};
\draw ( 5.703125 , 2.12765957447  ) -- ( 5.75 , 2.12765957447 );
\path ( 5.75 , 2.12765957447  ) node[right] {\small $ 150 $};
\draw ( 5.703125 , 2.83687943262  ) -- ( 5.75 , 2.83687943262 );
\path ( 5.75 , 2.83687943262  ) node[right] {\small $ 200 $};
\draw ( 5.703125 , 3.54609929078  ) -- ( 5.75 , 3.54609929078 );
\path ( 5.75 , 3.54609929078  ) node[right] {\small $ 250 $};
\begin{scope}[yshift= -1.15 cm,xshift= -1.2 cm]
\begin{scope}[yshift=1.125cm,xshift=0.0cm]
\draw[Eedges] (0.125,0.125) rectangle (0.375,0.375);
\draw[EedgesP] (0.125,0.125) rectangle (0.375,0.375);
\path (0.25,0.375) node[right,rotate=90] {\small  \sf Edges};
\end{scope}
\begin{scope}[yshift=2.55cm,xshift=0.0cm]
\draw[Euris] (0.125,0.125) rectangle (0.375,0.375);
\draw[EurisP] (0.125,0.125) rectangle (0.375,0.375);
\path (0.25,0.375) node[right,rotate=90] {\small  \sf URIs};
\end{scope}
\begin{scope}[yshift=3.75cm,xshift=0.0cm]
\draw[Eliterals] (0.125,0.125) rectangle (0.375,0.375);
\draw[EliteralsP] (0.125,0.125) rectangle (0.375,0.375);
\path (0.25,0.375) node[right,rotate=90] {\small  \sf Literals};
\end{scope}
\end{scope}
\begin{scope}[yshift= -1.15 cm,xshift= 6.5 cm]
\begin{scope}[yshift=0.725cm,xshift=0.0cm]
\draw[ETrivial] (0.125,0.125) rectangle (0.375,0.375);
\draw[ETrivialP] (0.125,0.125) rectangle (0.375,0.375);
\path (0.25,0.375) node[right,rotate=90] {\small  \sf Trivial};
\end{scope}
\begin{scope}[yshift=2.25cm,xshift=0.0cm]
\draw[EHybrid] (0.125,0.125) rectangle (0.375,0.375);
\draw[EHybridP] (0.125,0.125) rectangle (0.375,0.375);
\path (0.25,0.375) node[right,rotate=90] {\small  \sf Hybrid};
\end{scope}
\begin{scope}[yshift=3.75cm,xshift=0.0cm]
\draw[EOverlap] (0.125,0.125) rectangle (0.375,0.375);
\draw[EOverlapP] (0.125,0.125) rectangle (0.375,0.375);
\path (0.25,0.375) node[right,rotate=90] {\small  \sf Overlap};
\end{scope}
\path (0.75, 1.6 ) node[right,rotate=90] {\small  Execution time (sec.)};
\end{scope}
\draw[ Eedges ] ( 0.479166666667 ,0.0) rectangle ( 0.638541666667 , 2.03733333333 );
\draw[EedgesP] ( 0.479166666667 ,0.0) rectangle ( 0.638541666667 , 2.03733333333 );
\draw[ Euris ] ( 0.319791666667 ,0.0) rectangle ( 0.479166666667 , 0.616 );
\draw[EurisP] ( 0.319791666667 ,0.0) rectangle ( 0.479166666667 , 0.616 );
\draw[ Eliterals ] ( 0.319791666667 , 0.616 ) rectangle ( 0.479166666667 , 0.699466666667 );
\draw[EliteralsP] ( 0.319791666667 , 0.616 ) rectangle ( 0.479166666667 , 0.699466666667 );
\draw[ Eedges ] ( 1.4375 ,0.0) rectangle ( 1.596875 , 2.44533333333 );
\draw[EedgesP] ( 1.4375 ,0.0) rectangle ( 1.596875 , 2.44533333333 );
\draw[ Euris ] ( 1.278125 ,0.0) rectangle ( 1.4375 , 0.712 );
\draw[EurisP] ( 1.278125 ,0.0) rectangle ( 1.4375 , 0.712 );
\draw[ Eliterals ] ( 1.278125 , 0.712 ) rectangle ( 1.4375 , 0.816 );
\draw[EliteralsP] ( 1.278125 , 0.712 ) rectangle ( 1.4375 , 0.816 );
\draw[ Eedges ] ( 2.39583333333 ,0.0) rectangle ( 2.55520833333 , 2.59733333333 );
\draw[EedgesP] ( 2.39583333333 ,0.0) rectangle ( 2.55520833333 , 2.59733333333 );
\draw[ Euris ] ( 2.23645833333 ,0.0) rectangle ( 2.39583333333 , 0.746666666667 );
\draw[EurisP] ( 2.23645833333 ,0.0) rectangle ( 2.39583333333 , 0.746666666667 );
\draw[ Eliterals ] ( 2.23645833333 , 0.746666666667 ) rectangle ( 2.39583333333 , 0.8576 );
\draw[EliteralsP] ( 2.23645833333 , 0.746666666667 ) rectangle ( 2.39583333333 , 0.8576 );
\draw[ Eedges ] ( 3.35416666667 ,0.0) rectangle ( 3.51354166667 , 3.008 );
\draw[EedgesP] ( 3.35416666667 ,0.0) rectangle ( 3.51354166667 , 3.008 );
\draw[ Euris ] ( 3.19479166667 ,0.0) rectangle ( 3.35416666667 , 0.842666666667 );
\draw[EurisP] ( 3.19479166667 ,0.0) rectangle ( 3.35416666667 , 0.842666666667 );
\draw[ Eliterals ] ( 3.19479166667 , 0.842666666667 ) rectangle ( 3.35416666667 , 0.969066666667 );
\draw[EliteralsP] ( 3.19479166667 , 0.842666666667 ) rectangle ( 3.35416666667 , 0.969066666667 );
\draw[ Eedges ] ( 4.3125 ,0.0) rectangle ( 4.471875 , 3.344 );
\draw[EedgesP] ( 4.3125 ,0.0) rectangle ( 4.471875 , 3.344 );
\draw[ Euris ] ( 4.153125 ,0.0) rectangle ( 4.3125 , 0.898666666667 );
\draw[EurisP] ( 4.153125 ,0.0) rectangle ( 4.3125 , 0.898666666667 );
\draw[ Eliterals ] ( 4.153125 , 0.898666666667 ) rectangle ( 4.3125 , 1.03573333333 );
\draw[EliteralsP] ( 4.153125 , 0.898666666667 ) rectangle ( 4.3125 , 1.03573333333 );
\draw[ Eedges ] ( 5.27083333333 ,0.0) rectangle ( 5.43020833333 , 3.65066666667 );
\draw[EedgesP] ( 5.27083333333 ,0.0) rectangle ( 5.43020833333 , 3.65066666667 );
\draw[ Euris ] ( 5.11145833333 ,0.0) rectangle ( 5.27083333333 , 0.978666666667 );
\draw[EurisP] ( 5.11145833333 ,0.0) rectangle ( 5.27083333333 , 0.978666666667 );
\draw[ Eliterals ] ( 5.11145833333 , 0.978666666667 ) rectangle ( 5.27083333333 , 1.12933333333 );
\draw[EliteralsP] ( 5.11145833333 , 0.978666666667 ) rectangle ( 5.27083333333 , 1.12933333333 );
\draw[ ETrivial ] ( 0.852083333333 ,0.0) rectangle ( 1.06458333333 , 0.466765957447 );
\draw[ETrivialP] ( 0.852083333333 ,0.0) rectangle ( 1.06458333333 , 0.466765957447 );
\draw[ EHybrid ] ( 0.852083333333 , 0.466765957447 ) rectangle ( 1.06458333333 , 0.992468085106 );
\draw[EHybridP] ( 0.852083333333 , 0.466765957447 ) rectangle ( 1.06458333333 , 0.992468085106 );
\draw[ EOverlap ] ( 0.852083333333 , 0.992468085106 ) rectangle ( 1.06458333333 , 2.09690780142 );
\draw[EOverlapP] ( 0.852083333333 , 0.992468085106 ) rectangle ( 1.06458333333 , 2.09690780142 );
\draw[ ETrivial ] ( 1.81041666667 ,0.0) rectangle ( 2.02291666667 , 0.449120567376 );
\draw[ETrivialP] ( 1.81041666667 ,0.0) rectangle ( 2.02291666667 , 0.449120567376 );
\draw[ EHybrid ] ( 1.81041666667 , 0.449120567376 ) rectangle ( 2.02291666667 , 0.845304964539 );
\draw[EHybridP] ( 1.81041666667 , 0.449120567376 ) rectangle ( 2.02291666667 , 0.845304964539 );
\draw[ EOverlap ] ( 1.81041666667 , 0.845304964539 ) rectangle ( 2.02291666667 , 1.05858156028 );
\draw[EOverlapP] ( 1.81041666667 , 0.845304964539 ) rectangle ( 2.02291666667 , 1.05858156028 );
\draw[ ETrivial ] ( 2.76875 ,0.0) rectangle ( 2.98125 , 0.568014184397 );
\draw[ETrivialP] ( 2.76875 ,0.0) rectangle ( 2.98125 , 0.568014184397 );
\draw[ EHybrid ] ( 2.76875 , 0.568014184397 ) rectangle ( 2.98125 , 1.26439716312 );
\draw[EHybridP] ( 2.76875 , 0.568014184397 ) rectangle ( 2.98125 , 1.26439716312 );
\draw[ EOverlap ] ( 2.76875 , 1.26439716312 ) rectangle ( 2.98125 , 3.14530496454 );
\draw[EOverlapP] ( 2.76875 , 1.26439716312 ) rectangle ( 2.98125 , 3.14530496454 );
\draw[ ETrivial ] ( 3.72708333333 ,0.0) rectangle ( 3.93958333333 , 0.792581560284 );
\draw[ETrivialP] ( 3.72708333333 ,0.0) rectangle ( 3.93958333333 , 0.792581560284 );
\draw[ EHybrid ] ( 3.72708333333 , 0.792581560284 ) rectangle ( 3.93958333333 , 1.31158865248 );
\draw[EHybridP] ( 3.72708333333 , 0.792581560284 ) rectangle ( 3.93958333333 , 1.31158865248 );
\draw[ EOverlap ] ( 3.72708333333 , 1.31158865248 ) rectangle ( 3.93958333333 , 2.0619858156 );
\draw[EOverlapP] ( 3.72708333333 , 1.31158865248 ) rectangle ( 3.93958333333 , 2.0619858156 );
\draw[ ETrivial ] ( 4.68541666667 ,0.0) rectangle ( 4.89791666667 , 0.766113475177 );
\draw[ETrivialP] ( 4.68541666667 ,0.0) rectangle ( 4.89791666667 , 0.766113475177 );
\draw[ EHybrid ] ( 4.68541666667 , 0.766113475177 ) rectangle ( 4.89791666667 , 1.73730496454 );
\draw[EHybridP] ( 4.68541666667 , 0.766113475177 ) rectangle ( 4.89791666667 , 1.73730496454 );
\draw[ EOverlap ] ( 4.68541666667 , 1.73730496454 ) rectangle ( 4.89791666667 , 3.63916312057 );
\draw[EOverlapP] ( 4.68541666667 , 1.73730496454 ) rectangle ( 4.89791666667 , 3.63916312057 );
\end{tikzpicture}
  \caption{\label{fig:dbpedia}Evaluation time on a subset of DBpedia}  
\end{figure}
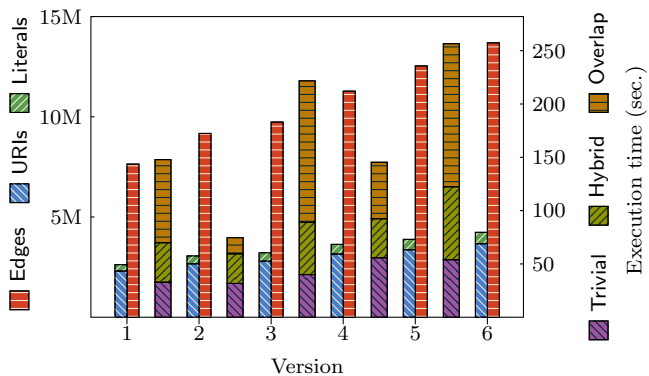

To evaluate scalability of our methods, we report in
Figure~\ref{fig:dbpedia} the running times on a subset of DBpedia
containing category information (including hierarchical information
and Wikipedia article categorization). We used versions 3.0 through
3.5, run our experiments on a MacBook Pro with 2.3 GHz Intel Core i7,
16 GB RAM, and 512 GB SSD, our (single-thread) implementation was in
Python 2.7.
The RDF graphs progressively grow from 2.6M nodes and 7.6M edges to
4.2M nodes and 13.7M edges. The performance of our methods
fluctuates mainly due to the varying number of overlapping nodes
between two consecutive versions. The general trend appears
proportional to the size of the input graphs. Furthermore, the
execution times are in line with those presented in~\cite{SNLPZ13},
which suggest that our methods should scale to larger datasets, using
methods such as MapReduce.
\section{Conclusions and future work}
\label{sec:concl}

We have presented an approach of identifying nodes corresponding to
the same entity in different versions of the same graph, a task whose
importance has recently been identified~\cite{LaTz14}. Our approach is
based on the classical notion of bisimulation, which essentially
defines the identity of a node based on the identity of its outbound
neighborhood. This approach is particularly suited to align the nodes of
two graphs that follow the same structure, and evolving RDF is one
such real-life scenario. We have
also presented a generalization of the basic bisimulation technique
that produces weighted partitions, which allows to approximate
similarity measures on nodes without incurring the high complexity of
computing similarity measures, a matter of obvious importance
when handling large RDF graphs.

While our methods are relatively straightforward, they have been
designed with simplicity and possible extensions in mind. In the
future, we would like to explore variants of our approach where only
selected parts of the outbound neighborhood are used, for instance
specified by a notion of a key for graph, possibly allowing to align
nodes of graphs following different structure, or even a broader
context of the node involving its inbound neighborhood and the triple
where the node is used as predicate, possibly allowing to better align
them. Our experiments show, however, that the presented methods
perform very well in the scenario of evolving RDF database.

An interesting question arises: can the (constructed) alignments be
used to construct compact representations of all
versions of an RDF database? One way of approaching this 
would be to decorate triples with intervals that represent versions
where the triple was present. Our preliminary observations suggest
that triples tend to enter and leave with their subject.
with its subject, and moving the interval information where possible
to the subject nodes could offer further
improvements on space requirements.

{\bf Acknowledgements} We are grateful to  Simon Jupp and Tony Burdett
for discussions on the EFO database and to Joanna Sharman and Jamie
Davies for the GtoPdb data. The referees also made many useful
comments. This work was funded by the EU DIACHRON project, the EPSRC
SOCIAM project and  NSF IIS 1302212:  Citing Structured and Evolving Data.

\end{document}